\documentclass{psp-rv975x65-arxiv}
\usepackage[square]{psp-rv-van}            
\usepackage{psp-index}             
\usepackage{subfigure}

\begin{document}

\chapter[Bohmian Pathways into Chemistry]
{Bohmian Pathways into Chemistry:\\ A Brief Overview \label{as:chap4}}

\author{\'Angel S. Sanz}

\address{Department of Optics, Faculty of Physical Sciences,\\
Universidad Complutense de Madrid,\\
Pza.\ Ciencias 1, Ciudad Universitaria 28040 - Madrid, Spain\\
a.s.sanz@fis.ucm.es}

\begin{abstract}
Perhaps because of the popularity that trajectory-based methodologies have
always had in Chemistry and the important role they have played, Bohmian
mechanics has been increasingly accepted within this community, particularly
in those areas of the theoretical chemistry based on quantum mechanics,
e.g., quantum chemistry, chemical physics, or physical chemistry.
From a historical perspective, this evolution is remarkably interesting,
particularly when the scarce applications of Madelung's former
hydrodynamical formulation, dating back to the late 1960s and the 1970s,
are compared with the many different applications available at present.
As also happens with classical methodologies, Bohmian trajectories are
essentially used to described and analyze the evolution of chemical systems,
to design and implement new computational propagation techniques,
or a combination of both.
In the first case, Bohmian trajectories have the advantage that they
avoid invoking typical quantum-classical correspondence to interpret
the corresponding phenomenon or process, while in the second case
quantum-mechanical effects appear by themselves, without the
necessity to include artificially quantization conditions.
Rather than providing an exhaustive revision and analysis of all these
applications (excellent monographs on the issue are available in the
literature for the interested reader, which can be consulted in the
bibliography here supplied), this Chapter has been prepared in a way
that it may serve the reader to acquire a general view (or impression)
on how Bohmian mechanics has permeated the different traditional
levels or pathways to approach molecular systems in Chemistry:
electronic structure, molecular dynamics and statistical mechanics.
This is done with the aid of some illustrative examples
--- theoretical developments in some cases and numerical simulations
in other cases.
\end{abstract}


\body


\section{Introduction}
\label{as:sec4.1}

Quantum mechanics has acquired a prominent role over the years in different
areas of the theoretical chemistry, including the quantum chemistry, the physical
chemistry or the chemical physics, as well as other areas at the borderline between
Chemistry and Physics.
It is not surprising that the idea of analyzing and understanding
quantum-mechanically chemical systems by means of a hydrodynamic language,
which dates back to the late 1960s and early 1970s \cite{as:wyatt2-1,as:wyatt2-2,as:wyatt2-3,%
as:hirschfelder1-1,as:hirschfelder1-2,as:hirschfelder1-3}, has also permeated
all these areas.
Bohmian mechanics \cite{as:bohm-1,%
as:bohm-2,as:bohm-3,as:takabayasi-1,as:takabayasi-2,as:holland,%
as:duerr-1,as:duerr-2,as:wyatt1,as:book-1,as:book-2}, the widespread
denomination that is now used for the hydrodynamic picture or reformulation of
the Schr\"odinger equation \cite{as:sanz:AJP:2017} --- formerly suggested by
Madelung in 1926 \cite{as:madelung} ---,
has thus made its own pathways throughout Chemistry
\cite{as:wyatt1,as:book-2,as:chattaraj-bk,as:hughes-bk}.
This is not a minor issue taking into account that Chemistry is at present a sort of
crossroad for a number of overlapping disciplines, such as atomic and molecular
physics, optics and quantum optics, solid state physics, condensed matter
physics, chemical physics, nanotechnology, biochemistry or molecular chemistry,
and more recently also different branches associated with the so-called quantum
technologies \cite{as:milburn}, such as quantum information and computation
\cite{as:nielsen-chuan-bk,as:NSF-report-2016}, chemicals and materials quantum design
\cite{as:eberhart02,as:eberhart04,as:maranas,as:caflisch,as:carter} or quantum
biology \cite{as:nori,as:schulten,as:fleming07,as:fleming10}.

A question that naturally arises is why Bohmian mechanics now receives more attention
than about half a century ago, when it was hardly considered a worthless hidden-variable
theory almost relegated to oblivion --- a source of much controversy within the quantum
foundations community (something that unfortunately has survived to date without
a truly deep physical justification).
Perhaps a suitable answer can be given taking on a pragmatic view on what Bohmian
mechanics is and what Bohmian mechanics is not in simple terms, that is, at a very basic
(that is, pragmatic) level, without entering a lengthy discussion on the issue.
In this sense, on the one hand, we find a hydrodynamic model that describes the
evolution (actually, {\it diffusion}) of a quantum system in time throughout the corresponding
configuration space --- just in the same way classical hydrodynamic systems behave ---,
in compliance with Madelung's view \cite{as:madelung}.
The particularity of the quantum system versus a classical hydrodynamic one is that,
a priori, if no further specifications are made, quantum pure states evolve as ideal,
inviscid fluids \cite{as:birula}.
This does not introduce any new interpretation for the quantum mechanics based on
the presence or not of external observers, since the mathematics of the Schr\"odinger
equation are just the mathematics of the diffusion equation with a pure imaginary
diffusion constant, as noticed by F\"urth \cite{as:furth} or Comisar \cite{as:comisar}.
Actually, the idea of devising quantum systems as quantum fluids has also been a
viewpoint shared by others apart from Madelung, as Landau \cite{as:landau1} or
London \cite{as:london}, and is of common use in solid state physics and condensed
matter physics (actually, in many cases, Madelung is not even credited for it).

On the other hand, we have de Broglie's pilot wave theory \cite{as:broglie-1,%
as:broglie-2,as:broglie-3}, which after the 5th Solvay Conference and further
(rather tough) discussions has reached us as the (oversimplified) concept of de Broglie's
hypothesis: electrons (and any kind of material system, in general, from tiny electrons or
neutrons to large, complex organic molecular compounds) display both
particle-like and wave-like properties in the same way that light is also characterized
by both properties (after Einstein's 1905 explanation for the photoelectric effect).
However, the idea formerly supported by de Broglie in 1924 (and taken as a basis
by Schr\"odinger in the derivation of his famous equation) is that quantum systems
consist of two subjects: a carrier (pilot) wave field and a carried singularity
(point-like) representing the electron.
Unlike Madelung's proposal, de Broglie's one already implies a certain way to
understand or ``interpret'' quantum systems, and therefore it found too much
opposition at the Solvay Conference, eventually being rejected in its former
version.
However, from the hydrodynamic perspective mentioned above, there is nothing
wrong with it, since de Broglie's approach is, in the end, only a way to describe
with a quantum language the evolution of, saying it in simple terms, a leaf released
on the surface of a stream: the leaf conveys information about the evolution of the
stream, but tells nothing about the evolution of the elementary compounds of such
a stream, which is still in compliance with the Copenhagen (Bohr's) view.
These streamlines thus allow a visualization of the evolution of the quantum
fluid just as the path pursued by a (classical) tracer particle
\cite{as:swinney,as:gilreath} (e.g., charcoal dust) is used to follow the flow
of a classical fluid \cite{as:sanz:AJP:2012,as:sanz:JPhysConfSer:2012}.
Since 2005, experiments with bouncing droplets in vibrating fluids, independently
performed by the groups of Yves Couder and Emmanuel Fort in Paris
\cite{as:couder:Nature:2005,as:couder:JFluidMech:2006,as:couder:PRL:2006,%
as:couder:PhysFluid:2006,as:couder:PRL:2009,as:couder:PNAS:2010,as:couder:EPN:2010}
and John Bush in Boston \cite{as:bush:PNAS:2010,as:bush:PRE:2013,as:bush:ARFM:2015},
have become nice illustrations of a classical analog de Broglie's proposal.

Madelung and de Broglie devised their approaches during the early days of quantum
mechanics, when the theory was still being developed.
About a quarter of a century afterwards, Bohm proposed another analogous model to
challenge the formal and conceptual grounds of quantum mechanics, which was already a widely
accepted theory.
It is interesting to note, though, that Bohm's approach, formerly suggested as a
counterproof to von Neumann's theorem on the impossibility of (local) hidden-variable
theories compatible with quantum mechanics \cite{as:bohm-1}, gathers the flavors of
both Madelung's hydrodynamic formulation and the insight of de Broglie's pilot-wave
theory.
The most remarkable aspect of Bohm's suggestion is the fact that it emphasizes very
clearly that quantum mechanics is intrinsically nonlocal, thus introducing a cutting-edge
view of this theory that transcends the understanding of physical systems according to
the principle of locality.
This was the main reason for its rejection by the mid-1950s, until Bell decided to analyze
such implications and formalize them about a decade later, producing his brilliant proof on
the nonlocal nature of quantum mechanics in terms of an inequality.
Some applications of Bohm's started appearing by the late 1970s and during the 1980s and
1990s, where different physical problems typical of the quantum mechanics were translated
into Bohm's trajectory-based model.
Nonetheless, the model was strongly attached to interpretational (metaphysical-like)
connotations related to the measurement problem in quantum mechanics \cite{as:zurek-bk},
and therefore it was often neglected or not taken too seriously by the community.
The landscape started changing by the late 1990s, when it underwent a remarkable boost
through its applications and potential interests in Chemistry, not only to describe and explain
the dynamics exhibited by quantum systems, but also as a source of new computational
tools, where the key element is precisely the Bohmian trajectory.

Also the negative metaphysical (``surrealistic'' \cite{as:scully}) connotations associated
with Bohm's trajectories started changing, since, in the end, they are properly defined
from a formal viewpoint.
For those familiarized with the treatment and solution of partial differential equations,
such trajectories are just the curves arising from the method of characteristics, well-known in
mathematical physics to solve finite-order partial differential equations
\cite{as:courant-hilbert-bk}.
In quantum mechanics, the method of quantum characteristics is applied, by analogy
to its classical counterpart, to the flows in the configuration space defined by the
Weyl-Wigner transform \cite{as:weyl,as:wigner}, which eventually gives rise to the
appearance of an infinite-order, with respect to the classical case, through the Moyal
product \cite{as:moyal} (in the classical limit, the bracket that defines this product
becomes the Poisson bracket, which implies an order reduction due to the disappearance
of nonlocal correlations).
Alternatively, by virtue of a suitable decoupling between the phase and the amplitude
of the wave function, Schr\"odinger's equation can be recast in the form of a
differential equation susceptible to also be solved by the method of characteristics,
either in terms of real-valued trajectories \cite{as:wyatt1,as:lopreore-1,as:lopreore-2,as:lopreore-3}
or complex-valued ones \cite{as:tannor:JCP:2006,as:tannor:JCP-2:2007,as:tannor:JPCA:2007,%
as:tannor:CP:2007,as:chou:JCP:2006,as:chou:JCP-1:2008,as:chou:JCP-3:2008,as:chou:JCP:2010}.

From a historical perspective, it is remarkable the fact that Bohmian mechanics
entered Chemistry from Madelung's approach instead of Bohm's one, more closely
connected, perhaps, to Physics due to its deep metaphysical (interpretational)
connotations discussed above, which are absent in Madelung's one.
Madelung's approach, moreover, is in compliance with the tradition in Chemistry
of appealing to pictorial representations of the systems under study (the
pictorial view enabled by Bohmian mechanics was also a positive feature
remarked by Bell \cite{as:bell1}).
Setting as our ``coordinate origin'' the two categories pointed out by Wyatt \cite{as:wyatt1},
namely {\it analytic} and {\it synthetic}, this acceptance and future development is even
more understandable.
Within the analytic approach \cite{as:book-1,as:book-2}, first the wave
function is propagated and then the Bohmian trajectories are obtained
``on the fly'' from its evolution.
The goal of this approach is essentially interpretational, in a sense
that it is used to understand what is going on with the system along
its evolution in time.
This kind of information complements the information obtained by means
of other standard quantum techniques, thus providing a wider
understanding of the underlying phenomenology.
The synthetic approach \cite{as:wyatt1}, on the other hand, is aimed at
computing quantum-mechanical quantities ``on the fly'' by directly
synthesizing the Bohmian trajectories, that is, acting the other way around, without
any need for pre-determining the wave function.
In this case, the equations of motion for the trajectories are
integrated alongside the quantum Hamilton-Jacobi equation (which rules
the trajectory dynamics) and the continuity equation (which rules the
``ensemble'' dynamics).

The applications revisited in this Chapter constitute a brief account on
both approaches, analytic and synthetic, within Chemistry.
Due to the many problems where it
has been applied and their different nature, the discussion will be
limited to some illustrative examples that cover a series of aspects
that are discussed with more detail in monographs on the issue
\cite{as:wyatt1,as:book-2,as:chattaraj-bk,as:hughes-bk} as well as in
the bibliography here provided.
With this in mind, the Chapter has been organized as follows.
An overview on the different levels at which molecular systems are
traditionally studied is presented in Sec.~\ref{as:sec4.2}, starting from
the Born-Oppenheimer approximation \cite{as:BOA} and then revisiting
(in general terms) the methodologies available to deal with both
electronic configuration
\cite{as:bransden,as:levine,as:szabo,as:fulde,as:koch,as:martin,as:kaplan}
and dynamics, distinguishin in this latter case between molecular dynamics
for a few degree-of-freedom systems \cite{as:zhang,as:tannor} and
statistical mechanics (many-body problems)
\cite{as:louisell,as:mcquarrie,as:weiss,as:breuer,as:carmichael,as:percival}.
Analogously, a discussion on some fundamental aspects of Bohmian
mechanics is introduced in Sec.~\ref{as:sec4.3}.
Some specific illustrations of the application of Bohmian mechanics to
problems typical of the levels mentioned before are described in
Sec.~\ref{as:sec4.4}.
Finally, a general, summarizing reflection on the matter in Sec.~\ref{as:sec4.5}
concludes this Chapter.


\section{Approaching molecular systems at different levels}
\label{as:sec4.2}

Molecular systems typically include a rather large number of degrees
of freedom, which makes unaffordable their direct study.
Different approximations and/or approaches have thus been
considered to cope with this inconvenience.
Usually, given the different time-scales involved in the dynamics of
the different components, namely nuclei and electrons, the first
important approximation is the so-called {\it Born-Oppenheimer
approximation} \cite{as:BOA}, which allows us to simplify the study
of any molecular system (this notion is taken here in a broad sense,
thus including simple molecules, solid surfaces, polymeric
chains, clusters, crystalline structures, etc.) by splitting it into its
nuclear and electronic parts.
This division has led to three levels or types of methodologies in
Chemistry to study and explain molecular systems:
electronic structure \cite{as:bransden,as:levine,as:szabo,%
as:fulde,as:koch,as:martin,as:kaplan}, dynamics \cite{as:zhang,%
as:tannor} and statistics \cite{as:louisell,as:mcquarrie,%
as:weiss,as:breuer,as:carmichael,as:percival}.

Electronic structure methods (e.g., valence bond
theory, {\it ab initio} methods, density functional theory, etc.) are aimed
at obtaining properties associated with the electronic configuration,
such as the chemical bonding and intermolecular forces \cite{as:kaplan}.
Basically all these methods are based on quantum mechanics and are
included under what is known as quantum chemistry.
Molecular dynamics and statistical mechanics methods are based either
in classical mechanics or in quantum mechanics, and essentially constitute
the scope of the chemical physics and physical chemistry.
Molecular dynamics methods, e.g., wave-packet propagation methods,
are used to to describe the properties associated with the nuclear motion
(dynamics), necessary to understand chemical reaction processes, for
example.
In this case, the role of electrons manifests through effective
potential functions, namely potential energy surfaces,
generated from their bonding to the nuclei.
At this point, it is worth highlighting that, for practical purposes, a
division has been considered between a few degree-of-freedom
treatments, which are henceforth denoted as molecular dynamics
methods, and those to deal with many-body problems, which will be
referred to as statistical mechanics methods.
This convenient classification allows us to distinguish between methods
and problems where the full system dynamics is relevant, from those
where we are more interested in statistical properties or we only need
to focus our attention on a part of the total system (neglecting the
remaining ``environmental'' degrees of freedom).
Bearing this in mind, when the number of nuclei involved is relatively large,
we shall talk of statistical methods (e.g., molecular dynamics, Monte Carlo,
path integrals, etc.), which include methodologies and theories
developed to tackle open quantum systems, complex systems, systems
far from equilibrium, etc., all of them approximate.

Of course, sometimes the Born-Oppenheimer approximation cannot be used
and the above division can no longer be considered.
In such cases, one needs to consider a full treatment of the molecular
system (electronic plus nuclear parts) using ``on the fly'' methods that
take into account the interaction between electrons and nuclei at the
same time in a sort of two-step feedback process: electrons accommodate
to a certain nuclei configuration, which is then used to evolve the nuclei and,
from the latter new rearrangement, a new electron configuration is
determined, and so on, until the full dynamical-configurational simulation
is completed.
Dealing with this kind of systems is often complicated and highly
computationally demanding, although different methods have been
developed to deal with them \cite{as:nakai-2002,as:nakai-2007,as:nakai-2016},
but this is a topic that goes beyond the scope of this Chapter.


\subsection{The Born-Oppenheimer approximation}
\label{as:sec4.2.1}

Molecular systems are usually characterized by a large number of
degrees of freedom, which makes complicated and even computationally
prohibitive their study.
To circumvent this inconvenience we can take advantage of the
remarkable mass difference between nuclei and electrons
($m_n \gg m_e$), which physically translates into very different
time scales ruling their respective dynamics (electrons are faster
than nuclei) as well as very different kinetic energies
($\langle p_n^2 \rangle/2m_n \ll \langle p_e^2 \rangle/2m_e$).
Accordingly, at a first level of approximation, it is reasonable to
assume that nuclei are almost at rest, while electrons move
under the action of the effective potential field generated by
point-like particles (the nuclei) that occupy definite positions.
This is the basic idea behind the Born-Oppenheimer approximation,
which we can also find in other contexts where there is an
important asymmetry between time scales and that serves us to
simplify the dynamical study of different sets of degrees of
freedom.
For example, in molecular dynamics it can be used to describe
some degrees of freedom according to Newton's equation
(classical trajectories) under the action of some effective
generalized potential functions, while other degrees of freedom
evolve quantum-mechanically (Bohmian trajectories).
A simple illustrative example of this splitting will be seen with more
detail in Sec.~\ref{as:sec4.4.5}.
The same idea can also be found in the context of the theory of
open quantum systems \cite{as:weiss,as:breuer,as:carmichael,as:percival},
where large systems can be split up into two subsystems, each ruled
by its own dynamical time scales (see Sec.~\ref{as:sec4.2.4}).
Depending on the difference between these scales and the
coupling strength between the two subsystems, more or less
complex treatment can be devised to tackle the corresponding
problem.

More specifically, the Born-Oppenheimer approximation arises in
Chemistry as a need for solving one of its major problems: finding
solutions of the non-relativistic, time-independent Schr\"odinger
equation that describes a molecular system,
\begin{equation}
 \hat{H} \Phi_\ell (\vec{R}_1, \vec{R}_2, \ldots, \vec{R}_M,
  q_1, q_2, \ldots, q_N) =
  E_\ell \Phi_\ell (\vec{R}_1, \vec{R}_2, \ldots, \vec{R}_M,
  q_1, q_2, \ldots, q_N) .
 \label{as:eq1}
\end{equation}
The energy eigenvalues that we extract from this equation serve
us to determine the electronic configuration of the system and,
therefore, chemical and physical properties, such as bonding,
electric, magnetic, phase transitions and states of matter, geometry, etc.
In Eq.~(\ref{as:eq1}), $\Phi_\ell$ stands for the wave function of the
$\ell$th state of the molecular system, which depends on the $3N$
space coordinates $\{\vec{r}_i\}_{i=1}^N$ and the $N$ spin coordinates
$\{s_i\}_{i=1}^N$ of its electrons, collectively denoted as
$\{q_i\}_{i=1}^N$, and the $3M$ space coordinates of its nuclei
$\{\vec{R}_A\}_{A=1}^M$; $E_\ell$ is the eigenenergy associated with
the state $\Phi_\ell$; and $\hat{H}$ is the Hamiltonian describing the
$M$ nuclei and $N$ electrons as well as their interactions.
In the absence of external fields (e.g., gravitational,
electromagnetic, etc.), $\hat{H}$ reads (in atomic units) as
\begin{equation}
 \hat{H} = - \frac{1}{2} \sum_{i=1}^N \nabla_{\vec{r}_i}^2
   + \sum_{i=1}^N \sum_{j>i}^N \frac{1}{r_{ij}}
   - \frac{1}{2} \sum_{A=1}^M \frac{\nabla_{\vec{R}_A}^2}{M_A}
   + \sum_{A=1}^M \sum_{B>A}^M \frac{Z_A Z_B}{R_{AB}}
   - \sum_{i=1}^N \sum_{A=1}^M \frac{Z_A}{r_{iA}} ,
 \label{as:eq2}
\end{equation}
where $r_{ij} = |\vec{r}_i - \vec{r}_j|$ is the distance between the
$i$th and $j$electrons (the same holds for $r_{iA}$ and $R_{AB}$ for
the electron-nucleus and nucleus-nucleus distances, respectively).
The first two terms in Eq.~(\ref{as:eq2}) are the total electron kinetic
energy and the electron-electron repulsive potential interaction,
respectively, and represent the (full) electronic contribution to
the Hamiltonian, which we generically denote as $\hat{h}_e$.
Similarly, the third and fourth terms describe the total nuclear
kinetic energy and the nucleus-nucleus repulsive potential interaction,
respectively, which we gather under the label $\hat{H}_n$.
Finally, the fifth term gives the attractive electron-nucleus
electrostatic interaction, which will be labeled as $\hat{V}_{en}$,
since it couples the electronic and nuclear motions.
Accordingly, Eq.~(\ref{as:eq2}) can be expressed in a more compact
form as
\begin{equation}
 \hat{H} = \hat{H}_n + \hat{h}_e + \hat{V}_{en} ,
 \label{as:eq3}
\end{equation}
which has the same form of the generic potentials in the theory of open
quantum systems \cite{as:weiss,as:breuer,as:carmichael,as:percival},
\begin{equation}
 \hat{H} = \hat{H}_S + \hat{H}_E + \hat{V}_{\rm int} ,
 \label{as:eq4}
\end{equation}
where the full system is partitioned into two subsystems: the system
of interest ($S$) and a surrounding environment or bath ($E$), with
their interaction being described by the term
$\hat{V}_{\rm int}$ ($= \hat{V}_{en}$).

At this level of approximation (it is here where the Born-Oppenheimer
approximation comes into play), the nuclei are assumed to have zero
motion.
Therefore, for a certain fixed position of all the involved nuclei, i.e.,
what we call a nuclear configuration, only the second and third terms
in (\ref{as:eq3}) are relevant, i.e.,
\begin{equation}
 \hat{H}_e = \hat{h}_e + \hat{V}_{en} .
 \label{as:eq5}
\end{equation}
The description of the molecular system thus only
depends on the distribution of the electrons ({\it electronic cloud})
around the nuclei, which play the role of a fixed environment that
acts upon the electrons through $\hat{V}_{en}$, but that do not
feel any back-action from the electrons.
The total wave function of the system can then be {\it factorized} and
represented as the product of two separate wave functions, one accounting
for the electronic cloud and the other for the ensemble of nuclei.
That is, we have
\begin{eqnarray}
 & & \Phi_\ell (\vec{R}_1, \vec{R}_2, \ldots, \vec{R}_M,
  q_1, q_2, \ldots, q_N) \nonumber \\
 & & \qquad \qquad \qquad \qquad \approx
  \varphi_{\rm n,M}^{(\ell)} (\vec{R}_1, \vec{R}_2, \ldots, \vec{R}_M)
  \otimes \phi_{\rm e,N}^{(\ell)} (q_1, q_2, \ldots, q_N) ,
 \label{as:eq6}
\end{eqnarray}
where $\varphi_{\rm n,M}^{(\ell)}$ and $\phi_{\rm e,N}^{(\ell)}$
denote $M$- and $N$-particle wave functions, respectively.
Notice that each electronic coordinate depends parametrically on all the
nuclear ones, i.e., $q_i = q_i(\vec{R}_1, \vec{R}_2, \ldots, \vec{R}_M)$,
for $i = 1, 2, \ldots, N$, but not vice versa.
After this factorization, the total energy of the system, obtained from the
partitioned Hamiltonian (\ref{as:eq2}), reads as
\begin{equation}
 E_{\rm tot} = E_n + E_e ,
 \label{as:eq8}
\end{equation}
which, as can be seen, consists of the sum of two separate terms.
The first term,
\begin{equation}
 E_n = \sum_{A=1}^M \sum_{B>A}^M \frac{Z_A Z_B}{R_{AB}} ,
 \label{as:eq7}
\end{equation}
is a constant associated with the nucleus-nucleus Coulomb-type repulsive
interaction, which can be determined either classically or
quantum-mechanically, depending on the degree of sophistication
of the method considered (and the accuracy requested).
It is a constant, because it only depends on the position of the nuclei,
which is assumed to be, by hypothesis, fixed.
The second term,
\begin{equation}
 E_e^{(\ell)} = \langle\phi_e^{(\ell)}|\hat{H}_e |\phi_e^{(\ell)}\rangle ,
\end{equation}
corresponds to the expectation value of the Hamiltonian (\ref{as:eq5}) in
the electronic state $\phi_e^{(\ell)}$.
Unlike the nuclear term, the electronic one changes each time the
nuclear configuration changes; all values generated in this way provide
us with the potential energy surface associated with the molecular
system under study, from which we eventually determine dynamical
properties (e.g., ionization and photoionization, isomerization, reaction
and scattering rates, electron transfer properties, diffraction, etc.).


\subsection{Electronic configuration}
\label{as:sec4.2.2}

As seen above, once the nuclear and electronic parts of the full molecular
Hamiltonian are specified and split up by virtue of the Born-Oppenheimer
approximation, the next step consists in solving the Schr\"odinger
equation for the electronic Hamiltonian (\ref{as:eq5}).
The main goal of electronic structure methods is to obtain ``exact''
quantum energy states by further factorizing the problem, recasting the
total polyelectronic wave function $\phi_{e,N}^{(\ell)}$ as a product of
uncorrelated, single-electron wave functions, $\phi_{e,1}^{(\ell)}$:
\begin{equation}
 \phi_{\rm e,N}^{(\ell)} (q_1, q_2, \ldots, q_N) \quad
 \longrightarrow \quad \Pi_{j=1}^N \phi_{\rm e,1,j}^{(\ell)} (q_j) ,
 \label{as:eq9}
\end{equation}
where the products combine in the form of Slater determinants in order
to preserve the corresponding electron antisymmetry properties.
These wave functions are plugged into their corresponding {\it reduced}
Schr\"odinger equations, characterized by a single-particle Hamiltonian,
$\hat{h}_{e,j}$, which is a function of an effective potential where
the interaction with the other $N$-1 electrons is accounted for as
a sort of average or mean effective field --- in this sense, the $N$
electrons are commonly regarded as noninteracting.
Then, after a variational process in which the energy, assumed to be
a functional of the single-particle probability density, is required to
be an extremum, both the ``correct'' wave function
(\ref{as:eq9}) and energy (\ref{as:eq8}) are found.

After proceeding in that way, the electronic structure of the
molecular system is determined, which is the same to say how
electrons distribute throughout the system.
With this information we can obtain the system chemical properties,
such as chemical bonding, intermolecular interactions, electronic charge
distributions, dipole and multipole moments or vibrational/rotational
frequencies, for instance.
Now, only in the case of the hydrogen atom, the corresponding
Schr\"odinger equation can be exactly solved; for any other atomic
or molecular system, involving three or more particles, different
approximate approaches have been proposed in the
literature \cite{as:levine,as:szabo,as:fulde,as:koch,as:martin} to tackle
the problem of determining their electronic configurations.
These approaches or methodologies arise basically from two models
\cite{as:headgordon}: the {\it wave model} and the {\it density
matrix model}.

In the {\it wave model} each atom is assumed to be a point-like, positively
charged nucleus surrounded by an electronic cloud, which is associated
with a certain orbital and represents the electron probability
distribution in configuration space.
The two types of theories arisen from this model are the {\it valence
bond theories} and the {\it molecular orbital theories}.
Valence bond theories \cite{as:shaik} are based on the assumption that
orbitals primarily localize on each nuclei (atomic orbitals), thus they
essentially focus on pairwise interactions, and therefore establish a
direct link with the classical chemical models based on the drawings of
bonds typical of Lewis structures.
Though not very common nowadays, the concepts such as
{\it resonance} \cite{as:pauling1} (chemical aromaticity) and orbital
{\it hybridization} \cite{as:pauling2} arise from this theory.

Molecular orbital theories \cite{as:levine,as:szabo,as:fulde}, on the
contrary, consider delocalized orbitals covering the whole molecule
(molecular orbitals).
Thus, the molecular orbital wave function is expressed as a linear
combination of atomic orbitals (Slater determinants).
These atomic orbitals can be hydrogen-like or Gaussian functions.
The basic method within the molecular orbital approach is the
Hartree-Fock (HF) method, an {\it ab initio} approach based on assuming that
$\phi_{\rm e,N}^{(\ell)}$ is given by a Slater determinant, which leads
to a set of $N$ coupled monoelectronic Schr\"odinger equations.
In these equations, the electron-electron repulsion interactions
involved in $\hat{h}_e$ are accounted for by a {\it mean field}
interaction (i.e., the averaged action of the $N$-1 remaining
electrons), thus neglecting {\it electron correlation}.
In those cases where large molecules are considered and the HF method
becomes inoperative, one can apply the so-called semiempirical
molecular quantum-mechanical methods (e.g., the H\"uckel and extended
H\"uckel methods or the Pariser-Parr-Pople method), where a simple
Hamiltonian plus a set of fitting parameters are used to adjust the
experimental data.
On the other hand, within the {\it ab initio} stream, there are different
routes to tackle the problem of the electron correlation, and therefore to
improve the HF results, such as the post-Hartree-Fock methods (e.g.,
configuration interaction, coupled cluster or M{\o}ller-Plesset), the
multi-configurational self-consistent field or the multireference
configuration interaction.

Regarding the {\it density matrix model} \cite{as:koch,as:blum}, the interest
in developing this type of methodology arises as a need from the very
high computational demand involved in the study of large molecular
systems by means of {\it ab initio} methodologies.
Thus, although the latter are more accurate physically, this accuracy is
sacrificed in order to get computational advantages.
The first attempt in this direction is the Thomas-Fermi model
\cite{as:thomas,as:fermi}, where the electron kinetic energy of an
atom is expressed as a functional of the atom electronic distribution.
This model is the precursor of the modern {\it density functional
theory} \cite{as:koch} (DFT), where the fundamental physical
information about the molecular system is obtained from a
single-particle density in a three-dimensional space, derived
variationally either within a time-independent framework \cite{as:koch}
(ground state) or a time-dependent one \cite{as:marques,as:botti}
(excited states).
Nowadays, DFT is a very popular calculation method in quantum chemistry,
as well as in solid state physics, due to its computational simplicity
combined with its (numerical) accuracy.
Also within the density matrix model it is worth stressing the role
of those approaches which are aimed at including the electronic
correlation, utilizing for such a purpose a {\it reduced density
matrix} formalism \cite{as:nakatsuji-1,as:nakatsuji-2,as:nakatsuji-3,%
as:valdemoro-1,as:valdemoro-2,as:valdemoro-3,as:valdemoro-4,%
as:valdemoro-5,as:rdm-reviews-1,as:rdm-reviews-2,as:rdm-reviews-3,%
as:mazziotti-1,as:mazziotti-2,as:mazziotti-3,as:mazziotti-4}.

By changing the positions of the nuclei, one obtains different values
of the electronic energy, thus describing a multidimensional energy
surface parameterized in terms of the nuclei positions,
$\{\vec{R}_A\}_{A=1}^M$.
This is the so-called {\it potential energy surface} or
{\it Born-Oppenheimer surface}, which reflects the configuration of
the molecular system and determines its dynamical behavior (i.e., its
physical/chemical properties).
An alternative approach to obtain the potential energy surface of a
molecular system is by using {\it molecular mechanics} or {\it force
field methods}.
Unlike the previously described methods, molecular mechanics is not
based on quantum mechanics, but on a classical treatment of systems.
In this case, the system is modeled by a collection of atoms (by
``atom'' it is assumed a single unity rather than a true physical
atom) held together by spring-like force functions (force fields)
and a corresponding set of fitting parameters.
Each atom is characterized by a radius (usually, the van der Waals
one), a charge and polarizability, and the potential energy surface
is obtained by minimizing some energy functional at different nuclei
positions using some optimization method (e.g., the Metropolis
algorithm or the Monte Carlo method).
Strictly speaking, these methods cannot be considered electronic
structure ones, since electrons are not explicitly taken into account.
However, they are assumed to be in an optimal distribution around the
fixed nuclei positions, and therefore are implicitly responsible for
the nuclear structural arrangement.
Furthermore, in order to improve the performance of these methods,
hybrid quantum mechanical molecular mechanics approaches are considered
\cite{as:sherwood}, which combine ``on the fly'' molecular mechanics
with forces evaluated quantum-mechanically from either {\it ab initio} or
DFT calculations.

Finally, it is also worth mentioning the availability of schemes
based on topology, devised to analyze the structures (potential energy
surfaces) obtained and determine, for example, eventual {\it reaction
paths} or {\it minimum energy paths}, i.e., the paths connecting
reactants with products following steepest descent curves
\cite{as:born1} (curves with the fastest descent from a potential
hill to a valley).
These paths are used to understand and interpret chemical reactions as
well as to determine kinetic constants, and therefore reaction rates
without involving any dynamical simulation.
This is part of what is known as mathematical chemistry.


\subsection{Dynamics of ``small'' molecular systems}
\label{as:sec4.2.3}

Once the electronic structure --- or, equivalently, the effective interaction
among nuclei due to the combination of the electromagnetic interaction
and the quantum nature of the electrons --- is determined and characterized
by a potential energy surface, the next step consists in obtaining properties
associated with the motion of the nuclei, which occurs at slower timescales
--- actually, many times it happens that the motion refers to some associated
effective degrees of freedom, e.g., vibrational or rotational modes, rather
than the nuclei themselves, which allows to further simplify the study of
the full system.
This procedure not only provides dynamical information about the system,
but also allows us to test the accuracy of the potential energy surfaces,
previously obtained by means of the electronic structure methods described
in previous section, by comparing the corresponding simulations with the
experimental data recorded (e.g., the potential energy surface characterizing a certain
metal surface can be compared with the real surface by computing the
associated Bragg peaks and then comparing this result with the
experimental diffraction pattern).
And, through this procedure, we also test the accuracy and suitability of the
electronic configurations computed.

To tackle the problem of the quantum dynamics, there are basically
two approaches \cite{as:sanz1,as:sanz2}: time-independent and
time-dependent.
Due to the complexity involved in solving the time-dependent
Schr\"odinger equation, historically the time-independent approach was
first developed.
By the 1960s,
different time-dependent classical and quantum-mechanical numerical
schemes started to developed, this also being the starting point of
computational chemistry.
This is the reason why former calculations and numerical schemes in
electronic structure as well as in dynamics, with the obtention of
spectral lines or collision cross sections, were mostly developed
within the first 25 years of quantum mechanics.
However, in the case of the dynamics, it is worth stressing that part
of the work was already done: the time-independent Schr\"odinger
equation is analogous to the Helmholtz equation that appears in
classical wave theory and electromagnetism, so many solutions were
already known since the $XIX$th century \cite{as:morse}.
In the case of hard-wall like potential models, one deals with standard
boundary condition problems, either in bound systems to find spectral
lines, or in the continuum to determine scattering cross sections.

For space-dependent potentials, such as those describing bound
systems or tunneling problems \cite{as:razavy,as:ankerhold}, also
electromagnetism provided a well-known methodology: the JWKB
method \cite{as:JWKB-1,as:JWKB-2,as:JWKB-3,as:JWKB-4,as:JWKB-5}.
More recently, different numerical methods have been developed in
the literature to solve directly the time-independent Schr\"odinger
equation and determine the bound levels associated with (bound)
potentials \cite{as:cerjan,as:giordano}, which are based either in the
so-called ``shooting and matching'' strategies or in variational
techniques.
In either case one can discretize the space or consider some basis
functions (e.g., Gaussians, sinc-functions) and express the solution
as a linear combination of these functions.
Also, as an analytical tool, classical trajectories have been employed,
taking into account the correspondence between eigenstates (bound
states) and classical periodic orbits \cite{as:gutzwiller,as:heller}.

In the case of processes and phenomena in the continuum, soon after the
establishment of quantum mechanics, a quantum scattering theory was
developed \cite{as:QST-1,as:QST-2}.
Essentially, in this theory an incoming particle is represented by a
coherent plane wave, $e^{i\vec{k}\cdot\vec{r}}$, at some initial time
at $t \to -\infty$.
At $t=0$, the particle collides with a target and gets diffracted,
which translates into an outgoing wave function consisting of a
superposition of scattering states (also coherent plane waves) at
$t \to +\infty$.
Apart from the analytical partial waves approach, there is also
a number of numerical methods which have been developed in the
literature \cite{as:sanz1,as:sanz2}, such as the close-coupling
method, which basically consists of splitting the time-independent
Schr\"odinger equation into a set of coupled one-dimensional
Schr\"odinger equations accounting for each of the eventual
diffraction channels.
Also, as in the case of bound systems, classical trajectories have
been used to determine, for example, diffraction probabilities
\cite{as:bowman-1,as:bowman-2}; although diffraction is a quantum
concept, in principle it is possible to obtain good estimates in
scattering \cite{as:scatt-1,as:scatt-2,as:scatt-3} and pre-ionization
\cite{as:delgado-1,as:delgado-2,as:delgado-3}.

Regarding the time-dependent domain, the strategies
developed \cite{as:tannor,as:sanz2,as:leforestier,as:sanz4} are aimed
at solving the time-dependent Schr\"odinger equation directly, which
allow to monitor the system evolution along time from the initial to
the final asymptotic state, instead of passing directly from the
former to the latter.
Usually, wave-packet propagation methods are the most commonly used to
to carry out these studies, although other approximate techniques have
also been developed in order to find a balance between accuracy and
computational demand.
These approaches result interesting, for example, to monitor the
passage from reactants to products in a chemical reaction (e.g., in
proton transfer reactions, isomerization reactions or coherent-control
schemes), and are usually accompanied and complemented by classical,
trajectory-based studies (quantum-classical correspondence schemes).


\subsection{Statistical approach to large (complex) molecular systems}
\label{as:sec4.2.4}

Typically, the dynamical studies mentioned above are related to
systems which are considered as isolated from a surrounding medium
or {\it environment}.
However, real physical and chemical systems interact, exchanging
energy, with their surroundings and deviations from their isolated
behavior are observed.
A very remarkable effect, in this regard, is the broadening undergone
by atomic and molecular spectral lines, for example, which is a
signature of the life-time associated with the corresponding
quasi-bound (stationary) states.
But we also find such situations when measures involving bulk
properties (e.g., the viscosity of liquid water or its crystallization)
are involved, and a connection between the microscopic description and
the macroscopic observations has to be established.
Obviously, a full quantum-mechanical calculation involving the system
of interest plus its environment results unaffordable computationally.
Thus, different approximated statistical approaches have been developed
in the literature, based on the theory of open quantum systems, with
the main analysis tool being the correlation function, which allows
to extract the most relevant properties associated with the system.
The correct way to tackle a statistical problem in quantum
mechanics is through the Liouville-von Neumann equation, which is the equivalent of
Schr\"odinger's equation, but for the {\it density matrix operator} or,
simply, {\it density matrix}, $\hat{\rho}(t) = \sum_\alpha
|\Psi_\alpha(t)\rangle \langle \Psi_\alpha (t)|$, instead of the wave
function, $|\Psi (t)\rangle$.
However, full many degree-of-freedom, quantum-mechanical dynamical
problems result prohibitive computationally and different strategies
have been developed in the literature to deal with them.
Within the different methodologies commonly employed, there are
basically three schemes, which correspond to the three levels at which
a quantum statistical problem can be solved: quantum-mechanical,
mixed/hybrid quantum-classical and fully classical.
Nevertheless, independently of the level of accuracy of our statistical
description, there is always a very clear distinction between system
and environment, this being the starting point in any of these
approaches.
In this regard, it is worth mentioning that this is also the
partitioning philosophy used in some recent {\it ab initio} approaches:
instead of studying in detail a full macro-molecular structure, the
total system is partitioned into the reactive (or more active) part
and the surrounding, which is described in terms of an effective,
{\it embedding} potential.

It is also worth mentioning that, depending on the properties of the
environment, we can refer to it in different ways \cite{as:breuer}.
For example, usually, when it consists of a few degrees of freedom,
it is commonly referred as environment, while the term {\it reservoir}
is used when there is a large number of degrees of freedom.
If it is in thermal equilibrium, i.e., its properties remain stationary
with time (except for small fluctuations) --- in other words, this is
the limit where the environment ``forgets'' information about its
initial state and stays in equilibrium at a certain temperature ---,
then we employ the term {\it thermal bath}.
For example, a harmonic oscillator coupled to another harmonic
oscillator acts as an environment with respect to the latter because
there is an energy and coherence transfer from the latter to the
former.
If instead of one harmonic oscillator, we couple an infinite number
of oscillators, the environment becomes a reservoir and, if such
collection of oscillators is at thermal equilibrium, as in the
so-called {\it Caldeira-Leggett model} \cite{as:caldeira-1,%
as:caldeira-2,as:caldeira-3}, we will have a thermal bath acting
on the system oscillator.

Within the first level, the quantum-mechanical one, one starts from the
Liouville-von Neumann equation.
Proceeding algebraically according to some assumptions on the
timescales associated with system and environment, one can reformulate
the problem in terms of a master equation describing only the system
evolution, and where the effects of the environment appear in terms of
some dissipative operators or dissipators after assuming the Markovian
approximation (i.e., the detailed dynamics of the environment becomes
irrelevant, and only its effects are important regarding the system
dynamics).
This is achieved by partially tracing over the environment states,
which leads to the so-called {\it reduced density matrix},
$\hat{\rho}_S$, describing the system (or subsystem) of interest or
{\it reduced system}.
Note that this splitting between system and environment as a function
of their corresponding timescales (i.e., $\hat{\rho} = \hat{\rho}_S
\otimes \hat{\rho}_E$) is not other than the Born-Oppenheimer
approximation considered in {\it ab initio} methods to separate the
electronic motion from the nuclear one (see Sec.~\ref{as:sec4.2.1}).
This procedure thus gives rise to very well known master equations for
the system reduced density matrix, such as the {\it Lindblad equation}
or its version in the Markovian weak-coupling limit, namely the
so-called {\it Redfield equation} \cite{as:redfield-1,as:redfield-2,%
as:redfield-3,as:redfield-4}.
But, also, one can obtain master equations to describe the evolution
of system expectation values starting from the Heisenberg representation,
such as the {\it quantum Langevin equation} \cite{as:louisell,%
as:ford-1,as:ford-2}, which is the Heisenberg equation of motion for
the coordinate of a Brownian particle coupled to a thermal bath.
At this stage, apart from analytical derivations, there are basically
two methods to solve these equations quantum-mechanically: the
{\it path integral method} \cite{as:feynman-1,as:feynman-2,%
as:feynman-3,as:schulman} and the {\it quantum state diffusion theory}
\cite{as:percival}.
In the first case, one directly attacks the problem of the density
matrix, while in the latter one solves a stochastic Sch\"odinger
equation, obtained from the transformation $\hat{\rho}_S = (1/N)
\sum_{k=1}^N |\Psi_k\rangle \langle \Psi_k |$, where the $\Psi_k$ are
(stochastic) solutions of such an equation \cite{as:sanz:CJC:2014}.

Whenever the environment dynamics is relevant, the Markov approximation
cannot be applied.
Then, we are in the next level, that of the mixed or hybrid
quantum-classical methods, where usually the system dynamics is
accounted for quantum-mechanically, while the environment dynamics is
described by means of classical mechanics, this splitting also being
based in the Born-Oppenheimer approximation, as mentioned above.
Different methods and techniques have been developed in the literature
to tackle this problem, such as mean-field theories \cite{as:billing},
surface-hopping methods \cite{as:tully} or semiclassical initial value
representations \cite{as:miller1-1,as:miller1-2,as:miller1-3,%
as:miller1-4,as:miller2-1,as:miller2-2,as:others-ivr-1,as:others-ivr-2,%
as:pollak-1,as:pollak-2,as:pollak-3}, which allow to describe both the
effects of the environment over the system and vice-versa in spite of
their different (classical or quantum) mechanical evolution.
In this regard, it is worth stressing that, depending on whether
interatomic interactions can be described by a potential energy surface
or not, we may have adiabatic or non-adiabatic dynamics.
In the latter case, the interaction between different coupled potential
energy surfaces (corresponding to different electronic molecular
states) has to be included into the corresponding dynamical
description \cite{as:kapral}, thus not satisfying the Born-Oppenheimer
approximation, which implies finding good descriptors of the coupling
between the electronic and nuclear parts of the Hamiltonian.
Now, for certain cases, one can further proceed and represent the
system also classically, thus leading to the third level, which is
constituted by purely classical methodologies (i.e., based on Newton's
equations), such as simple molecular dynamics simulations, or
refinements of them, such as the classical Wigner method or the
linearized initial value representation, where a particular choice
of the initial conditions is considered (though the propagation is
fully classical).
Sometimes, molecular dynamics calculations are combined with DFT in
order to obtain ``on the fly'' the potential energy surfaces governing
the nuclear motion, as happens in the {\it Car-Parrinello
method} \cite{as:carpar-1,as:carpar-2}.


\section{Bohmian Mechanics}
\label{as:sec4.3}


\subsection{Fundamentals}
\label{as:sec4.3.1}

For simplicity, but without loss of generality, we are going to start with
the Schr\"odinger equation associated with a non-relativistic particle
(system) of mass $m$.
The general solutions to this equation are given in the form of
time-dependent complex-valued fields, which we generally call wave
functions and represent the state of the system (quantum states) at
a given time throughout the full accessible configuration space
associated with the physical system studied.
From these solutions we extract all relevant physical information about
the quantum system, that is, intensities and coherence-based
properties (interference effects).
Without entering here a discussion on the well-known debate about the
completeness of the wave function (actually, not necessary at all),
but taking into account that such information is encoded in the
probability density (i.e., the square of the amplitude of the wave
function) and the phase of the wave function, we can recast the
latter in terms of two real-valued fields.
This is actually a closer approach to Schr\"odinger's derivation of
his equation from Hamilton's formulation of classical mechanics
and de Broglie's novel idea of associating a wave to nonzero mass
systems (the so-called {\it Hamiltonian analogy} \cite{as:born1}).

From a formal viewpoint,
in principle, if we consider time-independent, eigenvalue problems,
Schr\"odinger's equation acquires the form of a general wave
equation (Helmholtz equation), which has exactly the same
solutions as the homologous equations we find in electromagnetism
or classical wave mechanics \cite{as:courant-hilbert-bk,as:morse}.
However, in the time-dependent case this analogy is not direct.
If the Helmholtz equation is rewritten in paraxial form, it is true that
it resembles Schr\"odinger's equation, although with one of the
coordinates playing the role of the evolution parameter (time).
However, in general, what we really observe is a closer resemblance
to the heat equation, since it is also a parabolic partial
differential equation, as diffusion equations are, although with the
particularity that the diffusion coefficient is a pure imaginary constant.
This fact was already stressed by F\"urth \cite{as:furth} shortly after
the equation appeared and later on by Comisar \cite{as:comisar}.
Nelson also introduced \cite{as:nelson:pr:1966} a Markovian approach from
which Schr\"odinger's equation arose as a result of diffusion process
in a subquantum medium.

Taking into account the previous discussion, let us consider the
nonlinear transformation
\begin{equation}
 \Psi (\vec{r},t) = R(\vec{r},t)\ e^{i S(\vec{r},t)/\hbar} ,
 \label{as:eq10}
\end{equation}
which allows us to recast the complex-valued field $\Psi$ in terms of
two real-valued fields, one accounting for the amplitude of such a
complex field, and the other one for the phase that any complex
quantity introduces.
The same transformation is also considered for the conjugate complex
(we go from two field variables to another two field variables), with
the only difference being that the phase acquires a minus sign.
The equations of motion for the amplitude ($R$) and phase ($S$) fields arise
after substitution of the wave fields ($\Psi$ and $\Psi^*$) into the
time-dependent Schr\"odinger equation and its complex form,
respectively.
After a brief rearrangement of terms, adding and subtracting these
equations, we obtain their imaginary and real parts, which read,
respectively, as
\begin{eqnarray}
 \frac{\partial \rho}{\partial t} & + & \vec{\nabla} \cdot
  \left( \rho \ \frac{\vec{\nabla} S}{m} \right) = 0 ,
 \label{as:eq11} \\
 \frac{\partial S}{\partial t} & + &
  \frac{(\vec{\nabla} S)^2}{2m} + V_{\rm eff} = 0 .
 \label{as:eq12}
 \end{eqnarray}
In Eq.~(\ref{as:eq12}), the third term on the left-hand side,
\begin{eqnarray}
 V_{\rm eff} = V + Q & = & V - \frac{\hbar^2}{2m}
  \frac{\nabla^2 R}{R} \nonumber \\
 & = & V - \frac{\hbar^2}{4m} \left[
  \frac{\nabla^2 \rho}{\rho} - \frac{1}{2}
  \left( \frac{\vec{\nabla} \rho}{\rho} \right)^2 \right] ,
 \label{as:eq13}
\end{eqnarray}
is an {\it effective potential} resulting from the sum of the external
potential function $V$ (particle-particle interactions in electronic structure
problems and potential energy surfaces in dynamics ones; other
external contributions, such as the application of external laser fields
or gravity, are also included), and the {\it quantum potential}, $Q$.
The latter is responsible for transferring the {\it nonlocality}
property of quantum mechanics \cite{as:bell2,as:sanz3} (which manifests
through the phenomenon of {\it quantum entanglement} \cite{as:schro-1,%
as:schro-2}) to the particle's motion due to its dependence on the
quantum state (through the probability density $\rho$).

As it can be noticed, Eq.~(\ref{as:eq11}) is the well-known probability
conservation equation, which states that for a pure state the probability
must preserve along time, but also that if we specify a certain region of
the configuration space, the probability contained in it must also
remain constant along time if its boundaries vary in time in a prescribed
way \cite{as:sanz:AnnPhys:2013}.
In classical mechanics \cite{as:tannor}, this variation takes place according to
the evolution in phase space of (Newtonian) trajectories with initial conditions
selected on the boundary; the evolution of the trajectories generates tubes
in phase space where the probability is confined all the way through.
Quantum-mechanically the same also holds \cite{as:sanz:AnnPhys:2013} if we
choose as the confining trajectories the ones that arise from the equation of
motion
\begin{equation}
 \vec{v} = \frac{\vec{J}}{\rho} = \frac{\vec{\nabla} S}{m} ,
 \label{as:eq14}
\end{equation}
although the preservation of the probability takes places in the system
configuration space instead of a generalized phase space.
This transport equation arises naturally in the standard formulations of
quantum mechanics if we divide the quantum flux or probability current
\cite{as:bohm-bk,as:schiff-bk},
\begin{equation}
 \vec{J} = \frac{\hbar}{2mi} \left( \Psi^* \nabla \Psi - \Psi \nabla \Psi^* \right) ,
\end{equation}
by the probability density.
In other words, physically it means that the transport or diffusion of probability
through the system configuration space can be described as a velocity vector
field multiplying the probability density, in compliance with usual diffusion
(transport) equations.
Nonetheless, the way how Bohm introduced that quantity in 1952 was
postulating it once the quantum or generalized Hamilton-Jacobi equation
(\ref{as:eq12}) was settled \cite{as:bohm-1}.
That is, he established a direct analogy with classical mechanics, where an
analogous equation arises within the Hamilton-Jacobi formulation, namely
the Jacobi momentum law.
In the classical case, $S$ denotes the classical (mechanical) action, which is
a hypersurface in configuration space associated with the evolution of the
system directly linked to the geometric phase in the {\it eikonal approximation}
\cite{as:born1}; in quantum mechanics, the phase field $S$ is a direct
analog of the classical action, although it is specified in the configuration
space of the system.
It is by virtue of this constraint (i.e., to restrict the motion to the
configuration space instead of a phase space with coordinates and momenta
independent of each other) that nonlocality arises (or manifests) in quantum
mechanics.

Notice that the trajectories that one obtains by integration (in time) of
Eq.~(\ref{as:eq14}) describe the flow of the probability throughout the
corresponding configuration space.
That is, such trajectories are just probability streamlines, analogous to those
that we find in classical hydrodynamics (this is precisely the point emphasized
by Madelung in 1926).
However, one can also be tempted to identify such trajectories with the
actual motion of real particles in order to establish a (alternative) quantum
theory without observers.
Although this has been a major issue of debate over years and, actually, one
of the main reasons why Bohmian mechanics has been generally rejected,
there is nothing in the theory, not even current experiments, supporting
such a link.
The only accessible information to date in an experiment is the quantum
flux, as it has been recently shown \cite{as:kocsis:Science:2011}, which is
in compliance with the usual theory \cite{as:sanz:AnnPhysPhoton:2010}.

Equation (\ref{as:eq14}) together with the time-dependent Schr\"odinger
equation are the two basic equations within the analytic approach
discussed above \cite{as:wyatt1}, for they allow us to obtain the evolution
of the wave function $\Psi$ as well as the associated Bohmian trajectories,
necessary to interpret the flow of the probability or, equivalently,
the motion of a swarm of {\it non-interacting} quantum
particles \cite{as:book-1,as:book-2,as:sanz4,as:sanz5}.
On the other hand, from a computational viewpoint, schemes aimed at
integrating partial equations are always affected by numerical instabilities,
as happens with Eqs.~(\ref{as:eq11}) and (\ref{as:eq12}).
Hydrodynamically speaking \cite{as:fluidmech-1,as:fluidmech-2}, these
equations are expressed within an {\it Eulerian framework}, i.e., from
a fixed space point we will observe the quantum fluid passing by.
So, an easier way to proceed consists in taking advantage of
Eq.~(\ref{as:eq14}) to define the Lagrange time derivative,
\begin{equation}
 \frac{d}{dt} = \frac{\partial}{\partial t} + \vec{v} \cdot \vec{\nabla} .
\end{equation}
Recasting Eqs.~(\ref{as:eq11}), (\ref{as:eq12}) and (\ref{as:eq14})
in terms of this operator, which renders the set of coupled equations of
motion
\begin{eqnarray}
 \frac{d\rho}{dt} & = & - \rho \vec{\nabla} \cdot \vec{v} ,
 \label{as:eq15} \\
 \frac{d \vec{v}}{dt} & = & - \frac{\vec{\nabla} V_{\rm eff}}{m} ,
 \label{as:eq16} \\
 \frac{dS}{dt} & = & \frac{1}{2}\  m v^2 - V_{\rm eff} ,
 \label{as:eq17}
\end{eqnarray}
or, equivalently,
\begin{eqnarray}
 \frac{d\rho}{dt} & = & - \rho \vec{\nabla} \cdot \vec{v} ,
 \label{as:eq18} \\
 \frac{dS}{dt} & = & \frac{1}{2}\ m v^2 - V_{\rm eff} ,
 \label{as:eq19} \\
 \frac{d \vec{r}}{dt} & = & \vec{v} = \frac{\vec{\nabla} S}{m} .
 \label{as:eq20}
\end{eqnarray}
This set of equations constitutes the core of the so-called
(Bohmian) {\it quantum trajectory method} \cite{as:wyatt1}, an algorithm
based on Bohmian mechanics to obtain quantum-mechanical results.
It was firstly proposed \cite{as:lopreore-1,as:lopreore-2,%
as:lopreore-3} following the scheme of Eqs.~(\ref{as:eq15}),
(\ref{as:eq16}) and (\ref{as:eq17}).
However, evaluating the gradient of the force is an important source
of numerical error and the second scheme was considered.
Alternatively, to avoid these drawbacks, one can re-derive this set of
equations in complex form
\cite{as:tannor:JCP:2006,as:tannor:JCP-2:2007,as:tannor:JPCA:2007,%
as:tannor:CP:2007,as:chou:JCP:2006,as:chou:JCP-1:2008,as:chou:JCP-3:2008,%
as:chou:JCP:2010}, by considering a complex phase field, which already include
information about the amplitude of the wave function \cite{as:benseny:EPJD:2014}.


\subsection{Nonlocality and entanglement}
\label{as:sec4.3.2}

Dealing with many-body problems, as it is the case when molecular systems are
involved, unavoidably leads to the question of the appearance of {\it quantum
correlations} among the different degrees of freedom necessary to specify the
(quantum) state of the system.
Typically the series of approximations that are considered to tackle these problems
(starting from the Born-Oppenheimer one, whenever it is required) implies a
breach of such correlations.
Although this is an efficient route from a problem-solving viewpoint, physically
it means there is a certain loss of quantum coherence in the system, which is
not going to be shared anymore by the different separated parts.
In other words, the degree of {\it entanglement} \cite{as:schro-1,as:schro-2}
between such parts is totally lost -- whatever happens to one of the parts
of the full system will only have classical-like consequences on the others,
and vice versa.

To better understand the implications of entanglement and the
possibility to nonlocally share information among different parts of a quantum
system is precisely through Bohmian mechanics.
Consider the $N$-body wave function in the position space
\begin{equation}
 \Psi (\vec{r}_1,\vec{r}_2,\ldots,\vec{r}_N,t) =
   R(\vec{r}_1,\vec{r}_2,\ldots,\vec{r}_N,t)\
   e^{i S(\vec{r}_1,\vec{r}_2,\ldots,\vec{r}_N,t)/\hbar} .
 \label{as:eq21}
\end{equation}
The {\it reduced} single-particle probability density for the $i$th
particle (or, in general, degree of freedom) is defined as
\begin{equation}
 \tilde{\rho}_i(\vec{r}_i,t) = \int \!\! \cdots \!\! \int
   \rho (\vec{r}_1,\vec{r}_2,\ldots,\vec{r}_N,t)
    d\vec{r}_1 d\vec{r}_2 \cdots d\vec{r}_{i-1}
    d\vec{r}_{i+1} \cdots d\vec{r}_N ,
 \label{as:eq22}
\end{equation}
where $\rho = R^2$.
This reduced probability density describes the statistics of the $i$th particle
ignoring the statistics of the remaining particles, although due to the
correlation among them any change in the configuration of the system will
have consequences on this particle.
To this particle we can also assign an equation of motion,
\begin{equation}
 \vec{v}_i = \frac{\vec{\nabla}_{\vec{r}_i} S}{m} ,
 \label{as:eq23}
\end{equation}
which takes into account the full wave function and therefore the motion
of the other particles.
This procedure gives rise to a set of $N$ equations of motion coupled through
the ``wholistic'' phase $S$, where the evolution of a particle will be
strongly nonlocally influenced by the others (apart from other classical
like interactions through $V$).
This entanglement becomes more
apparent through the quantum potential,
\begin{equation}
 Q = - \frac{\hbar^2}{2m} \sum_{i=1}^N
 \frac{\nabla_{\vec{r}_i}^2 R}{R} ,
 \label{as:eq24}
\end{equation}
where $Q = Q(\vec{r}_1,\vec{r}_2,\ldots,\vec{r}_N,t)$, which is
nonseparable and therefore strongly nonlocal.

As a simple illustration of these ideas, consider the case of two
particles which interact, at some time, through a potential $V$.
The two-particle wave function describing this system is
\begin{equation}
 \Psi (x,y,t) = R(x,y,t)\ e^{i S(x,y,t)/\hbar} ,
 \label{as:eq25}
\end{equation}
with $\rho (x,y,t) = R^2(x,y,t)$ being the total probability density,
and $\tilde{\rho}_1(x,t)$ and $\tilde{\rho}_2(y,t)$ the reduced
probability densities associated with each particle.
The trajectories for these particles are obtained from
\begin{equation}
 \vec{v}_1 = \frac{1}{m} \frac{\partial S(x,y,t)}{\partial x} , \qquad
 \vec{v}_2 = \frac{1}{m} \frac{\partial S(x,y,t)}{\partial y} ,
 \label{as:eq26}
\end{equation}
which are implicitly influenced by the quantum potential
\begin{equation}
 Q (x,y,t) = - \frac{\hbar^2}{2m} \frac{1}{R(x,y,t)}
  \left[ \frac{\partial^2 R(x,y,t)}{\partial x^2} +
         \frac{\partial^2 R(x,y,t)}{\partial y^2} \right] .
 \label{as:eq27}
\end{equation}
According to Schr\"odinger \cite{as:schro-1,as:schro-2}, after the
interaction, even if the two particles are very far apart one from
another, the wave function (\ref{as:eq25}) becomes non-factorizable, i.e.,
\begin{equation}
 \Psi (x,y,t) \neq \psi (x,t) \otimes \psi (y,t) ,
 \label{as:eq28}
\end{equation}
and the motion of both particles will remain entangled.
Different works in the literature analyze the trajectory correlation
among entangled particles \cite{as:dewdney1-1,as:dewdney1-2,%
as:marchildon,as:wyatt3-1,as:wyatt3-2,as:sanz6-1,as:sanz6-2},
the most recent one within the many-body context of transport phenomena
in mesoscopic systems \cite{as:oriols:PRL:2007}.

This does not happen, however, when the wave function is factorizable
in terms of single-particle partial wave functions,
\begin{equation}
 \Psi (x,y,t) = \psi_1 (x,t) \psi_2 (y,t)
  = R_1(x,t) R_2(y,t)\ e^{i \left[ S_1(x,t) + S_2(y,t) \right]/\hbar} ,
 \label{as:eq29}
\end{equation}
since $\rho(x,y,t) = \tilde{\rho}_1(x,t)\tilde{\rho}_2(y,t)$ is a direct product of
(partial) densities, while the particles' equations of motion,
\begin{equation}
 \vec{v}_1 = \frac{1}{m} \frac{\partial S_1(x,t)}{\partial x} , \qquad
 \vec{v}_2 = \frac{1}{m} \frac{\partial S_2(y,t)}{\partial y}
 \label{as:eq30}
\end{equation}
are uncoupled.
This can also be readily inferred from the quantum potential,
which is fully separable:
\begin{eqnarray}
 Q (x,y,t) & = &
   - \frac{\hbar^2}{2m} \frac{1}{R_1(x,t)}
     \frac{\partial^2 R_1(x,t)}{\partial x^2}
   - \frac{\hbar^2}{2m} \frac{1}{R_2(y,t)}
     \frac{\partial^2 R_2(y,t)}{\partial y^2}
  \nonumber \\
  & = & Q_1 (x,t) + Q_2 (y,t) ,
 \label{as:eq31}
\end{eqnarray}
In general, for an $N$-particle wave function,
\begin{equation}
 \Psi (\vec{r}_1,\vec{r}_2,\ldots,\vec{r}_N,t) =
  \Pi_{i=1}^N \psi_i (\vec{r}_i,t) =
  \Pi_{i=1}^N R_i (\vec{r}_i,t)\ e^{i S_i(\vec{r}_i,t)/\hbar} ,
 \label{as:eq32}
\end{equation}
the $i$th-particle equation of motion is
\begin{equation}
 \vec{v}_i = \frac{\vec{\nabla}_{\vec{r}_i} S_i}{m} ,
 \label{as:eq33}
\end{equation}
and the total quantum potential will be given by the direct addition of
the quantum potentials associated with each separate particle,
\begin{equation}
 Q = - \frac{\hbar^2}{2m} \sum_{i=1}^N
  \frac{\nabla_{\vec{r}_i}^2 R_i}{R_i}
  = \sum_{i=1}^N Q_i .
 \label{as:eq34}
\end{equation}

After using the Born-Oppenheimer approximation, splitting electronic from
nuclear degrees of freedom, the essence of electronic structure methodologies
is precisely finding the electronic configuration of molecular systems by
appealing to the idea of separability, from {\it ab initio} methods (wave function
approach) to the density functional theory (probability density approach).
Of course, once an optimal basis set is found, correlations will appear when
appealing to the appropriate symmetrization conditions.
For example, in the case of {\it ab initio} methodologies, Slater determinants are
considered to provide the correct symmetry to the linear combinations of orbitals
formed to describe the configuration of the corresponding molecular system
(see Sec.~\ref{as:sec4.2.2}).

There is another interesting issue concerning correlation and the
Born-Oppenheimer approximation.
Notice that although Eq.~(\ref{as:eq6}) is separable, due to the dependence
of the electronic coordinates, $\vec{r}_i$, on the nuclear ones, $\vec{R}_A$,
the eventual trajectories for the electrons will depend on the nuclei
positions (in agreement with the fact that the potential energy surface
varies throughout space), but the same will not happen for the nuclear
trajectories with respect to the electrons.
Again here we find the problem of the back-reaction, although this time it
is within the Bohmian context.


\subsection{Weak values and equations of change}
\label{as:sec4.3.3}

In the recent years, the concept of {\it weak value} \cite{as:aharonov:PRL:1988} has been a
subject of much interest, getting some popularity within the quantum metrology and
quantum foundations communities.
For example, by means of measurements of the weak value it is possible to perform tomography
of the quantum state associated with a single photon \cite{as:lundeen:Nature:2011} and also to
determine the streamlines describing the (averaged) flow of probability in Young's two slit
experiment \cite{as:kocsis:Science:2011}.
However, this idea, now in fashion, is directly related to the well-known quantum flux
\cite{as:bohm-bk,as:schiff-bk} and to the averaging of two-state transition probabilities
\cite{as:wiseman,as:hiley}.
However, the essence behind this idea can be traced back to 1970s, to a work published
by Hirschfelder \cite{as:hirschfelder2}, where we also find a certain connection to the later
application of quantum hydrodynamics to time-dependent DFT (see Sec.~\ref{as:sec4.4.1}).
According to this author, the framework presented in this work
facilitates the study of problems involving external electric and
magnetic fields (as seen in previous section) by grounding it on the
concept of {\it equation of change} for arbitrary quantum-mechanical
properties in configuration space.
These equations can be used to study the energy flow from one part
of a molecule to another, the nature of molecular collisions or the
magnetic properties of molecules.
The first two cases are somehow summarized by the applications which
will be discussed in next section regarding the molecular dynamics.
The third type of application, on the other hand, has been exploited
in the literature by different authors \cite{as:bader,as:gomes-1,%
as:gomes-2,as:lazzeretti-1,as:lazzeretti-2,as:lazzeretti-3,%
as:lazzeretti-4,as:lazzeretti-5,as:lazzeretti-6,as:lazzeretti-7,%
as:lazzeretti-8} to understand the magnetic properties of molecules
within a framework that encompasses electronic structure and topology.

Although in a brief manner, let us here consider how the ideas underlying
quantum hydrodynamics emerge in a very elegant fashion within the
framework proposed by Hirschfelder \cite{as:hirschfelder2}.
To start with, let us define an arbitrary property $\mathcal{S} \equiv
\Psi_1^* \hat{\mathcal{O}} \Psi_2$ (do not confuse this calligraphic
$\mathcal{S}$ denoting a certain property with the normal $S$ referring
to the phase of a wave function), where both $\Psi_1$ and $\Psi_2$
satisfy the same time-dependent Schr\"odinger equation (i.e., both are
governed by the same Hamiltonian) and $\hat{\mathcal{O}}$ is an
(arbitrary) operator associated with a quantum observable.
If $\Psi_2 = \Psi_1$, the integral of $\mathcal{S}$ over configuration
space, denoted by $\bar{\mathcal{S}}$, describes the
{\it expectation value} associated with $\mathcal{O}$, while for $\Psi_2 \neq
\Psi_1$, $\bar{\mathcal{S}}$ gives a {\it transition probability} or {\it rate}
(from the state $\Psi_1$ to a state $\Psi_2$) governed by $\mathcal{O}$.
It can be easily shown that $\mathcal{S}$ satisfies the equation of
change
\begin{equation}
 \frac{\partial}{\partial t} \ (\Psi_1^* \hat{\mathcal{O}} \Psi_2) =
  \Psi_1^* \left\{ \frac{\partial \hat{\mathcal{O}}}{\partial t}
   - \frac{i}{\hbar} \ [\hat{\mathcal{O}},\hat{H}] \right\} \Psi_2
   - \frac{i}{\hbar} \left[
     \Psi_1^* \hat{H} \hat{\mathcal{O}} \Psi_2
   - (\hat{H} \Psi_1)^\dag \hat{\mathcal{O}} \Psi_2 \right] .
 \label{hhir-1}
\end{equation}
Specifically, for a Hamiltonian
\begin{equation}
 \hat{H} = \frac{\hat{p}^2}{2m} + \hat{V}(\hat{\vec{r}}) ,
 \label{hhir-2}
\end{equation}
it can be shown \cite{as:hirschfelder2} that Eq.~(\ref{hhir-1}) takes
the functional form
\begin{equation}
 \frac{\partial}{\partial t} \ (\Psi_1^* \hat{\mathcal{O}} \Psi_2)
  + \vec{\nabla} \cdot \vec{F} = \mathcal{K} ,
 \label{hhir-3}
\end{equation}
where
\begin{equation}
 \vec{F} = \frac{1}{2} \left[
    \Psi_1^* \dot{\hat{\vec{r}}} \hat{\mathcal{O}} \Psi_2
  - (\dot{\hat{\vec{r}}} \Psi_1)^\dag \hat{\mathcal{O}} \Psi_2 \right]
 \label{hhir-4}
\end{equation}
and
\begin{equation}
 \mathcal{K} =
  \Psi_1^* \left\{ \frac{\partial \hat{\mathcal{O}}}{\partial t}
   - \frac{i}{\hbar} \ [\hat{\mathcal{O}},\hat{H}] \right\} \Psi_2
 \label{hhir-5}
\end{equation}
are, respectively, the flux of $\mathcal{S}$ and its rate of production,
and
\begin{equation}
 \dot{\hat{\vec{r}}} = - \frac{i}{\hbar}\  [\hat{\vec{r}},\hat{H}]
  = \frac{\hat{\vec{p}}}{m} .
\end{equation}

Now, if $\hat{\mathcal{O}}$ is a linear Hermitian operator, physically
the most significant form of $\mathcal{S}$ will be
\begin{equation}
 \mathcal{S} = {\rm Re} [\Psi_1^* \hat{\mathcal{O}} \Psi_2]
  = \frac{1}{2} \left[
      \Psi_1^* \hat{\mathcal{O}} \Psi_2
    + (\hat{\mathcal{O}} \Psi_1)^\dag \Psi_2 \right] .
 \label{hir-1}
\end{equation}
This is because $\mathcal{S}$, as given by (\ref{hir-1}), can be
considered itself an observable associated with a certain linear
Hermitian operator, $\hat{\mathcal{O}}^H$, in such a way that the
integral over configuration space of $\Psi_1^* \hat{\mathcal{O}}^H
\Psi_2$ is precisely (\ref{hir-1}).
In particular, this associated operator reads as
\begin{equation}
 \hat{\mathcal{O}}^H (\vec{r},\vec{r}') = \frac{1}{2} \left[
   \hat{\mathcal{O}}(\vec{r}') \delta (\vec{r}' - \vec{r})
 + \delta (\vec{r}' - \vec{r}) \hat{\mathcal{O}}(\vec{r}') \right] ,
 \label{eq-hirsch1}
\end{equation}
and therefore one readily notices
\begin{equation}
 \int \Psi_1^* (\vec{r}') \hat{\mathcal{O}}^H (\vec{r},\vec{r}')
   \Psi_2 (\vec{r}') d\vec{r}'
  = \frac{1}{2} \
    {\rm Re}[\Psi_1^* \hat{\mathcal{O}} \Psi_2 ] .
 \label{eq-hirsch2}
\end{equation}
Because $\mathcal{S}$ is an observable with respect to a linear
Hermitian operator, it is called a {\it subobservable}, which obeys
the equation of change
\begin{equation}
 \frac{\partial}{\partial t} \
  {\rm Re} [ \Psi_1^* \hat{\mathcal{O}} \Psi_2 ] =
  {\rm Re} \left[ \Psi_1^* \
   \frac{\partial \hat{\mathcal{O}}}{\partial t} \ \Psi_2 \right]
   + \frac{1}{\hbar} \ {\rm Im} \left[
     \Psi_1^* \hat{\mathcal{O}} \hat{H} \Psi_2
   - (\hat{H} \Psi_1)^\dag \hat{\mathcal{O}} \Psi_2 \right] .
 \label{hhir-8}
\end{equation}
Note that this equation is precisely the expectation value of
$\dot{\hat{\mathcal{O}}}^H$.

Let us now consider the expression (\ref{hir-1}) for $\mathcal{S}$
and the particular case $\Psi_2 = \Psi_1 = \Psi$.
If $\hat{\mathcal{O}} = \mathbb{I}$, the subobservable corresponds
to the probability density $\mathcal{S} = \Psi^* \Psi = \rho$.
Substituting this condition into Eq.~(\ref{hhir-8}), we find the
well-known continuity equation,
\begin{equation}
 \frac{\partial \rho}{\partial t} + \vec{\nabla} \cdot \vec{J} = 0 ,
 \label{hhir-9}
\end{equation}
where
\begin{equation}
 \vec{J} \equiv {\rm Re} [ \Psi^* \dot{\hat{\vec{r}}} \Psi ]
  = \left( \frac{\hbar}{2mi} \right)
    \left( \Psi^* \vec{\nabla} \Psi - \Psi \vec{\nabla} \Psi^* \right)
 \label{hhir-10}
\end{equation}
is the quantum probability density current or quantum flux (see
Sec.~\ref{as:sec4.3}).
The equation of change for this latter quantity is obtained assuming
$\hat{\mathcal{O}} = \hat{\vec{p}}/m$:
\begin{equation}
 \frac{\partial \vec{J}}{\partial t} + \vec{\nabla} \cdot {\rm {\bf F}} =
  - \frac{1}{m} \ \rho \vec{\nabla} V ,
 \label{hhir-11}
\end{equation}
where
\begin{equation}
 {\rm {\bf F}} = \frac{1}{4m^2}
  \left[ \Psi^* \vec{p} \vec{p} \Psi + \Psi \vec{p} \vec{p} \Psi^*
   + (\vec{p} \Psi)^\dag (\vec{p} \Psi)
   + (\vec{p} \Psi) (\vec{p} \Psi)^\dag \right]
 \label{hhir-12}
\end{equation}
is a tensor quantity accounting for the flux of the quantum probability
density current.
According to London \cite{as:london}, the velocity $\vec{v}$, as
defined in Sec.~\ref{as:sec4.3}, cannot be a subobservable since
there is no linear Hermitian operator $\hat{\mathcal{O}}^H$ for
which $\vec{v}$ is an expectation value.
However, an equation of motion for this quantity can be obtained using
both (\ref{hhir-9}) and (\ref{hhir-11}), which reads as
\begin{equation}
 m\rho\ \frac{d \vec{v}}{dt} = - \rho \vec{\nabla} V
  - \vec{\nabla} \cdot {\rm {\bf P}} .
 \label{hhir-13}
\end{equation}
As can be readily noticed, this equation is (\ref{as:eq16}), which
arises when its right-hand side is conveniently rearranged in terms
of the effective potential.
As it is given, (\ref{hhir-13}) describes an incompressible fluid,
with
\begin{equation}
 {\rm {\bf P}} = - \left( \frac{\hbar^2}{4m} \right) \rho
  \vec{\nabla} \vec{\nabla} \ln \rho
  = m {\rm {\bf F}} - \frac{m}{\rho}\ \vec{J} \vec{J}
 \label{hhir-14}
\end{equation}
being the {\it quantum pressure tensor} introduced by Takabayasi in the
1950s \cite{as:takabayasi-1,as:takabayasi-2}, whose effects will be discussed in
Sec.~\ref{as:sec4.4.3}.


\section{Applications}
\label{as:sec4.4}

Covering all aspects of the topic under consideration would take more
than a single chapter due to the many different applications that have
been developed in the literature.
Thus, instead of trying to go for such an ambitious programme, the
following sections are aimed at providing an illustration on how Bohmian
mechanics has been or can be used to tackle different problems
related to the aspects described in Sec.~\ref{as:sec4.2}.


\subsection{Time-dependent DFT: The quantum hydrodynamic route}
\label{as:sec4.4.1}

Although it is not well known in the Bohmian literature, one of
the former (and earlier) applications of Bohmian mechanics in its
hydrodynamic version was the design of time-dependent DFT
algorithms
\cite{as:bloch,as:bartolotti1,as:bartolotti2,as:bartolotti3,as:runge,%
as:deb,as:deb-chat-1,as:deb-chat-2,as:deb-chat-3,as:deb-chat-4,%
as:dey-deb-1,as:dey-deb-2,as:dey-deb-3,as:dey-deb-4,as:march-1,%
as:march-2,as:march-3,as:march-4,as:march-5,as:march-6,as:mcclendon}.
The formal grounds of this approach, known as {\it quantum fluid
dynamics DFT} (QFD-DFT), rely on a set of hydrodynamical
equations \cite{as:bartolotti1,as:bartolotti2,as:bartolotti3,%
as:runge,as:deb}, having the advantage of being able to deal with
dynamical processes evolving in time in terms of single-particle
time-dependent equations \cite{as:march-1,as:march-2,as:march-3,%
as:march-4,as:march-5,as:march-6}.
As happens with time-independent DFT, the density is also determined by
solving a single-particle non-interacting time-dependent Schr\"odinger
equation \cite{as:runge}, which approximately, but rather accurately,
deals with both static and dynamic correlations between electrons.
This methodology thus facilitates the computation of {\it ab initio}
electron densities for $N$-electron systems, with the advantage of
delivering both density and energy with a superior accuracy to HF
methods at a comparable computational overload.
For a more detailed historical account on this approach and its
achievements, the interested reader may consult Ref.~\cite{as:sanz7}.

In order to understand how QFD-DFT works, consider the polynuclear and
polyelectronic time-dependent Schr\"odinger equation,
\begin{equation}
 \hat{H} \Psi (\vec{R}_1, \vec{R}_2, \ldots, \vec{R}_M,
  q_1, q_2, \ldots, q_N, t) =
  i\hbar \frac{\partial \Psi (\vec{R}_1, \vec{R}_2, \ldots, \vec{R}_M,
  q_1, q_2, \ldots, q_N, t)}{\partial t} .
 \label{as:eq35}
\end{equation}
As in time-independent DFT, the purpose of time-dependent DFT is to
provide solutions to the electronic part of $\Psi$, also assumed to
be separable, as (\ref{as:eq6}), although now
\begin{equation}
 |\phi_{\rm e,N} (q_1, q_2, \ldots, q_N, t)|^2
  dq_1 dq_2 \cdots dq_N dt .
 \label{as:eq36}
\end{equation}
represents the probability that electrons $1,2,\ldots,N$ will be found
simultaneously within the volume elements $dq_1, dq_2, \ldots, dq_N$,
respectively, during the time interval $dt$.
Taking this into account and then further proceeding as in DFT, one
finds that the time-dependent polyelectronic wave function can be
expressed as a combination of product of single monoelectronic
time-dependent states,
\begin{equation}
 \phi_{\rm e,N} (q_1, q_2, \ldots, q_N, t) \quad
 \longrightarrow \quad \Pi_{i=1}^N \phi_{\rm e,1,i} (q_i,t) .
 \label{as:eq37}
\end{equation}
and time plays the role of a parameter.
In this regard, considering different times could be associated, within
the time-independent DFT approach, with assuming different states, each
labeled with a different index $k$.
The process to obtain the electronic density is thus similar to
that in time-independent DFT, but considering subsequent times, which
is done within the quantum hydrodynamics framework, where the quantum
probability is understood as a quantum fluid.

As an illustration of the method, we are going to analyze the
calculation of a density for an $N$-electron system under the influence
of an external time-dependent periodic force, formerly studied by
Bartolotti \cite{as:bartolotti1,as:bartolotti2,as:bartolotti3}.
In the time-independent case, assuming the electrons are
noninteracting, we need $N$ orbitals to describe the electrons, this
being equivalent to assume that each electron can be described by $N$
different orthonormal orbitals $\phi_i$ (i.e., $\langle\phi_i
|\phi_k\rangle = \delta_{ik}$), and therefore
\begin{equation}
  \sum_{i=1}^N |\phi_i(\vec{r})|^2 = \rho(\vec{r})
 \label{as:eq38}
\end{equation}
is the ``exact'' density of the system.
As in standard DFT \cite{as:kohn}, the $\phi_i$ are obtained after
minimization of the kinetic energy of the $N$-electron system,
\begin{equation}
 T_s [\{\phi_i\}] = - \frac{1}{2} \sum_{i=1}^N
  \int \phi_i(\vec{r}) \left(\nabla^2\phi_i (\vec{r})\right) d\vec{r} ,
 \label{as:eq39}
\end{equation}
which renders the Euler-Lagrange equation
\begin{equation}
 - \frac{1}{2} \nabla^2 \phi_i + v_{\rm eff} \phi_i
  = \epsilon_i \phi_i .
 \label{as:eq40}
\end{equation}
As can be noticed, (\ref{as:eq40}) is a single-particle
time-independent Schr\"odinger equation, where $v_{\rm eff}$ is an
effective potential including the external (nuclear) interaction as
well as the averaged action of the remaining $N$-1 electrons, and
$\epsilon_i$ is the associated eigenenergy --- $\epsilon_i$ can also be
interpreted as the Lagrange multiplier that insures (\ref{as:eq38}) is
satisfied by the $\phi_i$.
Also note in (\ref{as:eq40}) that, if this equation is divided by
$\phi_i$, it will read as
\begin{equation}
 Q_i + v_{\rm eff} = \epsilon_i ,
 \label{as:eq41}
\end{equation}
where
\begin{equation}
 Q_i (\vec{r}) = - \frac{1}{2}
  \frac{\nabla^2 \phi_i (\vec{r})}{\phi_i (\vec{r})}
 \label{as:eq42}
\end{equation}
is the {\it effective quantum potential} associated with the state
$\phi_i$.
This means that the sum of this quantum potential to the effective one
results in the bound-state energy~$\epsilon_i$.

Now, in the time-dependent case, consider the system is described by
the (time-dependent) orbitals $\phi_i (\vec{r},t)$, from which we want
to obtain the (time-dependent) density $\rho(\vec{r},t)$.
In polar form this orbitals read as
\begin{equation}
 \phi_i (\vec{r},t) = R_i(\vec{r},t)\ {\rm e}^{iS_i (\vec{r},t)} ,
 \label{as:eq43}
\end{equation}
with
\begin{equation}
 \int_t \int R_i(\vec{r},t) R_k(\vec{r},t) d\vec{r} = \delta_{ik} ,
 \label{as:eq44}
\end{equation}
where $\int_t$ denotes a time-averaged integration over one period of
time.
The kinetic energy associated with this (noninteracting) $N$-electron
system reads \cite{as:bartolotti2} as
\begin{equation}
 T_s [\{R_i,S_i\}]_t = - \frac{1}{2}
  \sum_{i=1}^N \int_t \int
  \left\{ R_i(\vec{r},t) \nabla^2 R_i(\vec{r},t)
  - R_i^2(\vec{r},t) \left(\vec{\nabla} S_i(\vec{r},t)\right)^2
    \right\} d\vec{r} .
 \label{as:eq45}
\end{equation}
Here, the constraints are
\begin{equation}
 \sum_{i=1}^N R_i^2 (\vec{r},t) = \rho (\vec{r},t) ,
 \label{as:eq46}
\end{equation}
as in the time-independent case, and the conservation of the number of
particles,
\begin{equation}
 \sum_{i=1}^N \frac{\partial R_i^2}{\partial t}
  \left( = \frac{\partial \rho}{\partial t} \right) =
  - \vec{\nabla} \cdot \vec{J} ,
 \label{as:eq47}
\end{equation}
where $\vec{J}$ is the single-particle quantum probability density
current vector.
Thus, after minimizing (\ref{as:eq45}) with respect to the $R_i$
(subjected to the previous constraints), we find the Euler-Lagrange
equation
\begin{equation}
 - \frac{1}{2} \nabla^2 R_i + v_{\rm eff} R_i = \epsilon_i R_i .
 \label{as:eq48}
\end{equation}
where $v_{\rm eff} (\vec{r},t)$ is the Lagrange multiplier associated
with the constraint defined in (\ref{as:eq46}), and the $\epsilon_i
(\vec{r},t)$ are the Lagrange multipliers associated with the
conservation of the number of particles, given by (\ref{as:eq44}) and
(\ref{as:eq47}).
The $\epsilon_i (\vec{r},t)$ can be split out as a sum of two terms,
\begin{equation}
 \epsilon_i (\vec{r},t) = \epsilon_i^{(0)}
  + \epsilon_i^{(1)} (\vec{r},t) ,
 \label{as:eq49}
\end{equation}
where $\epsilon_i^{(0)}$ arise from the normalization constraint and
$\epsilon_i^{(1)}$ are the Lagrange multipliers associated with the
charge-current conservation defined by (\ref{as:eq47}).
Also note that, analogously to (\ref{as:eq40}), (\ref{as:eq48}) can
now be expressed as
\begin{equation}
 Q_i (\vec{r},t) =
  - \frac{1}{2} \frac{\nabla^2 R_i(\vec{r},t)}{R_i(\vec{r},t)}
 \label{as:eq50}
\end{equation}
if we divide by $R_i$, where $Q_i$ is the time-dependent effective
quantum potential associated with the state $\phi_i$.

On the other hand, minimizing $T_s [\{R_i,S_i\}]_t$ with respect to
$S_i$, subject to the constraint
\begin{equation}
 \frac{\partial S_i}{\partial t} = - \epsilon_i ,
 \label{as:eq51}
\end{equation}
gives rise to the Euler-Lagrange equation
\begin{equation}
 \frac{\partial R_i^2}{\partial t}
  + \vec{\nabla} \cdot (R_i^2 \vec{\nabla} S_i) = 0 .
 \label{as:eq52}
\end{equation}
The scheme based on the coupled equations (\ref{as:eq48}) and
(\ref{as:eq52}) thus provides a means of determining the exact
time-dependent density of the system, noticing that
\begin{equation}
 \vec{J} (\vec{r},t) =
  \sum_{i=1}^N R_i^2 (\vec{r},t) \vec{\nabla} S_i (\vec{r},t) .
 \label{as:eq53}
\end{equation}
Moreover, if the time-dependence is turned off, the QFD-DFT approach
correctly reduces to the standard time-independent DFT results: since
$\vec{\nabla} S_i$ vanishes, (\ref{as:eq47}), (\ref{as:eq51}) and
(\ref{as:eq52}) will be identically satisfied and (\ref{as:eq45}) will
reduce to the time-independent kinetic energy of the $N$-electron
system.

More recently, Elsayed {\it et al.}\ \cite{as:molmer:arxiv:2017} have proposed a Bohmian-based
many-body approach to study the breathing dynamics of a many-boson
system enclosed in a trap with both long and short range interactions,
without appealing to typical mean-field approximations
\cite{as:cederbaum:PhysRev:2008,as:schmelcher:JPhysB:2017}.
This approach is based on the concept of conditional wave function
\cite{as:oriols:PRL:2007} and its accuracy has been tested by
comparing with the standardized, numerically exact multiconfigurational
time-dependent Hartree (MCDTH) method \cite{as:meyer:PhysRep:2000}.


\subsection{Bound system dynamics: Chemical reactivity}
\label{as:sec4.4.2}

\begin{figure}[t]
 \centering
 \includegraphics[width=12.5cm]{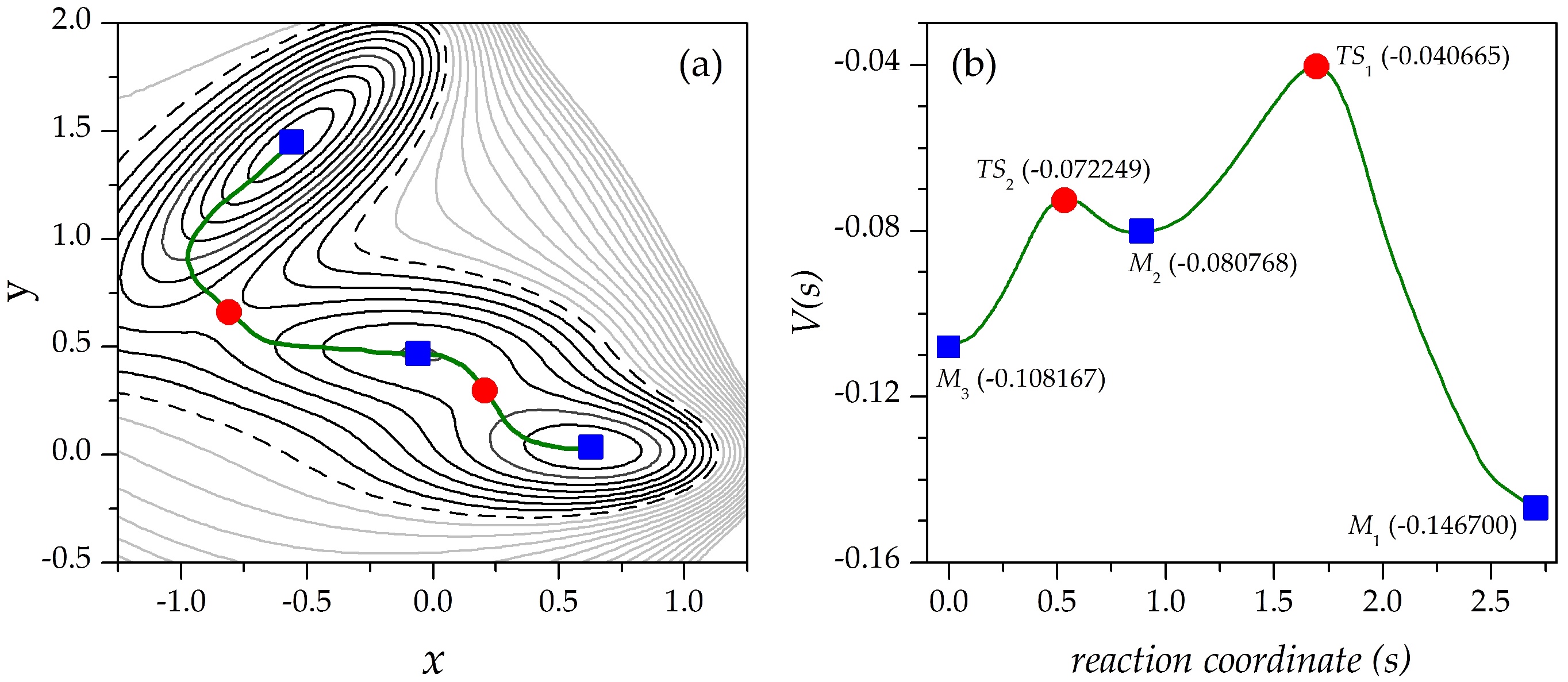}
 \caption{(a) Contour plot of the M\"uller-Brown potential energy surface
  and reaction path (green thick line) joining the reactants minimum with the
  products one.
  In the way how the surface is defined, the black contours represent
  equipotential lines for negative values of the energy, while the gray ones
  stand for positive energies; the zero-energy separatrix appears as a
  dashed line.
  (b) Energy profile along the reaction path.
  In both panels, the maxima indicate the position of transition states ($TS$)
  and are denoted by red circles, while the minima associated with the different
  molecular states ($M$) are indicated with blue squares.}
 \label{as:fig1}
\end{figure}

As an example of bound dynamics describing a chemical reaction, let us
consider a prototype potential model in reaction-path finding,
namely the M\"uller-Brown potential energy surface
\cite{as:muellerbrown,as:bofill1,as:bofill2,as:quapp,as:sanz8}.
As seen in Fig.~\ref{as:fig1}(b), this potential energy surface has
three minima (blue squares), $M_1$, $M_2$ and $M_3$, corresponding to
the products, intermediate and reactants states, respectively, and two
transition states (red circles), $TS_1$ and $TS_2$, separating products
from pre-equilibrium and the latter from reactants, respectively.
The physical meaning of these singular points is better understood by
inspecting the energy profile displayed in Fig.~\ref{as:fig1}(b).
This profile is taken along the reaction path (green line), plotted as a
function of the arc-length, which is defined in simple (and standard)
terms as
\begin{equation}
 s(x,y) \approx \sum_{i=1}^N \sqrt{\Delta x_i^2 + \Delta y_i^2}
  = \sum_{i=1}^N \sqrt{(x_i - x_{i-1})^2 + (y_i - y_{i-1})^2} ,
 \label{as:eq54}
\end{equation}
where the initial point of the curve, $(x_0,y_0)$, is $M_3$, the final
point is $M_1$, and $(x,y) = (x_N,y_N)$ is some intermediate point
along the path.
Furthermore, the reaction path in this case corresponds to the steepest
descent curve from the relative maxima ($TS$s) to the adjacent minima
($M$s).

Regarding the dynamical analysis, the M\"uller-Brown potential energy surface is assumed to
describe a process governed by a proton transfer, for instance, Schr\"odinger's
equation is applied to a particle with mass $m = 1,\!836$ (atomic
units, where $m_e = 1$ is the electron mass).
Quantum-mechanically, the initial state is specified by an initial Gaussian
wave packet,
\begin{equation}
 \Psi_0(x,y) = \frac{1}{\sqrt{2\pi \sigma_x \sigma_y}}
 \ e^{-(x-x_0)^2/4\sigma_x^2
  -(y-y_0)^2/4\sigma_y^2 + i p_{x,0} (x - x_0)/\hbar
  + i p_{y,0} (y - y_0)/\hbar} ,
 \label{as:eq55}
\end{equation}
with $\sigma_x^2 = \sigma_y^2 = \sigma_0^2 = 0.0125$ and
$(p_{x,0},p_{y,0}) = (-p_0,p_0)$.
Physically, from a classical view point, this wave packet is assumed to
represent a Gaussian random distribution of initial positions around the
minimum of the products state, and a (Gaussian) Boltzmann distribution
of initial momenta [which can be readily obtained from the momentum
representation of the wave packet (\ref{as:eq55})], with dispersion
$\hbar/2\sigma_0$ along both directions and average momenta
$(p_{x,0},p_{y,0})$.

The value $p_0$ is considered as
a control parameter to determine the effectiveness of the transfer
process as a function of the initial energy carried by the proton.
The initial conditions for the Bohmian trajectories are obtained by
randomly sampling $\rho_0 = |\Psi_0|^2 \delta(p_x - p_{x,0}) \delta
(p_y - p_{y,0})$.
To compare with, alternative classical statistical simulations are also
considered in order to emphasize different quantum-classical aspects.
To this end, two samples of classical trajectories are considered.
In one of them, trajectories are initially distributed according to
the Wigner distribution associated with (\ref{as:eq55}),
\begin{eqnarray}
 \rho_{0,cl}^{(1)} & = & \rho_W(x,p_x,y,p_y) \nonumber \\
 & \propto &
  e^{-(x-x_0)^2/2\sigma_x^2 - \sigma_x^2 (p_x-p_{x,0})^2/\hbar^2
     -(y-y_0)^2/2\sigma_y^2 - \sigma_y^2 (p_y-p_{y,0})^2/\hbar^2} .
 \label{as:eq56}
\end{eqnarray}
This distribution introduces a dispersion in momenta such that the
classical average energy,
\begin{eqnarray}
 \bar{E}_{cl} & = & \int E(x,p_x,y,p_y) \rho_W(x,p_x,y,p_y) dx dy dp_x dp_y
  \nonumber \\
  & = & \frac{p_0^2}{m} + \bar{V} + \bar{\delta} \nonumber \\
  & \approx & \frac{1}{N} \sum_{i=1}^N E_i (x_0^i,y_0^i) ,
 \label{as:eq57}
\end{eqnarray}
coincides with the (quantum) expectation value of the energy,
$\bar{E}_{\rm q} = \langle \hat{H} \rangle$.
In Eq.~(\ref{as:eq57}), $\bar{V} = \sum_{i=1}^N V_i(x_0^i,y_0^i)/N$ is
the average value of the potential energy,
$\bar{\delta} = \hbar^2/4m\sigma^2$ is a sort of internal energy
related to the {\it spreading ratio} of a Gaussian wave packet
\cite{as:sanz9}, and the last sum runs over all (classical/quantum)
particles considered [with ($x_0^i, y_0^i$) denoting their
corresponding initial positions].
In the second ensemble, the classical trajectories are distributed
according to $\rho_{0,cl}^{(2)} = \rho_0$, just as in the Bohmian case.
By means of a simple calculation, it can be shown that $\bar{\delta}$
vanishes and therefore the average (ensemble) energy becomes
smaller.

The first calculation considered is a measure of
reaction probabilities, which provides information about the amount of
products formed with time, as well as the formation rate or reaction
velocity if we measure the slope of such a function.
The reaction probability is defined in the form of a
restricted norm \cite{as:sanz8,as:sanz10},
\begin{equation}
 \mathcal{P}(t) \equiv \int_\Sigma |\Psi(x,y,t)|^2 dx dy ,
 \label{as:eq58}
\end{equation}
where $\Sigma$ is the space region above the border line separating
products from pre-equilibrium/reactants, here chosen to be
\begin{equation}
 y_{R \to P}(x) = 0.8024 x + 1.2734 .
\end{equation}
From a Bohmian viewpoint, Eq.~(\ref{as:eq58}) has to be interpreted
\cite{as:sanz8,as:sanz10} as the fraction of trajectories $N_\Sigma$
that penetrate into $\Sigma$ at a time $t$ with respect to the total
number $N$ initially considered,
\begin{equation}
 \mathcal{W}(t) \equiv \frac{N_\Sigma (t)}{N} .
 \label{as:eq59}
\end{equation}
This quantity will approach $\mathcal{P}(t)$ in the limit $N\to\infty$
whenever the initial conditions of the trajectories will be sampled
according to $\rho_0$.
Classically, the analog of Eq.~(\ref{as:eq59}) is also considered to
determine a classical rate of production of products, using a
subscript ``cl'' to distinguish this quantity from its quantum-mechanical
counterpart [i.e., $\mathcal{W}_{cl}(t)$ for the classical fraction].
Probability can flow backwards from products to reactants
\cite{as:wyatt2-1,as:wyatt2-2,as:wyatt2-3}, mainly in bound potentials,
thus leading to multiple crossings of $y_{R \to P}(x)$ by the same
quantum/classical trajectory.
However, working with individual trajectories brings in an advantageous
feature: one can determine uniquely when a single particle is in the
products region, and therefore neglect its count in (\ref{as:eq59})
at subsequent times.
Hence, another interesting quantity is the fraction of trajectories
going from reactants to products without recrossing the border line at
subsequent times,
\begin{equation}
 \bar{\mathcal{W}}(t) \equiv \frac{\bar{N}_\Sigma (t)}{N} .
 \label{as:eq60}
\end{equation}
Assuming one could extract the products formed during the reaction
by some chemical or physical procedure, $\bar{\mathcal{W}}(t)$ would
provide the maximum amount of products at each time and, at
$t \to \infty$, it would render the maximum amount of products which
can be extracted from the reaction given a certain initial state.

\begin{figure}[t]
 \centering
 \includegraphics[width=12.5cm]{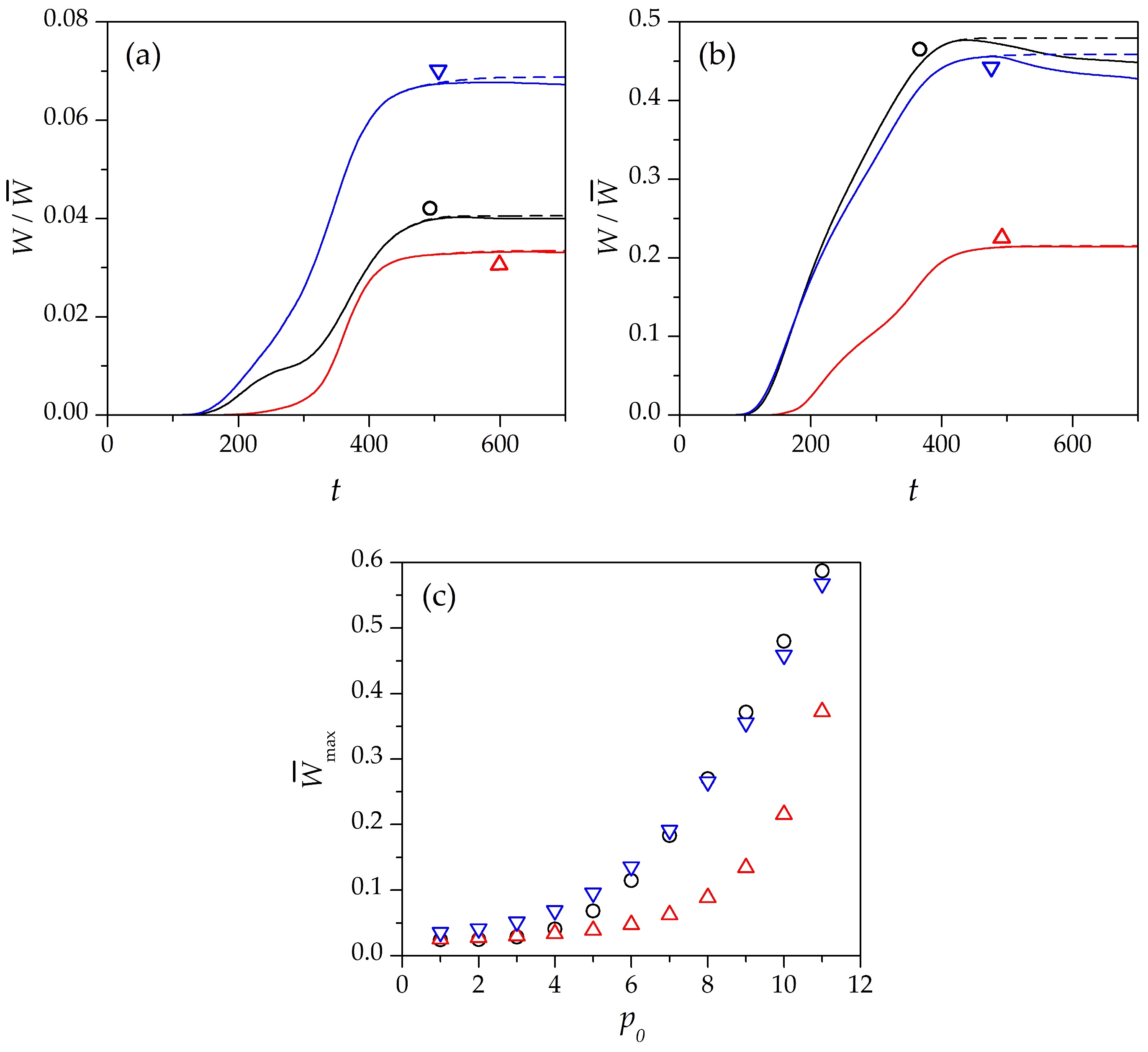}
 \caption{Reaction probabilities $\mathcal{W}$ (solid curve) and
  $\bar{\mathcal{W}}$ (dashed curve) for: (a) $p_0 = 4$ and (b)
  $p_0 = 10$, and three different initial distributions: quantum
  trajectories distributed according to $\rho_0$ (circles) and
  classical trajectories distributed according to $\rho_0$ (triangles)
  and $\rho_W$ (inverted triangles).
  In panel (c), $\bar{\mathcal{W}}$ at $t \approx 700$ as a function
  of $p_0$: $\bar{\mathcal{W}}_{\rm Bohm}$ (circles),
  $\bar{\mathcal{W}}_{\rm Bohm}^{cl}$ (triangles) and
  $\bar{\mathcal{W}}_{\rm Wigner}^{cl}$ (inverted triangles).}
 \label{as:fig2}
\end{figure}

The fractions $\mathcal{W}$ (solid line) and $\bar{\mathcal{W}}$ (dashed line)
are displayed in Figs.~\ref{as:fig2}(a) and (b) for $p_0 = 4$ and $p_0 = 10$,
respectively.
In each panel the three different initial distributions mentioned above are
shown: Bohmian trajectories according to $\rho_0$ (circles) and classical
trajectories according to $\rho_{0,cl}^{(1)}$ (triangles) and $\rho_{0,cl}^{(2)}$
(inverted triangles), using a total of $5 \times 10^4$ trajectories in all cases.
For $p_0 = 4$, $\bar{E}-\bar{\delta}$ is well below $TS_2$ and $TS_1$,
and therefore dynamics should mainly proceed via tunneling according
to standard quantum mechanics.
This means that one would expect $\mathcal{W}$ to
be larger asymptotically than $\mathcal{W}_{cl}^{(1)}$ and
$\mathcal{W}_{cl}^{(2)}$.
However, in Fig.~\ref{as:fig2}(a) we observe that
$\mathcal{W}$ is between $\mathcal{W}_{cl}^{(1)}$
and $\mathcal{W}_{cl}^{(2)}$, i.e., there is an excess of classical
trajectories which can pass the transition states and reach
products.
The reason for this behavior is that classical distributions can
explore many initial conditions, which eventually may imply energies of
individual trajectories higher than the energy of the transition states.
Eventually this leads to the formation of products even in cases where
the average (ensemble) energy is below the energy of the transition states.
This effect will be more relevant in the case of trajectories
distributed according to $\rho_{0,cl}^{(1)}$ than in the case of
$\rho_{0,cl}^{(2)}$.
On the other hand, for Bohmian trajectories the dynamics is very
different due to the presence of the quantum potential: for low $p_0$,
the wave packet spreads faster than it propagates \cite{as:sanz9},
thus favoring the formation of ripples by interference which will
hinder the passage of Bohmian trajectories to products --- remember that
in Bohmian mechanics tunneling cannot be understood as in standard
quantum mechanics, but as a process mediated by an ``effective''
time-dependent barrier \cite{as:dewdney2}, $V_{\rm eff}$.
Nevertheless, in both cases, quantum and classical, we note that the
maximum formation of products, $\bar{\mathcal{W}}$ is almost the same
as $\mathcal{W}$.

For $p_0 = 10$, however, $\bar{E}-\bar{\delta}$ is above $TS_1$, and
hence a larger amount of products is expected, as it can be
seen in Fig.~\ref{as:fig2}(b), where we observe that the Bohmian
distribution and the classical one $\rho_{0,cl}^{(1)}$ provide
similar values (although the latter goes below the former).
In this case, although tunneling may still be active, the direct passage
is going to control the dynamics in both cases, classical and
quantum-mechanical.
Note that in the Bohmian case the propagation is now faster than the
spreading of the wave packet, and therefore more trajectories can be
promoted to products before interference starts to influence the
dynamics.
Regarding $\bar{\mathcal{W}}$, we find a trend similar to
$\mathcal{W}$, but the difference between the asymptotic values of
these magnitudes has increased due to the larger energy (in average)
carried by the particles, which favors the recrossing.
Only for the distribution $\rho_{0,cl}^{(2)}$ such a difference is
negligible, since there is not much energy in excess.

\begin{figure}[!ht]
 \centering
 \includegraphics[height=17cm]{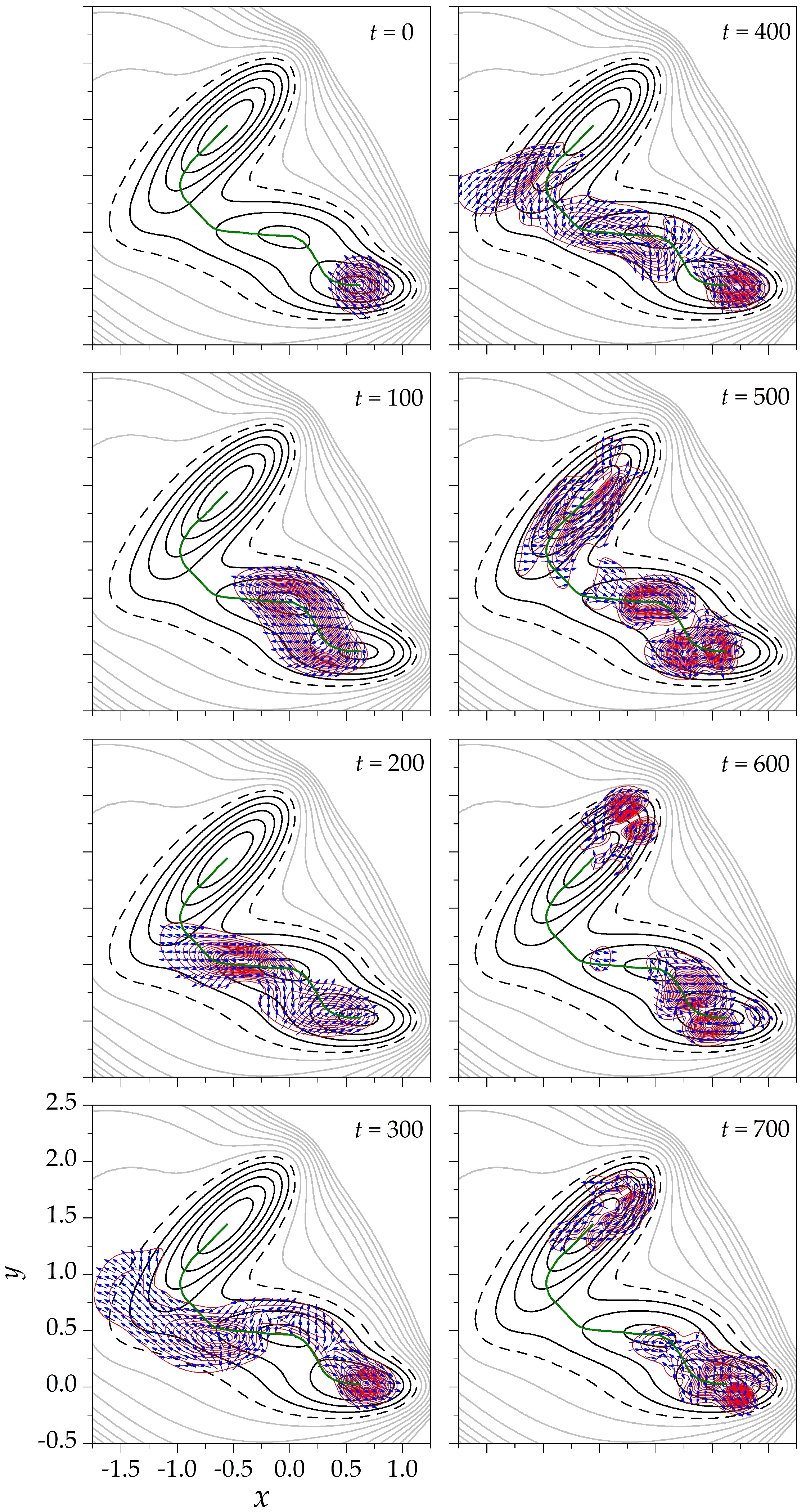}
 \caption{Time-evolution of the probability density (red contours) associated
  with an initial Gaussian wave packet starting on the
  reactants minimum with $p_0 = 9$ (see text for details).
  The superimposed (blue) arrows show the directionality of the quantum flux,
  thus providing a clue on the ensemble (Bohmian) trajectory dynamics.
  Several contours of the M\"uller-Brown potential energy surface (with the
  same meaning as in Fig.~\ref{as:fig3}) as well as the reaction path have
  also been included for a better understanding of the time-evolution of
  the probability density and the associated flux.}
 \label{as:fig3}
\end{figure}

In order to compare the maximum amount of products formed in a more
general way, in Fig.~\ref{as:fig2}(c) there is a comparison among the
three initial distributions for a range of $p_0$.
As can be noticed, the formation of products is more efficient
classically than quantum-mechanically for low values of $p_0$, while
this behavior changes as $p_0$ becomes larger (the switching
appears \cite{as:sanz8} for $p_0 \sim 8$, just when $\bar{E}$
approximately coincides with the energy of the transition state
$T_1$, which connects the pre-equilibrium with products).

The dynamics associated with the process here described is illustrated
in Fig.~\ref{as:fig3}, where a series of snapshots display the
time-evolution of the probability density associated with a Gaussian
wave packet (equally spaced darker contours) starting on the reactants
minimum with $p_0 = 9$.
As time proceeds the wave packet evolves essentially along the
direction indicated by the reaction path (thicker solid line).
However, as can be seen particularly in the plots at $t = 100$, 200 and
300, the motion of the wave packet is quite similar to that shown by a water
stream when flowing along a river bed, trying to burst its banks each
time it reaches a meander.
Specifically, at $t = 300$ we observe that the main stream deviates
remarkably from the reaction path, although later this quantum fluid gets
back and moves towards the products region (see plot at $t = 500$).
This deviation beyond the reaction path, which becomes larger with
$p_0$, was called {\it quantum bobsled effect} by
Marcus \cite{as:marcus}.
This effect, observed by McCullough and Wyatt \cite{as:wyatt2-1,%
as:wyatt2-2,as:wyatt2-3} when studying the dynamics of the collinear
H+H$_2$ reaction dynamics, arises from a direct analogy to the
deviation undergone by a bobsled from the center of the track where it
moves through due to its inertia (a {\it positive centrifugal effect},
according to Marcus \cite{as:marcus}).
Following with the analogy of the river, we observe that, after some
time, when the passage to products has occurred the quantum fluid
evolves more slowly and only shape changes are noticeable, just the
same case after a region of ``brave waters''.
At this stage, if the dynamics continues, there can be some evolution
from products to reactants (and vice versa), but there are no important
inertial effects.

The information provided by the probability density results very
valuable in order to determine the evolution of the system.
However, Bohmian trajectories present some advantages which cannot be
noticed studying only the wave packet dynamics.
The case analyzed here is a clear example.
Instead of displaying trajectories, which would result quite messy
after some time, we have superimposed an arrow map on the region occupied
by the wave packet at each time.
The direction of the arrows indicate at each particular point of space
the directionality of the vector field $\vec{v}$, and therefore the
local direction of the flow (in this regard, Bohmian trajectories will
be tangent to these vectors).
According to these maps, we observe how from a situation where all the
arrows are aligned along the same direction at $t = 0$, we pass to
another at $t = 100$ where part of the arrows point downwards and part
along the direction indicated by the reaction path.
This happens precisely because the wave packet has collided with the
first ``meander''; something similar can be seen at $t = 200$, when
the wave packet collides with the second ``meander'', and so on.
The arrows also allow us to detect the presence of {\it quantum
vorticity} \cite{as:holland,as:sanz13,as:sanz15}, i.e., the vortical
motion that appears whenever the wave function displays a node.
In this case, the arrows twist around the node, giving rise to a
{\it quantum whirlpool effect} \cite{as:wyatt2-1,as:wyatt2-2,%
as:wyatt2-3}.


\subsection{Scattering dynamics: Young's two-slit experiment}
\label{as:sec4.4.3}

\begin{figure}[t]
 \centering
 \includegraphics[width=12.5cm]{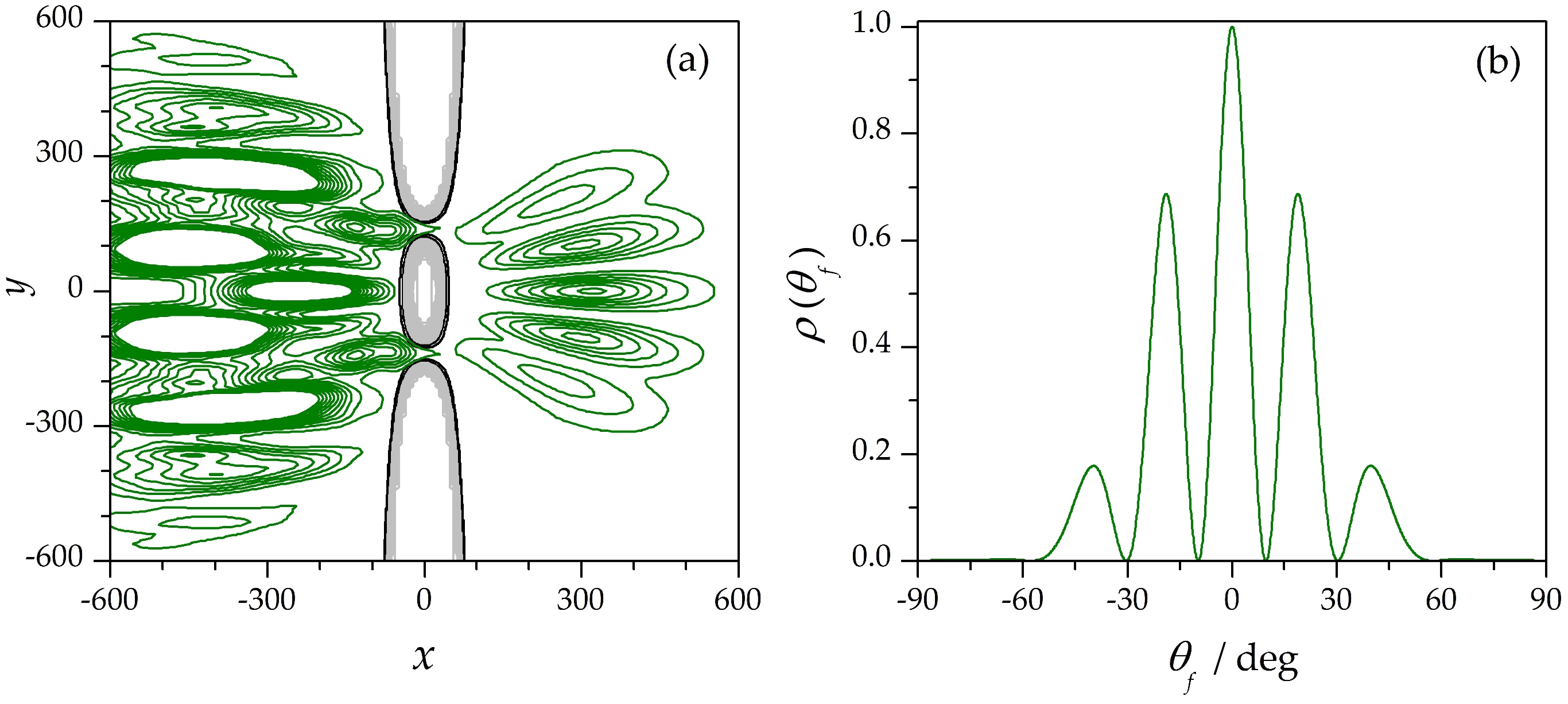}
 \caption{(a) Contour plot of the probability density associated with an
  electron crossing the two slits modeled by the potential energy surface
  (\ref{as:eq61}) at the end of the simulation, i.e., the time when the
  interference pattern is already formed \cite{as:sanz1,as:sanz5}.
  Given the high-backscattering component, for a better understanding the
  contours are truncated to a certain value that allows an optimal visualization
  of the transferred density.
  Moreover, the contour plot of the potential energy surface is also displayed,
  with black contours for equipotentials lesser than or equal to
  $\langle E \rangle_i \simeq 500$, and gray for higher values.
  (b) Angular distribution of the intensity computed at the end of the
  simulation (see text for details).}
 \label{as:fig4}
\end{figure}

In order to illustrate a practical application of Bohmian mechanics
now in the case of a scattering system, we are going to consider a
model of Young's experiment with electrons where the
interaction between the electrons and the two slits is described
by a repulsive potential energy surface (barrier) \cite{as:sanz1,as:sanz5},
\begin{equation}
 V(x,y) = \left( V_0 - \frac{1}{2} \ m \omega^2 y^2 +
  \frac{m^2 \omega^4 y^4}{16 V_0} \right) e^{-x^2/\alpha^2} ,
 \label{as:eq61}
\end{equation}
where $\alpha = 25$, $\omega = 600$, $V_0 = 8000$, and $m$ is the
electron mass.
As the ingoing wave function we consider a quasi-plane or
quasi-monochromatic initial wave function, with energy $\langle E
\rangle_i \simeq 500$, which is constructed by linearly superimposing
a series of identical Gaussian wave packets along the direction
parallel to the $y$-axis.
This wave function is then launched perpendicularly to the double-slit
from a distance $\langle x \rangle_0 = - 400$ (far enough from the
potential energy surface, where $V \approx 0$).
The interest in this kind of models relies on the fact that physical
diffracting systems are constituted by atoms or molecules which
interact with the diffracted electrons, and therefore
a description of the experiment including the interaction potentials
or potential energy surfaces results closer to the real experiment
\cite{as:toennies-1,as:toennies-2,as:toennies-3,as:toennies-4}.
In Fig.~\ref{as:fig4}(a), a contour plot of the probability density
after the collision and diffraction from the two slits is shown.
As can be noted, for this incidence energy, there is a large portion
of the wave function which is back-scattered, while an angular
distribution of very well-defined diffracted (forward-scattered)
peaks appear behind the slits (for $y > 0$).
The corresponding angular diffraction pattern is represented in
Fig.~\ref{as:fig4}(b), where the Gaussian-like envelope can be
associated with the particular form chosen for the ``holes'' of the
slits \cite{as:sanz4,as:sanz5}.

\begin{figure}[t]
 \centering
 \includegraphics[width=12.5cm]{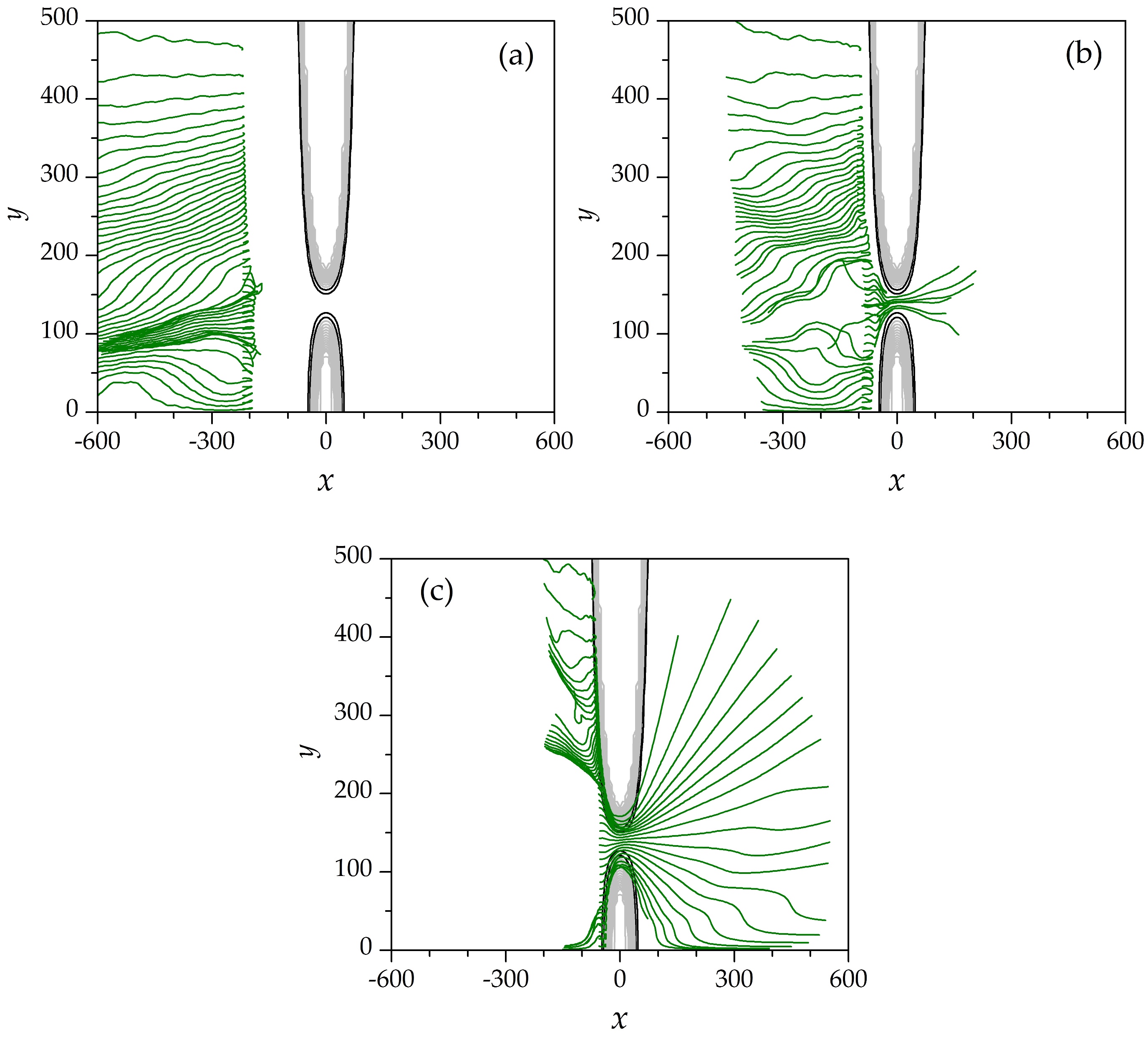}
 \caption{Bohmian trajectories illustrating the probability flow in the
  Young-type experiment with the potential energy surface
  (\ref{as:eq61}).
  Three different sets of initial conditions have been chosen, which are shown
  in the three corresponding panels.
  In these panels, the initial value along the $y$ direction is always the same,
  with $x_0$ ranging from 0 to about 500, while the three specific values have
  been chosen for $y_0$ with respect to the initial probability density: (a) at the
  rearmost part, (b) at the center, and (c) in the foremost part (see text for
  details).
  The propagation is up to the final time in the simulation, which corresponds to
  the case shown in Fig.~\ref{as:fig4}.
  For a better visualization in panels (a) and (b), only the back-scattered or diffracted
  part of the trajectories has been plotted.
  Moreover, the contour plot of the potential energy surface is also displayed,
  with black contours for equipotential lines lesser than or equal to
  $\langle E \rangle_i \simeq 500$, and gray for higher values.}
 \label{as:fig5}
\end{figure}

In order to understand the dynamics that is taking place within this
diffraction scenario (but that can also be extrapolated to any other
scattering scenario), now we consider it under a Bohmian or quantum
hydrodynamical, where the probability density describes a swarm of
noninteracting particles.
Thus, consider the three panels of Fig.~\ref{as:fig5}, where only a
half of the trajectories (those corresponding to the upper slit) has
been represented to make clearer the figures (moreover, their incident
parts are not shown either) taking advantage of the reflection symmetry
with respect to $y=0$ --- in Bohmian mechanics, the trajectories coming
from each slit cannot intersect because of the {\it non-crossing flux}
property \cite{as:sanz9} (see below).
The ensembles of trajectories in each panel have been chosen to sample
three regions of the initial probability density, i.e., varying their
initial position $y_0$, but with the same $x_0$-position: (a) $x_0 =
\langle x \rangle_0 - 100$, (b) $x_0 = \langle x \rangle_0$, and (c)
$x_0 = \langle x \rangle_0 + 100$.
As can be readily seen, the initial position is fundamental in the
quantum dynamics displayed by the trajectories, for it leads to very
different behaviors due to the action of $Q$, something which does not
happen in classical mechanics (provided classical trajectories are
launched far enough from the interaction region, their behavior would
be independent on their $x_0$-position \cite{as:sanz10}).
Note that those trajectories launched at distances further away from
the double-slit barrier will not reach it [see Fig.~\ref{as:fig5}(a)],
while only those started at closer distances will be able to surmount
it [see Fig.~\ref{as:fig5}(c)].
In the case of the trajectories started at intermediate positions [see
Fig.~\ref{as:fig5}(b)], they will remain near the barrier until the
``pressure'' exerted by those coming behind will diminish and either
they will be able to move backwards or pass through the slits.
This is a general effect which can also be observed in other scattering
problems \cite{as:sanz10,as:sanz13}, which can be understood in terms
of a sort of {\it quantum pressure} (within a hydrodynamical viewpoint).
The presence of a quantum pressure can be better understood when the
quantum Euler equation (\ref{as:eq16}) is recast in terms of the
velocity components.
Then, the quantum force (arising from the quantum potential) can be
rearranged \cite{as:takabayasi-1,as:takabayasi-2,as:holland} as
\begin{equation}
 F_i = \frac{1}{\rho} \frac{\partial \sigma_{ik}}{\partial r_k} ,
\end{equation}
where $i,k = x, y, z$, with
\begin{equation}
 \sigma_{ik} \equiv - \frac{\hbar^2 \rho}{4m}
  \frac{\partial^2 \ln \rho}{\partial r_i \partial r_k}
\end{equation}
being a quantum pressure tensor, in analogy to the classical pressure
tensor $p \delta_{ik}$.
Thus, taking into account this hydrodynamical scenario, it is clear
that electrons will tend to move towards those regions with lower
values of the quantum pressure.

The number of particles passing through the slits is a function of the
energy $E_i$ as well as the particular form of the potential energy
surface.
However, given a configuration of the potential energy surface, it is
clear that for a certain incidence energy, those electrons associated
with the rearmost part of the incident probability density (swarm)
will be more likely to get diffracted.
The fraction of diffracted electrons will then increase as the
incidence energy also increases, which means that electrons coming
from layers behind will also start to get diffracted.
But, more importantly, by means of this Bohmian picture we are able
to determine which electrons are going to contribute to diffraction
and, more specifically, to which diffraction peak \cite{as:sanz14},
something that cannot be known within the standard quantum framework.
As seen in Fig.~\ref{as:fig5}(c), the interference behind the slits
manifests by grouping or channeling the trajectories along the
diffraction angles (along which electrons display an essentially free
motion) undergoing a strong ``repulsive''-like behavior whenever they
approach regions of fast variation of the quantum potential.
The formation of these channels within the so-called Fraunhofer
region \cite{as:sanz14} is a direct consequence of the information that
the quantum potential transmits to particles about the status of each
slit (either open or close).
This is also related with the non-crossing flux
property \cite{as:sanz9} of Bohmian mechanics mentioned above, which
arises from the fact that two (Bohmian) trajectories cannot coincide on
the same space point at the same time due to the single-valuedness of
$S$ (except on a nodal point).
In this way, for the symmetric (with respect to the $y=0$) double slit
we are considering, all those trajectories started above $y=0$ will
not be able to cross to the region $y<0$, for this would violate
the non-crossing flux property.


\subsection{Effective dynamical treatments: Decoherence and reduced Bohmian trajectories}
\label{as:sec4.4.4}

An interesting description of Bohmian mechanics arises from the field
of decoherence and the theory of open quantum systems.
Here, in order to extract useful information about the system of
interest, one usually computes its associated reduced density matrix
by tracing the total density matrix, $\hat{\rho}(t)$, over the
environment degrees of freedom.
In the configuration representation and for an environment constituted
by $N$ particles, the system reduced density matrix is obtained after
integrating $\hat{\rho}(t) \equiv |\Psi(t)\rangle\langle\Psi(t)|$
over the 3$N$ environment degrees of freedom, $\{\vec{r}_i\}_{i=1}^N$,
\begin{equation}
 \tilde{\rho} (\vec{r},\vec{r}',t) =
  \int \langle \vec{r},\vec{r}_1, \vec{r}_2, \ldots , \vec{r}_N |
  \Psi (t)\rangle
  \langle \Psi (t)| \vec{r}', \vec{r}_1, \vec{r}_2,
  \ldots , \vec{r}_N \rangle\ {\rm d}\vec{r}_1 {\rm d}\vec{r}_2
  \cdots {\rm d}\vec{r}_N .
 \label{eq7}
\end{equation}
The system (reduced) quantum density current can be derived from this
expression, being
\begin{equation}
 \tilde{\vec{J}} (\vec{r},t) \equiv \frac{\hbar}{m}
  \ {\rm Im} [ \vec{\nabla}_{\vec{r}} \tilde{\rho} (\vec{r},\vec{r}',t)]
  \Big\arrowvert_{\vec{r}' = \vec{r}} \  ,
 \label{eq12}
\end{equation}
which satisfies the continuity equation
\begin{equation}
 \dot{\tilde{\rho}} + \vec{\nabla} \tilde{\vec{J}} = 0 .
 \label{eq13}
\end{equation}
In Eq.~(\ref{eq13}), $\tilde{\rho}$ is the diagonal element [i.e.,
$\tilde{\rho} \equiv \tilde{\rho} (\vec{r},\vec{r},t)$] of the
reduced density matrix and gives the measured intensity \cite{decoh}.

Taking into account Eqs.~(\ref{eq12}) and (\ref{eq13}), now we define
the velocity field, $\dot{\vec{r}}$, associated to the (reduced) system
dynamics as
\begin{equation}
 \tilde{\vec{J}} = \tilde{\rho} \dot{\vec{r}} ,
 \label{vfield}
\end{equation}
which is analogous to the Bohmian velocity field.
Now, from Eq.~(\ref{vfield}), we define a new class of quantum
trajectories as the solutions to the equation of motion
\begin{equation}
 \dot{\vec{r}} \equiv \frac{\hbar}{m}
  \frac{{\rm Im} [ \vec{\nabla}_{\vec{r}} \tilde{\rho} (\vec{r},\vec{r}',t)]}
       {{\rm Re} [ \tilde{\rho} (\vec{r},\vec{r}',t)]}
   \Bigg\arrowvert_{\vec{r}' = \vec{r}} .
 \label{eq14}
\end{equation}
These new trajectories are the so-called {\it reduced quantum
trajectories} \cite{as:sanz6-1,as:sanz6-2}, which are only related
to the system reduced density matrix.
As shown in \cite{as:sanz6-1,as:sanz6-2}, the dynamics described by
Eq.~(\ref{eq14}) leads to the correct intensity [with its time-evolution
being described by Eq.~(\ref{eq13})] when the statistics of a large number
of particles is considered.
Moreover, it is also straightforward to show that Eq.~(\ref{eq14})
reduces to the well-known expression for the velocity field in Bohmian
mechanics when there is no interaction with the environment.
An application of this approach to the analysis of dephasing by incoherence
in Young's experiment and its influence on the transverse momentum transfer
\cite{as:sanz:AOP:2015-1} can be found in \cite{as:sanz:AOP:2015-2}.


\subsection{Pathways to complex molecular systems: Mixed Bohmian-classical mechanics}
\label{as:sec4.4.5}

As mentioned in Sec.~\ref{as:sec4.2.4}, there are different hybrid
approaches to deal with many degree-of-freedom systems, where the
system degrees of freedom are described quantum-mechanically while
the bath ones are accounted for classically.
Among them, we find methods such as the mean-field
approximation \cite{as:billing} or the surface hopping
trajectories \cite{as:tully}.
In all these methods, the key point is the implementation of the
so-called {\it backreaction} \cite{as:prezhdo}, i.e., the action of the
system over the bath, since the contrary is simple and it is usually
done in terms of a time-dependent potential which is function of the
bath coordinates (following the Born-Oppenheimer approximation scheme).
Here, we are going to analyze this problem within the framework of the
so-called mixed quantum-Bohmian approach
\cite{as:beswick-1,as:beswick-2,as:beswick-3,as:beswick-4}.

Consider two interacting systems $X$ and $Y$ (or degrees of freedom), each
specified by the coordinates $x$ and $y$, respectively, and with associated
masses $m_x$ and $m_y$.
Proceeding as in Sec.~\ref{as:sec4.3}, with the polar ansatz
(\ref{as:eq10}) and the corresponding two-dimensional Schr\"odinger
equation, Eqs.~(\ref{as:eq11}) and (\ref{as:eq12}) read as
\begin{eqnarray}
 \frac{\partial S}{\partial t} +
  \frac{1}{2m_x} \left( \frac{\partial S}{\partial x} \right)^2 +
  \frac{1}{2m_y} \left( \frac{\partial S}{\partial y} \right)^2 & = &
  - V_{\rm eff} ,
 \label{eq:2.5.22} \\
 \frac{\partial R^2}{\partial t} +
  \frac{1}{m_x} \frac{\partial S}{\partial x}
   \frac{\partial R^2}{\partial x} +
  \frac{1}{m_y} \frac{\partial S}{\partial y}
   \frac{\partial R^2}{\partial y}
 & = & - R^2 \left( \frac{1}{m_x} \frac{\partial^2 S}{\partial x^2}
               + \frac{1}{m_y} \frac{\partial^2 S}{\partial y^2}
      \right) ,
 \label{eq:2.5.23}
\end{eqnarray}
where the quantum potential has the form
\begin{equation}
 Q(x,y,t) =
  - \frac{\hbar^2}{2m_x} \frac{1}{R}
      \frac{\partial^2 R}{\partial x^2}
  - \frac{\hbar^2}{2m_y} \frac{1}{R}
      \frac{\partial^2 R}{\partial y^2} ,
 \label{eq:2.5.24}
\end{equation}
which is non-factorizable (see Sec.~\ref{as:sec4.3.2}).
Let us recast Eq.~(\ref{eq:2.5.22}) in its Eulerian form by applying
the operators $\partial/\partial x$ and $\partial/\partial y$ to both
sides of this equation.
This gives rise to two coupled equations,
\begin{eqnarray}
 \frac{\partial}{\partial t}
 \left( \frac{\partial S}{\partial x} \right) +
    v_x\ \frac{\partial^2 S}{\partial x^2} +
    v_y\ \frac{\partial^2 S}{\partial y \partial x} & = &
  - \frac{\partial V_{\rm eff}}{\partial x} ,
 \label{eq:2.5.26} \\
 \frac{\partial}{\partial t}
 \left( \frac{\partial S}{\partial y} \right) +
   v_y\ \frac{\partial^2 S}{\partial y^2} +
    v_x\ \frac{\partial^2 S}{\partial x \partial y} & = &
   - \frac{\partial V_{\rm eff}}{\partial y} ,
 \label{eq:2.5.27}
\end{eqnarray}
where $v_x = p_x/m_x = (1/m_x)(\partial S/\partial x)$ and
$v_y = p_y/m_y = (1/m_y)(\partial S/\partial y)$.
Taking into account the definition of the Lagrange time
derivative (see Sec.~\ref{as:sec4.3}), these two equations
can be recast in a more familiar Newtonian form,
\begin{eqnarray}
 m_x\ \frac{d^2 x}{dt^2} & = &
   - \frac{\partial V_{\rm eff}}{\partial x} , \\
 \label{eq:2.5.27bb-1}
 m_y\ \frac{d^2 y}{dt^2} & = &
   - \frac{\partial V_{\rm eff}}{\partial y} .
 \label{eq:2.5.27bb-2}
\end{eqnarray}

So far no approximation has been invoked.
Let us now consider, for instance, that $m_y \gg m_x$, which physically
means that for the timescale ruling the behavior of $X$, the system $Y$
behaves almost classically.
This slower evolution along the $y$ coordinate means that the second
space-derivatives of $S$ and $R$ with respect to this coordinate will be
negligible.
Accordingly, Eqs.~(\ref{eq:2.5.26}) and (\ref{eq:2.5.27}) can be recast as
\begin{eqnarray}
 \frac{\partial}{\partial t}
 \left( \frac{\partial \tilde{S}}{\partial x} \right) +
    \frac{1}{m_x} \frac{\partial \tilde{S}}{\partial x}
     \frac{\partial^2 \tilde{S}}{\partial x^2} +
    \frac{1}{m_y} \frac{\partial \tilde{S}}{\partial y}
     \frac{\partial^2 \tilde{S}}{\partial y \partial x} & = &
   - \frac{\partial \tilde{V}_{\rm eff}}{\partial x} ,
 \label{eq:2.5.28} \\
 \frac{\partial}{\partial t}
 \left( \frac{\partial \tilde{S}}{\partial y} \right) +
    \frac{1}{m_x} \frac{\partial \tilde{S}}{\partial x}
     \frac{\partial^2 \tilde{S}}{\partial x \partial y} & = &
   - \frac{\partial \tilde{V}_{\rm eff}}{\partial y} ,
 \label{eq:2.5.29}
\end{eqnarray}
where $\tilde{S}$ and $\tilde{R}$ denote the approximate values of
$S$ and $R$, respectively.
In these equations, $\tilde{V}_{\rm eff}$ is the corresponding
approximate effective potential, with
\begin{equation}
 \tilde{Q}(x,t|y) =
  - \frac{\hbar^2}{2m_x} \frac{1}{\tilde{R}}
      \frac{\partial^2 \tilde{R}}{\partial x^2} ,
 \label{eq:2.5.24bb}
\end{equation}
where $(x,t|y)$ means that $\tilde{Q}$ depends on $y$ implicitly
through a sort of parametrization --- as it also happens in the Born-Oppenheimer
approximation (see Sec.~\ref{as:sec4.2.1}).
On the other hand, (\ref{eq:2.5.23}) becomes the approximate continuity
equation,
\begin{eqnarray}
 \frac{\partial \tilde{R}^2}{\partial t} & + &
  \frac{\partial}{\partial x} \,
   \left( \frac{\tilde{R}^2}{m_x}
    \frac{\partial \tilde{S}}{\partial x} \right) +
  \frac{1}{m_y} \frac{\partial \tilde{S}}{\partial y}
   \frac{\partial \tilde{R}^2}{\partial y} = 0 .
 \label{eq:2.5.30}
\end{eqnarray}

Evaluating (\ref{eq:2.5.28}) and (\ref{eq:2.5.30}) along the
quasi-classical trajectory $y(t)$ allows us to define the
pseudo-Lagrangian time derivative operator
\begin{equation}
 \frac{d\ }{dt} = \frac{\partial \ }{\partial t}
  + v_y\ \frac{\partial \ }{\partial y} ,
 \label{eq:2.5.31}
\end{equation}
and therefore to recast those equations as
\begin{eqnarray}
 \frac{d\ }{dt} \left( \frac{\partial \tilde{S}}{\partial x} \right)
 + \left( \frac{1}{m_x} \frac{\partial \tilde{S}}{\partial x} \right)
   \left( \frac{\partial^2 \tilde{S}}{\partial x^2} \right) & = &
 - \frac{\partial \tilde{V}_{\rm eff}}{\partial x} ,
 \label{eq:2.5.32} \\
 \frac{d \tilde{R}^2}{dt} + \frac{\partial}{\partial x}
   \left( \tilde{R}^2 \ \, \frac{1}{m_x}
    \frac{\partial \tilde{S}}{\partial x} \right) & = & 0 ,
 \label{eq:2.5.33}
\end{eqnarray}
which satisfy the pseudo-Schr\"odinger equation
\begin{equation}
 i\hbar \ \, \frac{d \tilde{\Psi} (x,y(t),t)}{dt}
  = \left[ - \frac{\hbar^2}{2m_x} \frac{\partial^2}{\partial x^2}
  + V(x,y(t)) \right] \tilde{\Psi} (x,y(t),t) ,
 \label{eq:2.5.34}
\end{equation}
where $\tilde{\Psi} = \tilde{R} e^{i\tilde{S}/\hbar}$.
As can be noticed, the dimensionality of the full quantum problem
reduces to the (subspace) dimensionality associated with $X$, since the
classical-like coordinate $y$ acts as a time-dependent parameter (the
external potential $V$ has become time-dependent in virtue of this
parametrization).
On the other hand, $Y$ evolves according to a quasi-classical Newtonian
equation,
\begin{equation}
 m_y\ \frac{d^2 y}{dt^2} =
  \frac{\partial }{\partial y}
   \left( V(x,y,t) + \widetilde{Q}(x,t|y) \right) ,
 \label{eq:2.5.35}
\end{equation}
which arises from Eq.~(\ref{eq:2.5.29}) after applying the
pseudo-Lagrangian operator [note that it can also be obtained
from Eq.~(\ref{eq:2.5.27bb-2}) after the corresponding approximation is
considered] and is integrated after getting the solution $\tilde{\Psi}
(x,y(t),t)$ from (\ref{eq:2.5.34}).

As briefly mentioned above, the method here described constitutes a sort of
translation of the Born-Oppenheimer rule to the case of slow and fast degrees
of freedom.
This quantum-classical approach is commonly used when dealing with quantum
open systems, whenever some degrees of freedom display a slower
dynamics than other, which are faster and in many cases can be
treated even as a surrounding noise function (thermal bath).
Within the Bohmian literature, approaches to deal with the many-body problem
have been suggested, for instance, by Garashchuk and Rassolov \cite{as:garashchuk:CPL:2002,as:garashchuk:JCP:2004,as:garashchuk:RCC:2011},
Bittner \cite{as:bittner:JCP:2003}, Makri \cite{as:makri:JCP:2003,as:makri:JPCA-1:2004,as:makri:JPCA-2:2004}
and, more recently, Franco \cite{as:franco:JCP:2017}.


\section{Concluding remarks}
\label{as:sec4.5}

Looking back, if there is a healthy and increasing interest at present in
Bohmian mechanics, it could be said it partly comes from the important numerical
work developed by Wyatt and colleagues since the end of the 1990s.
Up to that time Bohmian mechanics had been applied to some physical cases of
interest, disproving that trajectories were not incompatible at all with
quantum mechanics beyond just metaphysical discussions.
From a pragmatic viewpoint, leaving aside interpretational issues concerning
hidden variables, the reality of the wave function and the quantum realm, such
important contributions were probably marginalized because, in the end, they were
only another way to describe what was already known.
If a formulation renders the same, what is it worth for?
Beyond quantum foundations aspects, this might have been a
wondering question --- a rather reasonable and understandable one, after all ---
in people's minds (reviewers, many of them) that made them reluctant to accept
a trajectory-based description of quantum phenomena.
Notice that something similar also happened when the first time-dependent quantum-mechanical
simulations started appearing by the end of the 1960s \cite{as:goldberg:AJP:1968},
making possible the impossible, namely providing a neat and unambiguous visualization
of the full quantum process, since the initial state up to the final one.
Quantum mechanics was not a ``black box'' anymore, as claimed by Bohr and followers
over decades.
With time it was understood that these pictures detailing the process could be
used to extract valuable information about it, which could not be done (or not, at least,
in an easy manner) with the other more widespread time-independent techniques.
Nowadays this type of simulations constitute a consolidated standard tool to
approach quantum-mechanical systems, from simple academic-type problems to very
complex ones.
Something similar has also happened with Bohmian mechanics as soon as it has started
being perceived as a worth exploring and exploiting numerical methodology, particularly
in Chemistry, as Wyatt and other pioneers in the field have shown along the about last
twenty years.
In this regard, it is possible to find Bohmian-based algorithms to determine electronic
structure as well as to describe the dynamics of many-body systems.

Besides the computational aspect, the interpretational one has also gained in relevance
over the last decades, just as soon as relatively complex (more realistic) problems have
started being considered, from atomic and molecular scattering to nanodevices, from
nonlinear optics to electron microscopy.
For years, molecular dynamics and statistics have mainly relied on the computation
of classical trajectories (based on both Newtonian and Hamiltonian schemes) and,
in general, the tools of classical mechanics.
Good for the purpose, but approximate (at different degrees, depending on the
method and/or problem considered) because they lack the essence of quantum
mechanics, namely coherence and all the related properties, such as interference,
tunneling, or entanglement.
With the extensive use and application of quantum mechanics in the development
of electronic structure methods, first, and then in molecular dynamics through
wave-packet propagation methods, better and more accurate solutions were
obtained, of course.
However, interpretations of the corresponding outcomes still had to make use
of classical argumentations based on the longstanding quantum-classical
correspondence thinking established by Bohr back in the 1920s.
Bohmian mechanics has allowed to change the landscape, providing a correct
trajectory-based interpretation for quantum phenomena (compared to classical
approaches), that is, in consonance with the evolution of the corresponding
wave functions and without introducing any extra {\it ad hoc} artifacts (e.g.,
quantization rules over classical ensembles).
Of course, this does not mean at all that we should now make the mistake of
only seeking for Bohmian trajectories to the detriment of trying to still establish
bridges between quantum and classical dynamics with the aid of classical
trajectories.
So far, the quantum world has been and still is disconcerting and challenging
to us (and perhaps it will remain so for long), hence any tool to explore it
should always be welcome rather than banned.

Getting back to Chemistry, it is perhaps pertinent to conclude this Chapter
quoting the very last paragraph of Walter Kohn's Nobel Lecture \cite{as:kohn-nobel}:
\begin{quote}
 Looking into the future I expect that wavefunction-based and density based
 theories will, in complementary ways, continue not only to give us
 quantitatively more accurate results, but also contribute to a better
 physical/chemical understanding of the electronic structure of matter.
\end{quote}
Probably Bohmian mechanics, that is, the hydrodynamic picture of quantum
mechanics will play a major role in this regard.
This is something yet to come, although some signs are already on the way,
principally the change in the perception we have at present of this approach.


\section*{Acknowledgments}

This work has been financially supported by the Spanish MINECO (grant No.
FIS2016-76110-P).




\begin{thebibliography}{200}

\bibitem{as:wyatt2-1}
 E. A. McCullough and R. E. Wyatt, {\it Quantum dynamics of the
 collinear (H, H$_2$) reaction}, J. Chem. Phys. {\bf 51}, 1253 (1969).

\bibitem{as:wyatt2-2}
 E. A. McCullough and R. E. Wyatt, {\it Dynamics of the collinear
 H + H$_2$ reaction. I. Probability density and flux},
 J. Chem. Phys. {\bf 54}, 3578 (1971).

\bibitem{as:wyatt2-3}
 E. A. McCullough and R. E. Wyatt, {\it Dynamics of the collinear
 H + H$_2$ reaction. II. Energy analysis},
 J. Chem. Phys. {\bf 54}, 3592 (1971).

\bibitem{as:hirschfelder1-1}
 J. O. Hirschfelder, A. C. Christoph, and W. E. Palke, {\it Quantum
 mechanical streamlines. I. Square potential barrier},
 J. Chem. Phys. {\bf 61}, 5435 (1974).

\bibitem{as:hirschfelder1-2}
 J. O. Hirschfelder, C. J. Goebel, and L. W. Bruch, {\it Quantized
 vortices around wavefunction nodes. II},
 J. Chem. Phys. {\bf 61}, 5456 (1974).

\bibitem{as:hirschfelder1-3}
 J. O. Hirschfelder and K. T. Tang, {\it Quantum mechanical
 streamlines. III. Idealized reactive atom-diatomic molecule
 collision}, J. Chem. Phys. {\bf 64}, 760 (1976).

\bibitem{as:bohm-1}
 D. Bohm, {\it A suggested interpretation of the quantum theory in
 terms of ``hidden'' variables. I}, Phys. Rev. {\bf 85}, 166 (1952).

\bibitem{as:bohm-2}
 D. Bohm, {\it A suggested interpretation of the quantum theory in
 terms of ``hidden'' variables. II}, Phys. Rev. {\bf 85}, 180 (1952).

\bibitem{as:bohm-3}
 D. Bohm, {\it Proof that probability density approaches $|\psi|^2$ in
 causal interpretation of the quantum theory},
 Phys. Rev. {\bf 89}, 458 (1953).

\bibitem{as:takabayasi-1}
 T. Takabayasi, {\it On the formulation of quantum mechanics associated
 with classical pictures}, Prog. Theor. Phys. {\bf 8}, 143 (1952).

\bibitem{as:takabayasi-2}
 T. Takabayasi, {\it Remarks on the formulation of quantum mechanics
 with classical pictures and on relations between linear scalar fields
 and hydrodynamical fields}, Prog. Theor. Phys. {\bf 9}, 187 (1953).

\bibitem{as:holland}
 P. R. Holland, {\ it The Quantum Theory of Motion}
 (Cambridge University Press, Cambridge 1993).

\bibitem{as:duerr-1}
 D. D\"urr, {\it Bohmsche Mechanik als Grundlage der Quanten-mechanik}
 (Springer, Berlin, 2001).

\bibitem{as:duerr-2}
 D. D\"urr and S. Teufel, {\it Bohmian Mechanics}
 (Springer, Berlin, 2009).

\bibitem{as:wyatt1}
 R. E. Wyatt, {\it Quantum Dynamics with Trajectories}
 (Springer, New York, 2005).

\bibitem{as:book-1}
 A. S. Sanz and S. Miret-Art\'es, {\it A Trajectory Description of
 Quantum Processes. I. Fundamentals}, Lecture Notes in Physics {\bf 850}
 (Springer, Berlin, 2012).

\bibitem{as:book-2}
 A. S. Sanz and S. Miret-Art\'es, {\it A Trajectory Description of
 Quantum Processes. II. Applications}, Lecture Notes in Physics {\bf 831}
 (Springer, Berlin, 2014).

\bibitem{as:sanz:AJP:2017}
 A. S. Sanz, {\it Bohm's approach to quantum mechanics: Alternative theory or practical picture?},
 preprint arXiv:1707.00609v1 (2017).

\bibitem{as:madelung}
 E. Madelung, {\it Quantentheorie in hydrodynamischer Form},
 Z. Physik {\bf 40}, 332 (1926).

\bibitem{as:chattaraj-bk}
 P. K. Chattaraj, {\it Quantum Trajectories} (CRC Taylor and Francis,
 New York, 2010).

\bibitem{as:hughes-bk}
 K. H. Hughes and G. Parlant, {\it Quantum Trajectories}
 (CCP6, Daresbury, UK, 2011).

\bibitem{as:milburn}
 J. P. Dowling and G. J. Milburn, {\it The second quantum revolution},
 Phil. Trans. R. Soc. Lond. A {\bf 361}, 1655 (2003).

\bibitem{as:nielsen-chuan-bk}
 M. A. Nielsen and I. L. Chuang, {\it Quantum Computation and Quantum
 Information} (Cambridge University Press, Cambridge, 2000).

\bibitem{as:NSF-report-2016}
 J. Olson, Y. Cao, J. Romero, P. Johnson, P.-L. Dallaire-Demers, N. Sawaya, P. Narang,
 I. Kivlichan, M. Wasielewski and A. Aspuru-Guzik, {\it Quantum information and computation
 for Chemistry}, NSF Workshop Report, arXiv:1706.05413 (2016).

\bibitem{as:eberhart02}
 M. E. Eberhart, {\it Quantum mechanics and molecular design in the
 twenty first century}, Found. Chem. {\bf 4}, 201 (2002).

\bibitem{as:eberhart04}
 M. E. Eberhart and D. P. Clougherty, {\it Looking for design in
 materials design}, Nat. Mater. {\bf 3}, 659 (2004).

\bibitem{as:maranas}
 A. Lehmann and C. D. Maranas, {\it Molecular design using quantum
 chemical calculations for property estimation}, Ind. Eng. Chem. Res.
 {\bf 43}, 3419 (2004).

\bibitem{as:caflisch}
 T. Zhou, D. Huang, and A. Caflisch, {\it Quantum mechanical methods for
 drug design}, Curr. Top. Med. Chem. {\bf 10}, 33 (2010).

\bibitem{as:carter}
 M. Pavone, A. M. Ritzmann, and E. A. Carter, {\it Quantum-mechanics-design
 principles for solid oxide fuel cell cathode materials}, Energy
 Environ. Sci. {\bf 4}, 4933 (2011).

\bibitem{as:nori}
 N. Lambert, Y.-N. Chen, Y.-C. Cheng, C.-M. Li, G.-Y. Chen, and F. Nori,
 {\it Quantum biology} Nat. Phys. {\bf 9}, 10 (2012).

\bibitem{as:schulten}
 Th. Ritz, S. Adem, and K. Schulten, {\it A model for photoreceptor-based
 magnetoreception in birds}, Biophys. J. {\bf 78}, 707 (2000).

\bibitem{as:fleming07}
 G. S. Engel, T. R. Calhoun, E. L. Read, T.-K. Ahn, T. Man\v cal,
 Y.-C. Cheng, R. E. Bankenship, and G. R. Fleming, {\it Evidence for
 wavelike energy transfer through quantum coherence in
 photosynthetic systems}, Nature {\bf 446}, 782 (2007).

\bibitem{as:fleming10}
 M. Sarovar, A. Ishizaki, G. R. Fleming, and K. B. Whaley, {\it Quantum
 entanglement in photosynthetic light-harvesting complexes},
 Nat. Phys. {\bf 6}, 462 (2010).

\bibitem{as:birula}
 I. Bialynicki-Birula, M. Cieplak, and J. Kamisnki, {\it Theory of
 Quanta} (Oxford University Press, Oxford, 1992).

\bibitem{as:furth}
 R. F\"urth, {\it \"Uber einige Beziehungen zwischen klassischer
 Statistik und Quantenmechanik}, Z. Physik {\bf 81}, 143 (1933).

\bibitem{as:comisar}
 G. G. Comisar, {\it Brownian-motion model of nonrelativistic quantum
 mechanics}, Phys. Rev. {\bf 138}, B1332 (1965).

\bibitem{as:landau1}
 L. Landau, {\it The theory of superfluidity of helium II},
 J. Phys. (USSR) {\bf 5}, 71 (1941).

\bibitem{as:london}
 F. London, {\it Planck's constant and low temperature transfer},
 Rev. Mod. Phys. {\bf 17}, 310 (1945).

\bibitem{as:broglie-1}
 L. de Broglie, {\it Remarques sur la nouvelle Mec\'anique
 ondulatoire}, Compt. Rend. {\bf 183}, 273 (1926).

\bibitem{as:broglie-2}
 L. de Broglie, {\it Sur la possibilit\'e de relier les ph\'enom\`enes
 d'interf\'erence et de diffraction \`a la th\'eorie des quanta de
 lumi\`ere}, Compt. Rend. {\bf 183}, 447 (1926).

\bibitem{as:broglie-3}
 L. de Broglie, {\it Sur le r\^ole des ondes continues $\Psi$ en
 M\'ecanique ondulatoire}, Compt. Rend. {\bf 185}, 380 (1927).

\bibitem{as:swinney}
 T. H. Solomon, E. R. Weeks, and H. L. Swinney, {\it Observation of
 anomalous diffusion and L\'{e}vy flights in a two-dimensional rotating
 flow}, Phys. Rev. Lett. {\bf 71}, 3975 (1993).

\bibitem{as:gilreath}
 J. C. Sommerer, H.-C. Ku, and H. E. Gilreath, {\it Experimental
 evidence for chaotic scattering in a fluid wake},
 Phys. Rev. Lett. {\bf 77}, 5055 (1996).

\bibitem{as:sanz:AJP:2012}
 A. S. Sanz and S. Miret-Art\'es, {\it Quantum phase analysis with
 quantum trajectories: A step towards the creation of a Bohmian
 thinking}, Am. J. Phys. {\bf 80}, 525 (2012).

\bibitem{as:sanz:JPhysConfSer:2012}
 A. S. Sanz, {\it Quantumness beyond quantum mechanics}, J. Phys.:
 Conf. Ser. {\bf 361}, 012016 (2012).

\bibitem{as:couder:Nature:2005}
 Y. Couder, S. Proti\`ere, E. Fort, and A. Boudaoud,
 {\it Dynamical phenomena: Walking and orbiting droplets}, Nature
 {\bf 437}, 208 (2005)

\bibitem{as:couder:JFluidMech:2006}
 S. Proti\`ere, A. Boudaoud, Y. Couder, {\it Particle-wave
 association on a fluid interface}, J. Fluid. Mech. {\bf 554}, 85 (2006).

\bibitem{as:couder:PRL:2006}
 Y. Couder and E. Fort, {\it Single-particle diffraction and
 interference at a macroscopic scale}, Phys. Rev. Lett. {\bf 97},
 154101 (2006)

\bibitem{as:couder:PhysFluid:2006}
 S. Proti\`ere and Y. Couder, {\it Orbital motion of bouncing drops},
 Phys. Fluid. {\bf18}, 091114 (2006).

\bibitem{as:couder:PRL:2009}
 A. Eddi, E. Fort, F. Moisy, and Y. Couder, {\it Unpredictable
 tunneling of a classical wave-particle association}, Phys. Rev. Lett.
 {\bf 102}, 240401 (2009).

\bibitem{as:couder:PNAS:2010}
 E. Fort, A. Eddi, A. Boudaoud, J. Moukhtar, and Y. Couder,
 {\it Path-memory induced quantization of classical orbits},
 Proc. Natl. Acad. Sci. USA {\bf 108}, 17515 (2010).

\bibitem{as:couder:EPN:2010}
 Y. Couder, A. Boudaoud, S. Proti\`ere, and E. Fort,
 {\it Walking droplets: A form of wave-particle duality at macroscopic
 scale?}, Europhys. News {\bf 41}, 14 (2010).

\bibitem{as:bush:PNAS:2010}
 J. W. M. Bush, {\it Quantum mechanics writ large}, Proc. Natl. Acad.
 Sci. USA {\bf 107}, 17455 (2010).

\bibitem{as:bush:PRE:2013}
 D. M. Harris, J. Moukhtar, E. Fort, Y. Couder, and J. W. M. Bush,
 {\it Wavelike statistics from pilot-wave dynamics in a circular corral},
 Phys. Rev. E {\bf 88}, 011001(R) (2013).
 year      = {2013},

\bibitem{as:bush:ARFM:2015}
 J. W. M. Bush, {\it Pilot-wave hydrodynamics}, Annu. Rev. Fluid Mech.
 {\bf 47}, 269 (2015).

\bibitem{as:zurek-bk}
 W. H. Zurek and J. A. Wheeler, {\it Quantum Theory of Measurement}
 (Princeton University Press, Princeton, NJ, 1983).

\bibitem{as:scully}
 B. G. Englert, M. O. Scully, G. S\"ussmann, and H. Walther, {\it Surrealistic Bohm
 trajectories}, Z. Naturforsch. {\bf 47a}, 1175 (1992).

\bibitem{as:courant-hilbert-bk}
 R. Courant and D. Hilbert, {\it Methods of Mathematical Physics}
 (Wiley-Interscience, New York, 1966).

\bibitem{as:weyl}
 H. Weyl, {\it Quantenmechanik und Gruppentheorie}, Z. Phys. {\bf 46}, 1 (1927).

\bibitem{as:wigner}
 E. P. Wigner, {\it On the quantum correction for thermodynamic equilibrium},
 Phys. Rev. {\bf 40}, 749 (1932).

\bibitem{as:moyal}
 J. E. Moyal, {\it Quantum mechanics as a statistical theory}, Math. Proc. Cambridge
 Philos. Soc. {\bf 45}, 99 (1949).

\bibitem{as:lopreore-1}
 R. E. Wyatt, {\it Quantum wavepacket dynamics with trajectories:
 wavefunction synthesis along quantum paths},
 Chem. Phys. Lett. {\bf 313}, 189 (1999).

\bibitem{as:lopreore-2}
 C. L. Lopreore and R. E. Wyatt, {\it Quantum wave packet dynamics with
 trajectories}, Phys. Rev. Lett. {\bf 82}, 5190 (1999).

\bibitem{as:lopreore-3}
 C. L. Lopreore and R. E. Wyatt, {\it Quantum wave packet dynamics with
 trajectories: reflections on a downhill ramp potential},
 Chem. Phys. Lett. {\bf 325}, 73 (2000).

\bibitem{as:tannor:JCP:2006}
 Y. Goldfarb, I. Degani, and D. J. Tannor, {\it Bohmian mechanics with complex action:
 A new trajectory-based formulation of quantum mechanics}, J. Chem. Phys. {\bf 125},
 231103 (2006).

\bibitem{as:tannor:JCP-2:2007}
 Y. Goldfarb and D. J. Tannor, {\it Interference in Bohmian mechanics with complex
 action}, J. Chem. Phys. {\bf 127}, 161101 (2007).

\bibitem{as:tannor:JPCA:2007}
 Y. Goldfarb, J. Schiff, and D. J. Tannor, {\it Unified derivation of Bohmian methods
 and the incorporation of interference effects}, J. Chem. Phys. {\bf 111}, 10416 (2007).

\bibitem{as:tannor:CP:2007}
 Y. Goldfarb, I. Degani, and D. J. Tannor, {\it Semiclassical approximation with zero
 velocity trajectories}, Chem. Phys. {\bf 338}, 106 (2007).

\bibitem{as:chou:JCP:2006}
 C.-C. Chou and R. E. Wyatt, {\it Computational method for the quantum Hamilton-Jacobi
 equation: Bound states in one dimension}, J. Chem. Phys. {\bf 125}, 174103 (2006).

\bibitem{as:chou:JCP-1:2008}
 C.-C. Chou and R. E. Wyatt, {\it Quantum trajectories in complex space: One-dimensional
 stationary scattering problems}, J. Chem. Phys. {\bf 128}, 154106 (2008).

\bibitem{as:chou:JCP-3:2008}
 C.-C. Chou and R. E. Wyatt, {\it Quantum streamlines within the complex quantum
 Hamilton-Jacobi formalism}, J. Chem. Phys. {\bf 129}, 124113 (2008).

\bibitem{as:chou:JCP:2010}
 C.-C. Chou and R. E. Wyatt, {\it Complex-extended Bohmian mechanics}, J. Chem. Phys.
 {\bf 132}, 134102 (2010).

\bibitem{as:bell1}
 J. S. Bell, {\it Speakable and Unspeakable in Quantum Mechanics}
 (Cambridge University Press, Cambridge, 1987).

\bibitem{as:BOA}
 M. Born and J. R. Oppenheimer, {\it Zur Quantentheorie der Molekeln},
 Ann. Physik {\bf 84}, 457 (1927).

\bibitem{as:bransden}
 B. H. Bransden and C. J. Joachain, {\it Physics of Atoms and
 Molecules} (Longman Scientific \& Technical, Essex, 1983).

\bibitem{as:levine}
 I. N. Levine, {\it Quantum Chemistry}
 (Prentice Hall, Upper Saddle River, NJ, 2000), 5th Ed.

\bibitem{as:szabo}
 A. Szabo and N. S. Ostlund, {\it Modern Quantum Chemistry}
 (Dover, Publications, Mineola, NY, 1996).

\bibitem{as:fulde}
 P. Fulde, {\it Electron Correlations in Molecules and Solids}
 (Springer, Berlin, 2002), 3rd Ed.

\bibitem{as:koch}
 W. Koch and M. C. Holthausen, {\it A Chemist's Guide to Density
 Functional Theory} (Wiley-VCH, Weinheim, 2001), 2nd Ed.

\bibitem{as:martin}
 R. M. Martin, {\it Electronic Structure. Basic Theory and Practical
 Methods} (Cambridge University Press, Cambridge, 2004).

\bibitem{as:kaplan}
 I. G. Kaplan, {\it Intermolecular Interactions: Physical Picture,
 Computational Methods and Model Potentials}
 (John Wiley \& Sons, Chichester, 2006).

\bibitem{as:zhang}
 J. Z. H. Zhang, {\it Theory and Application of Quantum Molecular
 Dynamics} (World Scientific, Singapore, 1999).

\bibitem{as:tannor}
 D. J. Tannor, {\it Introduction to Quantum Mechanics}
 (University Science Books, Sausalito, CA, 2006).

\bibitem{as:louisell}
 W. H. Louisell, {\it Quantum Statistical Properties of Radiation}
 (John Wiley \& Sons, New York, 1973).

\bibitem{as:mcquarrie}
 D. A. McQuarrie, {\it Statistical Mechanics}
 (Harper \& Row, New York, 1976).

\bibitem{as:weiss}
 U. Weiss, {\it Quantum Dissipative Systems}
 (World Scientific, Singapore, 2008), 3rd Ed.

\bibitem{as:breuer}
 H.-P. Breuer and F. Petruccione, {\it The Theory of Open Quantum Systems}
 (Oxford University Press, New York, 2002).

\bibitem{as:carmichael}
 H. Carmichael, {\it An Open Systems Approach to Quantum Optics}
 (Springer-Berlag, Berlin, 1993).

\bibitem{as:percival}
 I. Percival, {\it Quantum State Diffusion}
 (Cambridge University Press, Cambridge, 1998).

\bibitem{as:nakai-2002}
 H. Nakai, {\it Simultaneous determination of nuclear and electronic wave functions without Born–Oppenheimer
 approximation: Ab initio NO+MO/HF theory}, Int. J. Quantum Chem. {\bf 86}, 511 (2002).

\bibitem{as:nakai-2007}
 H. Nakai, {\it Nuclear orbital plus molecular orbital theory: Simultaneous determination of nuclear and
 electronic wave functions without Born–Oppenheimer approximation},
 Int. J. Quantum Chem. {\bf 107}, 2849 (2007).

\bibitem{as:nakai-2016}
 L. Veis, J. Vi\v s\v n\'ak, H. Nishizawa, H. Nakai, and J. Pittner, {\it Quantum chemistry beyond
 Born–Oppenheimer approximation on a quantum computer: A simulated phase estimation study},
 Int. J. Quantum Chem. {\bf 116}, 1328 (2016).

\bibitem{as:headgordon}
 M. Head-Gordon, {\it Quantum chemistry and molecular processes},
 J. Phys. Chem. {\bf 100}, 13213 (1996).

\bibitem{as:shaik}
 S. S. Shaik and P. C. Hiberty, {\it A Chemist's Guide to Valence Bond
 Theory} (Wiley-Interscience, New Jersey, 2007).

\bibitem{as:pauling1}
 L. Pauling, {\it The application of the quantum mechanics to the
 structure of the hydrogen molecule and hydrogen molecule-ion and
 to related problems}, Chem. Rev. {\bf 5}, 173 (1928).

\bibitem{as:pauling2}
 L. Pauling, {\it The nature of the chemical bond. Application of
 results obtained from the quantum mechanics and from a theory of
 paramagnetic susceptibility to the structure of molecules},
 J. Am. Chem. Soc. {\bf 53}, 1367 (1931).

\bibitem{as:blum}
 K. Blum, {\it Density Matrix Theory and Applications}
 (Plenum Press, New York, 1981).

\bibitem{as:thomas}
 L. H. Thomas, {\it The calculation of atomic fields},
 Proc. Cambridge Phil. Soc. {\bf 23}, 542 (1927).

\bibitem{as:fermi}
 E. Fermi, {\it Un metodo statistico per la determinazione di alcune
 priopriet\`a dell'atomo},
 Rend. Accad. Naz. Lincei {\bf 6}, 602 (1927).

\bibitem{as:marques}
 M. A. L. Marques and E. K. U. Gross, {\it Time-dependent density
 functional theory}, Annu. Rev. Phys. Chem. {\bf 55}, 427 (2004).

\bibitem{as:botti}
 S. Botti, A. Schindlmayr, R. Del Sole, and L. Reining,
 {\it Time-dependent density-functional theory for extended systems},
 Rep. Prog. Phys. {\bf 70}, 357 (2007).

\bibitem{as:nakatsuji-1}
 H. Nakatsuji, {\it Equation for direct determination of density
 matrix}, Phys. Rev. A {\bf 14}, 41 (1976).

\bibitem{as:nakatsuji-2}
 H. Nakatsuji and K. Yasuda, {\it  Direct determination of the
 quantum-mechanical density matrix using the density equation},
 Phys. Rev. Lett. {\bf 76}, 1039 (1996).

\bibitem{as:nakatsuji-3}
 K. Yasuda and H. Nakatsuji, {\it  Direct determination of the
 quantum-mechanical density matrix using the density equation. 2},
 Phys. Rev. A {\bf 56}, 2648 (1997).

\bibitem{as:valdemoro-1}
 C. Valdemoro, {\it Approximating the 2nd-order reduced density-matrix
 in terms of the 1st-order one}, Phys. Rev. A {\bf 45}, 4462 (1992).

\bibitem{as:valdemoro-2}
 C. Valdemoro, {\it Contracting and calculating traces over the
 $N$-electron space: Two powerful tools for obtaining averages},
 Int. J. Quantum Chem. {\bf 60}, 131 (1996).

\bibitem{as:valdemoro-3}
 C. Valdemoro, {\it Electron Correlation and Reduced Density Matrices},
 in {\it Correlation and Localization}, P. R. Surjan (Ed.)
 (Springer, Berlin, 1999), pp.~187-200.

\bibitem{as:valdemoro-4}
 C. Valdemoro, L. M. Tel, D. R. Alcoba, E. P\'erez-Romero, and
 F. J. Casquero, {\it Some basic properties of the correlation
 matrices}, Int. J. Quantum Chem. {\bf 90}, 1555 (2002).

\bibitem{as:valdemoro-5}
 D. R. Alcoba and C. Valdemoro, {\it  Spin structure and properties
 of the correlation matrices corresponding to pure spin states:
 Controlling the S-representability of these matrices},
 Int. J. Quantum Chem. {\bf 102}, 629 (2005).

\bibitem{as:rdm-reviews-1}
 E. R. Davidson, {\it Reduced Density Matrices in Quantum Chemistry}
 (Academic Press, New York, 1976).

\bibitem{as:rdm-reviews-2}
 A. J. Coleman and V. I. Yukalov, {\it Reduced Density Matrices:
 Coulson's Challenge} (Springer-Verlag, New York, 2000).

\bibitem{as:rdm-reviews-3}
 J. Cioslowski (Ed.), {\it Many-Electron Densities and Reduced Density
 Matrices} (Kluwer, Dordrecht, 2000).

\bibitem{as:mazziotti-1}
 D. A. Mazziotti, {\it Contracted Schr\"odinger equation: Determining
 quantum energies and two-particle density matrices without wave
 functions}, Phys. Rev. A {\bf 57}, 4219 (1998).

\bibitem{as:mazziotti-2}
 D. A. Mazziotti, {\it Pursuit of $N$-representability for the
 contracted Schr\"odinger equation through density-matrix
 reconstruction}, Phys. Rev. A {\bf 60}, 3618 (1999).

\bibitem{as:mazziotti-3}
 D. A. Mazziotti, {\it Anti-Hermitian contracted Schr\"odinger
 equation: Direct determination of the two-electron reduced density
 matrices of many-electron molecules},
 Phys. Rev. Lett. {\bf 97}, 143002 (2006).

\bibitem{as:mazziotti-4}
 D. A. Mazziotti (Ed.), {\it Reduced-Density-Matrix Mechanics with
 Applications to Many-Electron Atoms and Molecules},
 Adv. Chem. Phys. {\bf 134} (Wiley, New York, 2007).

\bibitem{as:sherwood}
 P. Sherwood, {\it Hybrid Quantum Mechanics/Molecular Mechanics
 Approaches}, in {\it Modern Methods and Algorithms of Quantum
 Chemistry}, J. Grotendorst (Ed.) (John von Neumann Institute for
 Computing, J\"ulich, 2000), 2nd Ed.

\bibitem{as:born1}
 M. Born and E. Wolf, {\it Principles of Optics}
 (Cambridge University Press, Cambridge, 1999), 7th Ed.

\bibitem{as:sanz1}
 R. Guantes, A. S. Sanz, J. Margalef-Roig, and S. Miret-Art\'es,
 {\it Atom-surface diffraction: a trajectory description},
 {\it Surf. Sci. Rep.} {\bf 53}, 199 (2004).

\bibitem{as:sanz2}
 A.S. Sanz and S. Miret-Art\'es, {\it Selective adsorption resonances:
 Quantum and stochastic approaches}, Phys. Rep. {\bf 451}, 37 (2007).

\bibitem{as:morse}
 P. M. Morse and H. Feshbach, {\it Methods of Theoretical Physics}
 (McGraw Hill, New York, 1953).

\bibitem{as:razavy}
 M. Razavy, {\it Quantum Theory of Tunneling}
 (World Scientific, New Jersey, 2003).

\bibitem{as:ankerhold}
 J. Ankerhold, {\it Quantum Tunneling in Complex Systems}
 in {\it Springer Tracts in Modern Physics} {\bf 224}
 (Springer, New York, 2007).

\bibitem{as:JWKB-1}
 H. Jeffreys, {\it On certain approximate solutions of linear
 differential equations of the second order},
 Proc. London Math. Soc. (2), {\bf 23}, 428 (1925).

\bibitem{as:JWKB-2}
 G. Wentzel, {\it Eine Verallgemeinerung der Quantenbedingungen f\"ur
 die Zwecke der Wellenmechanik}, Z. Physik {\bf 38}, 518 (1926).

\bibitem{as:JWKB-3}
 H. A. Kramers, {\it Wellenmechanik und halbzahlige Quantisierung},
 Z. Physik {\bf 39}, 828 (1926).

\bibitem{as:JWKB-4}
 L. Brillouin, {\it La m\'ecanique ondulatoire de Schr\"odinger;
 une m\'ethode g\'en\'erale de r\'esolution par approximations
 successives}, Comptes Rendus {\bf 183}, 24 (1926).

\bibitem{as:JWKB-5}
 L. Brillouin, {\it Sur un type g\`en\'eral de probl\`emes, permettant
 la s\'eparation des variables dans la m\'ecanique ondulatoire de
 Schr\"odinger}, Comptes Rendus {\bf 183}, 270 (1926).


\bibitem{as:cerjan}
 C. Cerjan (Ed.), {\it Numerical Grid Methods and Their Application
 to Schr\"odinger's Equation} (Kluwer Academic, Amsterdam, 1993).

\bibitem{as:giordano}
 N. J. Giordano, {\it Computational Physics}
 (Prentice Hall, Upper Saddle River, NJ, 1997).

\bibitem{as:gutzwiller}
 M. C. Gutzwiller, {\it Chaos in Classical and Quantum Mechanics}
 (Sprigner-Verlag, New York, 1990).

\bibitem{as:heller}
 E. J. Heller, {\it Bound-state eigenfunctions of classically chaotic
 Hamiltonian systems: Scars of periodic orbits},
 Phys. Rev. Lett. {\bf 53}, 1515 (1984).

\bibitem{as:QST-1}
 R. G. Newton, {\it Scattering Theory of Waves and Particles}
 (McGraw-Hill, New York, 1966).

\bibitem{as:QST-2}
 M. S. Child, {\it Molecular Collision Theory}
 (Dover, New York, 1996).

\bibitem{as:bowman-1}
 C. J. Ray and J. M. Bowman, {\it Quasiclassical trajectory
 calculations of He-LiF(001) diffraction scattering},
 J. Chem. Phys. {\bf 63}, 5231 (1975).

\bibitem{as:bowman-2}
 C. J. Ray and J. M. Bowman, {\it Quasiclassical studies of rigid
 rotor-solid surface diffraction scattering},
 J. Chem. Phys. {\bf 66}, 1122 (1976).

\bibitem{as:scatt-1}
 D. Far\'{\i}as, C. D\'{\i}az, P. Rivi\`ere, H. F. Busnego, P. Nieto,
 M. F. Somers, G. J. Kroes, A. Salin, and F. Mart\'{\i}n, {\it In-plane
 and out-of-plane diffraction of H$_2$ from metal surfaces},
 Phys. Rev. Lett. {\bf 93}, 246104 (2004).

\bibitem{as:scatt-2}
 C. D\'{\i}az, M. F. Somers, G. J. Kroes, H. F. Busnego, A. Salin,
 and F. Mart\'{\i}n, {\it Quantum and classical dynamics of H$_2$
 scattering from Pd(111) at off-normal incidence},
 Phys. Rev. B {\bf 72}, 035401 (2005).

\bibitem{as:scatt-3}
 C. D\'{\i}az, H. F. Busnego, P. Rivi\`ere, D. Far\'{\i}as, P. Nieto,
 M. F. Somers, G. J. Kroes, A. Salin, and F. Mart\'{\i}n, {\it A
 classical dynamics method for H$_2$ diffraction from metal surfaces},
 J. Chem. Phys. {\bf 122}, 154706 (2005).

\bibitem{as:delgado-1}
 G. Delgado-Barrio, P. Villarreal, P. Mareca, and G. Albelda,
 {\it Three-dimensional vibrational predissociation of the van der
 Waals complex He$\cdots$I$_2$($B$). A quasiclassical study},
 J. Chem. Phys. {\bf 78}, 280 (1983).

\bibitem{as:delgado-2}
 J. Rubayo-Soneira, A. Gar\'{\i}a-Vela, G. Delgado-Barrio, and
 P. Villarreal, {\it Vibrational predissociation of I$_2$-Ne. A
 quasiclassical dynamical study},
 Chem. Phys. Lett. {\bf 243}, 236 (1995).

\bibitem{as:delgado-3}
 A. Gar\'{\i}a-Vela, J. Rubayo-Soneira, G. Delgado-Barrio, and
 P. Villarreal, {\it Quasiclassical dynamics of the I$_2$-Ne$_2$
 vibrational predissociation: A comparison with experiment},
 J. Chem. Phys. {\bf 104}, 8405 (1996).

\bibitem{as:leforestier}
 C. Leforestier, R. H. Bisseling, C. Cerjan, M. D. Feit, R. Friesner,
 A. Guldberg, A. Hammerich, G. Jolicard, W. Karrlein, H.-D. Meyer,
 N. Lipkin, 0. Roncero, and R. Kosloff, {\it A comparison of different
 propagation schemes for the time dependent Schr\"odinger equation},
 J. Comp. Phys. {\bf 94}, 59 (1991).

\bibitem{as:sanz4}
 A. S. Sanz, F. Borondo, and S. Miret-Art\'es, {\it Particle
 diffraction studied using quantum trajectories},
 J. Phys.: Condens. Matter {\bf 14}, 6109 (2002).

\bibitem{as:caldeira-1}
 V. B. Magalinskii, {\it Dynamical model in the theory of the Brownian
 motion}, Sov. Phys. JETP-USSR {\bf 9}, 1381 (1959).

\bibitem{as:caldeira-2}
 A. O. Caldeira and A. J. Leggett, {\it Influence of dissipation on
 quantum tunneling in macroscopic systems},
 Phys. Rev. Lett. {\bf 46}, 211 (1981).

\bibitem{as:caldeira-3}
 A. O. Caldeira and A. J. Leggett, {\it Quantum tunneling in a
 dissipative system}, Ann. Phys. (NY) {\bf 149}, 374 (1983).

\bibitem{as:redfield-1}
 R. K. Wangsness and F. Bloch, {\it The dynamical theory of nuclear
 induction}, Phys. Rev. {\bf 89}, 728 (1953).

\bibitem{as:redfield-2}
 A. G. Redfield, {\it On the theory of relaxation processes},
 IBM J. Res. Develop. {\bf 1}, 19 (1957).

\bibitem{as:redfield-3}
 A. G. Redfield, {\it The theory of relaxation processes},
 Adv. Magn. Reson. {\bf 1}, 1 (1965).

\bibitem{as:redfield-4}
 W. T. Pollard, A. K. Felts, and R. A. Friesner, {\it The Redfield
 Equation in Condensed-Phase Quantum Dynamics},
 Adv. Chem. Phys. {\bf 93}, 77 (1996).

\bibitem{as:ford-1}
 G. W. Ford, M. Kac, and P. Mazur, {\it Statistical mechanics of
 assemblies of coupled oscillators}, J. Math. Phys. {\bf 6}, 504 (1965).

\bibitem{as:ford-2}
 G. W. Ford and M. Kac, {\it On the quantum Langevin equation},
 J. Stat. Phys. {\bf 46}, 803 (1987).

\bibitem{as:feynman-1}
 R. P. Feynman, {\it Space-time approach to non-relativitic quantum
 mechanics}, Rev. Mod. Phys. {\bf 20}, 367 (1948).

\bibitem{as:feynman-2}
 R. P. Feynman and A. R. Hibbs, {\it Quantum Mechanics and Path
 Integrals} (McGraw-Hill, NewYork, 1965).

\bibitem{as:feynman-3}
 R. P. Feynman, {\it Statistical Mechanics}
 (W. A. Benjamin, Reading, MA, 1972).

\bibitem{as:schulman}
 L. S. Schulman, {\it Techniques and Applications of Path Integrals}
 (Wiley, New York, 1981).

\bibitem{as:sanz:CJC:2014}
 A. S. Sanz, {\it Effective Markovian description of decoherence in
 bound systems}, Can. J. Chem., {\bf 92}, 168 (2014).

\bibitem{as:billing}
 G. D. Billing, {\it Classical path method in inelastic and reactive
 scattering}, Int. Rev. Phys. Chem. {\bf 13}, 309 (1994).

\bibitem{as:tully}
 J. C. Tully, {\it Nonadiabatic molecular-dynamics},
 Int. J. Quantum Chem. {\bf 25}, 299 (1991).

\bibitem{as:miller1-1}
 W. H. Miller, {\it Semiclassical theory of atom-diatom collisions:
 Path integrals and the classical $S$ matrix},
 J. Chem. Phys. {\bf 53}, 1949 (1970).

\bibitem{as:miller1-2}
 W. H. Miller, {\it Classical $S$ matrix: Numerical application to
 inelastic collisions}, J. Chem. Phys. {\bf 53}, 3578 (1970).

\bibitem{as:miller1-3}
 W. H. Miller, {\it Quantum and semiclassical theory of chemical
 reaction rates}, Faraday Discuss. {\bf 110}, 1 (1998).

\bibitem{as:miller1-4}
 W. H. Miller, {\it The semiclassical initial value representation: A
 potentially practical way for adding quantum effects to classical
 molecular dynamics simulations},
 J. Phys. Chem. A {\bf 105}, 2942 (2001).

\bibitem{as:miller2-1}
 W. H. Miller and B. M. D. D. Jansen op de Haar, {\it A new basis set
 method for quantum scattering calculations},
 J. Chem. Phys. {\bf 86}, 6213 (1987).

\bibitem{as:miller2-2}
 J. Z. H. Zhang, S.-I. Chu, and W. H. Miller, {\it Quantum scattering
 via the $S$-matrix version of the Kohn variational principle},
 J. Chem. Phys. {\bf 88}, 6233 (1988).

\bibitem{as:others-ivr-1}
 M. A. Sep\'ulveda and F. Grossmann, {\it Time-dependent semiclassical
 mechanics}, Adv. Chem. Phys. {\bf 96}, 191 (1996).

\bibitem{as:others-ivr-2}
 N. Makri, {\it Quantum dissipative dynamics: A numerically exact
 methodology}, J. Phys. Chem. A {\bf 102}, 4414 (1998).

\bibitem{as:pollak-1}
 J. Ankerhold, M. Salteer, and E. Pollak, {\it A study of the
 semiclassical initial value representation at short times},
 J. Chem. Phys. {\bf 116}, 5925 (2002).

\bibitem{as:pollak-2}
 E. Pollak and J. Shao, {\it Systematic improvement of initial value
 representations of the semiclassical propagator},
 J. Phys. Chem. A {\bf 107}, 7112 (2003).

\bibitem{as:pollak-3}
 E. Pollak and S. Miret-Art\'es, {\it Thawed semiclassical IVR
 propagators}, J. Phys. A {\bf 37}, 9669 (2004).

\bibitem{as:kapral}
 S. Nielsen, R. Kapral, and G. Ciccotti, {\it Non-Adiabatic dynamics in
 mixed quantum-classical systems}, J. Stat. Phys. {\bf 101}, 225 (2000).

\bibitem{as:carpar-1}
 R. Car and M. Parrinello, {\it Unified qpproach for molecular dynamics
 and density-functional theory}, Phys. Rev. Lett. {\bf 55}, 2471 (1985).

\bibitem{as:carpar-2}
T. D. Kuhne, M. Krack, F. R. Mohamed, and M. Parrinello, {\it
 Efficient and accurate Car-Parrinello-like approach to
 Born-Oppenheimer molecular dynamics},
 Phys. Rev. Lett. {\bf 98}, 066401 (2007).

\bibitem{as:nelson:pr:1966}
 E. Nelson, {\it Derivation of the Schr\"odinger equation from Newtonian mechanics},
 Phys. Rev. {\bf 150}, 1079 (1966).

\bibitem{as:bell2}
 J. S. Bell, {\it On the problem of hidden variables in quantum
 mechanics}, Rev. Mod. Phys. {\bf 38}, 447 (1966).

\bibitem{as:sanz3}
 A. S. Sanz and S. Miret-Art\'es, {\it Aspects of nonlocality from a
 quantum trajectory perspective: A WKB approach to Bohmian mechanics},
 Chem. Phys. Lett. {\bf 445}, 350 (2007).

\bibitem{as:schro-1}
 E. Schr\"odinger, {\it Discussion of probability relations between
 separated systems}, Proc. Cambridge Phil. Soc. {\bf 31}, 555 (1935).

\bibitem{as:schro-2}
 E. Schr\"odinger, {\it Probability relations between separated
 systems}, Proc. Cambridge Phil. Soc. {\bf 32}, 446 (1936).

\bibitem{as:sanz:AnnPhys:2013}
 A. S. Sanz and S. Miret-Art\'es, {\it On the unique mapping relationship between
 initial and final quantum states}, Ann. Phys. {\bf 339}, 11 (2013).

\bibitem{as:bohm-bk}
 D. Bohm, {\it Quantum Theory} (Prentice-Hall, New York, 1951).

\bibitem{as:schiff-bk}
 L. I. Schiff, {\it Quantum Mechanics} (McGraw-Hill, Singapore, 1968), 3rd Ed.

\bibitem{as:kocsis:Science:2011}
 S. Kocsis, B. Braverman, S. Ravets, M. J. Stevens, R. P. Mirin, L. K. Shalm and
 A. M. Steinberg, {\it Observing the average trajectories of single photons in a two-slit interferometer},
 Science, {\bf 332}, 1170 (2011).

\bibitem{as:sanz:AnnPhysPhoton:2010}
 A. S. Sanz, M. Davidovi\'c, M. Bo\v{z}i\'c, and S. Miret-Art\'es, {\it Understanding interference
 experiments with polarized light through photon trajectories}, Ann. Phys. {\bf 325}, 763 (2010).

\bibitem{as:sanz5}
 A. S. Sanz and S. Miret-Art\'es, {\it Atom-Surface Diffraction: A
 Quantum Trajectory Description}, in {\it Quantum Dynamics of Complex
 Molecular Systems}, D. A. Micha and I. Burghardt (Eds.)
 (Springer, Berlin, 2007).

\bibitem{as:fluidmech-1}
 P. M. Gerhardt, R. J. Gross, and J. I. Hochstein, {\it Fundamentals of
 Fluid Mechanics} (Addison-Wesley, New York, 1992).

\bibitem{as:fluidmech-2}
 J. H. Spurk, {\it Fluid Mechanics} (Springer-Verlag, Berlin, 1997).

\bibitem{as:benseny:EPJD:2014}
 A. Benseny, G. Albareda, A. S. Sanz, J. Mompart, and X. Oriols, {\it Applied Bohmian mechanics},
 Eur. Phys. J. D {\bf 68}, 286 (2014).

\bibitem{as:dewdney1-1}
 C. Dewdney, {\it Nonlocally correlated trajectories two-particle
 quantum mechanics}, Found. Phys. {\bf 18}, 867 (1988).

\bibitem{as:dewdney1-2}
 M. M. Lam and C. Dewdney, {\it Locality and nonlocality in correlated
 two-particle interferometry}, Phys. Lett. A {\bf 150}, 127 (1990).

\bibitem{as:marchildon}
 E. Guay and L. Marchildon, {\it Two-particle interference in standard
 and Bohmian quantum mechanics}, J. Phys. A {\bf 36}, 5617 (2003).

\bibitem{as:wyatt3-1}
 K. Na and R. E. Wyatt, {\it Quantum hydrodynamic analysis of
 decoherence: quantum trajectories and stress tensor},
 Phys. Lett. A {\bf 306}, 97 (2002).

\bibitem{as:wyatt3-2}
 K. Na and R. E. Wyatt, {\it Quantum hydrodynamic analysis of
 decoherence}, Phys. Scr. {\bf 67}, 169 (2003).

\bibitem{as:sanz6-1}
 A.S. Sanz and F. Borondo, {\it A quantum trajectory description of
 decoherence}, Eur. Phys. J. D {\bf 44}, 319 (2007).

\bibitem{as:sanz6-2}
 A.S. Sanz and F. Borondo, {\it Contextuality, decoherence and quantum
 trajectories}, Chem. Phys. Lett. {\bf 478}, 301 (2009).

\bibitem{as:oriols:PRL:2007}
 X. Oriols, {\it Quantum-trajectory approach to time-dependent
 transport in mesoscopic systems with electron-electron interactions},
 Phys. Rev. Lett. {\bf 98}, 066803 (2007).

\bibitem{as:aharonov:PRL:1988}
 Y. Aharonov, D. Z. Albert, and L. Vaidman, {\it How the result of a measurement of a component of the spin
 of a spin$-$$\frac{1}{2}$ particle can turn out to be 100}, Phys. Rev. Lett. {\bf 60}, 1351 (1988).

\bibitem{as:lundeen:Nature:2011}
 J. S. Lundeen, B. Sutherland, A. Patel, C. Stewart, and Ch. Bamber, {\it Direct measurement of the quantum
 wavefunction}, Nature {\bf 474}, 188 (2011).

\bibitem{as:wiseman}
 H. M. Wiseman, {\it Grounding Bohmian mechanics in weak values an bayesianism}, New J. Phys. {\bf 9},
 165 (2007).

\bibitem{as:hiley}
 B. J. Hiley, {\it Weak values: Approach through the Clifford and Moyal algebras}, J. Phys.: Conf. Ser.
 {\bf 361}, 012014 (2012).

\bibitem{as:hirschfelder2}
 J. O. Hirschfelder, {\it Quantum mechanical equations of change. I},
 J. Chem. Phys. {\bf 68}, 5151 (1978).

\bibitem{as:bader}
 R. F. W. Bader, {\it Quantum topology of molecular charge
 distributions. III. The mechanics of an atom in a molecule},
 J. Chem. Phys. {\bf 73}, 2871 (1980).

\bibitem{as:gomes-1}
 J. A. N. F. Gomes, {\it Delocalized magnetic currents in benzene},
 J. Chem. Phys. {\bf 78}, 3133 (1983).

\bibitem{as:gomes-2}
 J. A. N. F. Gomes, {\it Topological elements of the magnetically
 induced orbital current densities},
 J. Chem. Phys. {\bf 78}, 4585 (1983).

\bibitem{as:lazzeretti-1}
 P. Lazzeretti, {\it Ring currents},
 Prog. Nuc. Mag. Res. Spect. {\bf 36}, 1 (2000).

\bibitem{as:lazzeretti-2}
 S. Pelloni, F. Faglioni, R. Zanasi, and P. Lazzeretti, {\it Topology
 of magnetic-field-induced current-density field in diatropic
 monocyclic molecules}, Phys. Rev. A {\bf 74}, 012506 (2006).

\bibitem{as:lazzeretti-3}
 S. Pelloni, P. Lazzeretti, and R. Zanasi, {\it Spatial ring current
 model of the [2.2]paracyclophane molecule},
 J. Phys. Chem. A {\bf 111}, 3110 (2007).

\bibitem{as:lazzeretti-4}
 S. Pelloni, P. Lazzeretti, and R. Zanasi, {\it Topological models of
 magnetic field induced current density field in small molecules},
 Theor. Chem. Acc. {\bf 123}, 353 (2009).

\bibitem{as:lazzeretti-5}
 S. Pelloni and P. Lazzeretti, {\it Spatial ring current model for the
 prismane molecule}, J. Phys. Chem. A {\bf 112}, 5175 (2008).

\bibitem{as:lazzeretti-6}
 S. Pelloni and P. Lazzeretti, {\it Topology of magnetic-field induced
 electron current density in the cubane molecule},
 J. Chem. Phys. {\bf 128}, 194305 (2008).

\bibitem{as:lazzeretti-7}
 S. Pelloni and P. Lazzeretti, {\it Ring current models for acetylene
 and ethylene molecules}, Chem. Phys. {\bf 356}, 153 (2009).

\bibitem{as:lazzeretti-8}
 I. Garc\'{\i}a Cuesta, A. S\'anchez de Mer\'as, S. Pelloni, and
 P. Lazzeretti, {\it Understanding the ring current effects on magnetic
 shielding of hydrogen and carbon nuclei in naphthalene and anthracene},
 J. Comput. Chem. {\bf 30}, 551 (2009).

\bibitem{as:bloch}
 F. Bloch, {\it Bremsverm\"ogen von Atomen mit mehreren Elektronen},
 Z. Physik {\bf 81}, 363 (1933).

\bibitem{as:bartolotti1}
 L. J. Bartolotti and J.C. Mollmann, {\it 4th order time-dependent
 variation perturbation-theory based on the hydrodynamic analogy},
 Mol. Phys. {\bf 38}, 1359 (1979).

\bibitem{as:bartolotti2}
 L. J. Bartolotti, {\it Time-dependent extension of the
 Hohenberg-Kohn-Levy energy-density functional},
 Phys. Rev. A {\bf 24}, 1661 (1981)

\bibitem{as:bartolotti3}
 L. J. Bartolotti, {\it Time-dependent Kohn-Sham density-functional
 theory}, Phys. Rev. A {\bf 26}, 2243 (1982).

\bibitem{as:runge}
 E. Runge and E. K. U. Gross, {\it Density-functional theory for
 time-dependent systems}, Phys. Rev. Lett. {\bf 52}, 997 (1984).

\bibitem{as:deb}
 B. M. Deb and S. K. Ghosh, {\it Schr\"odinger fluid-dynamics of
 many-electron systems in a time-dependent density-functional
 framework}, J. Chem. Phys. {\bf 77}, 342 (1982).

\bibitem{as:deb-chat-1}
 B. M. Deb and P. K. Chattaraj, {\it How can density functional theory
 be excited from the ground state?},
 Proc. Indian Acad. Sci. {\bf 99}, 67 (1987).

\bibitem{as:deb-chat-2}
 B. M. Deb and P. K. Chattaraj, {\it Quantum fluid density functional
 theory of time-dependent phenomena -- Ion atom collisions}
 Chem. Phys. Lett. {\bf 148}, 550 (1988).

\bibitem{as:deb-chat-3}
 B. M. Deb and P. K. Chattaraj, {\it Density-functional and
 hydrodynamical approach to ion-atom collisions through a new
 generalized nonlinear Schr\"odinger equation},
 Phys. Rev. A {\bf 39}, 1696 (1989).

\bibitem{as:deb-chat-4}
 B. M. Deb, P. K. Chattaraj, and S. Mishra, {\it Time-dependent
 quantum-fluid density-functional study of high-energy proton-helium
 collisions}, Phys. Rev. A {\bf 43}, 1248 (1991).

\bibitem{as:dey-deb-1}
 B. Kr. Dey and B. M. Deb, {\it Time-dependent quantum fluid-dynamics
 of the photoionization of the he atom under an intense laser field},
 Int. J. Quant. Chem. {\bf 56}, 707 (1995).

\bibitem{as:dey-deb-2}
 B. Kr. Dey and B. M. Deb, {\it A theoretical study of the high-order
 harmonics of a 200 nm laser from H$^{-2}$ and HeH$^+$},
 Chem. Phys. Lett. {\bf 276}, 157 (1997).

\bibitem{as:dey-deb-3}
 B. Kr. Dey and B. M. Deb, {\it Stripped ion-helium atom collision
 dynamics within a time-dependent quantum fluid density functional
 theory}, Int. J. Quant. Chem. {\bf 67}, 251 (1998).

\bibitem{as:dey-deb-4}
 B. Kr. Dey and B. M. Deb, {\it Direct ab initio calculation of
 ground-state electronic energies and densities for atoms and molecules
 through a time-dependent single hydrodynamical equation},
 J. Chem. Phys. {\bf 110}, 6229 (1999).

\bibitem{as:march-1}
 G. P. Lawes and N. H. March, {\it Approximate differential-equation
 for calculating the electron-density in closed shell atoms and in
 molecules}, Phys. Scr. {\bf 21}, 402 (1980).

\bibitem{as:march-2}
 B. M. Deb and S.K. Ghosh, {\it New method for the direct calculation
 of electron-density in many-electron systems .1. application to
 closed-shell atoms}, Int. J. Quantum Chem. {\bf 23}, 1 (1983).

\bibitem{as:march-3}
 M. Levy, J. P. Pardew, and V. Sahni, {\it Exact differential equation
 for the density and ionization energy of a many-particle system},
 Phys. Rev. A {\bf 30}, 2745 (1984).

\bibitem{as:march-4}
 N. H. March, {\it The local potential determining the square root of
 the ground-state electron-density of atoms and molecules from the
 Schr\"odinger equation}, Phys. Lett. A {\bf 113}, 476 (1986).

\bibitem{as:march-5}
 G. Hunter, {\it The exact one-electron model of molecular-structure},
 Int. J. Quantum Chem. {\bf 29}, 197 (1986).

\bibitem{as:march-6}
 M. Levy and H. Ou-Yang, {\it Exact properties of the Pauli potential
 for the square root of the electron density and the kinetic energy
 functional}, Phys. Rev. A {\bf 38}, 625 (1988).

\bibitem{as:mcclendon}
 M. McClendon, {\it Real-space diffusion theory of multiparticle
 quantum systems}, Phys. Rev. A {\bf 38}, 5851 (1988).

\bibitem{as:sanz7}
 A. S. Sanz, X. Gim\'enez, J. M. Bofill, and S. Miret-Art\'es,
 {\it Time-dependent Density Functional Theory from a Bohmian
 Perspective}, in {\it Chemical Reactivity Theory},
 P. K. Chattaraj (Ed.) (Taylor \& Francis, New York, 2009).

\bibitem{as:kohn}
 W. Kohn and L. J. Sham, {\it Self-consistent equations including
 exchange and correlation effects}, Phys. Rev. {\bf 140}, A1133 (1965).

\bibitem{as:molmer:arxiv:2017}
 T. A. Elsayed, K. M\o lmer, and L. B. Madsen, {\it Solving the quantum many-body problem with Bohmian trajectories},
 preprint arXiv:1706.00818v1 (2017).

\bibitem{as:cederbaum:PhysRev:2008}
 O. E. Alon, A. I. Streltsov and L. S. Cederbaum, {\it Multiconfigurational time-dependent Hartree
 method for bosoms: Many-body dynamics of bosonic systems}, Phys. Rev. A {\bf 77}, 033613 (2008).

\bibitem{as:schmelcher:JPhysB:2017}
 V. Bolsinger, S. Kr\"onke and P. Schmelcher, {\it Beyond mean-field dynamics of ultra-cold bosonic
 atoms in higher dimensions: Facing the challenges with a multiconfigurational approach},
 J. Phys. B {\bf 50}, 034003 (2017).

\bibitem{as:meyer:PhysRep:2000}
 M. H. Beck, A. J\"ackle, G. Worth and H.-D. Meyer, {\it The multiconfiguration time-dependent Hartree
 (MCTDH) method: A high efficient algorithm for propagatin wavepackets},
 Phys. Rep. {\bf 324}, 1 (2000).

\bibitem{as:muellerbrown}
 K. M\"uller and L. D. Brown, {\it Location of saddle points and
 minimum energy paths by a constrained simplex optimization procedure},
 Theor. Chim. Acta {\bf 53}, 75 (1979).

\bibitem{as:bofill1}
 R. Crehuet and J. M. Bofill, {\it The reaction path intrinsic reaction
 coordinate method and the Hamilton-Jacobi theory},
 J. Chem. Phys. {\bf 122}, 234105 (2005).

\bibitem{as:bofill2}
 A. Aguilar-Mogas, X. Gim\'enez, and J. M. Bofill, {\it Finding
 reaction paths using the potential energy as reaction coordinate},
 J. Chem. Phys. {\bf 128}, 104102 (2008).

\bibitem{as:quapp}
 W. Quapp, {\it Chemical reaction paths and calculus of variations},
 Theor. Chem. Account. {\bf 121}, 227 (2008).

\bibitem{as:sanz8}
 A. S. Sanz, X. Gim\'enez, J. M. Bofill, and S. Miret-Art\'es,
 {\it Understanding chemical reactions within a generalized
 Hamilton-Jacobi framework}, Chem. Phys. Lett. {\bf 478}, 89 (2009);
 Erratum Chem. Phys. Lett. {\bf 488}, 235 (2010).

\bibitem{as:sanz9}
 A. S. Sanz and S. Miret-Art\'es, {\it A trajectory-based understanding
 of quantum interference}, J. Phys. A {\bf 41}, 435303 (2008).

\bibitem{as:sanz10}
 A. S. Sanz and S. Miret-Art\'es, {\it Quantum trajectories in
 elastic atom-surface scattering: Threshold and selective adsorption
 resonances}, J. Chem. Phys. {\bf 122}, 014702 (2005).

\bibitem{as:dewdney2}
 C. Dewdney and B. J. Hiley, {\it A quantum potential description of
 one-dimensional time-dependent scattering from square barriers and
 square wells}, Found. Phys. {\bf 12}, 27 (1982).


\bibitem{as:marcus}
 R. A. Marcus, {\it On the analytical mechanics of chemical reactions.
 Quantum mechanics of linear collisions},
 J. Chem. Phys. {\bf 45}, 4493 (1966).

\bibitem{as:sanz13}
 A. S. Sanz, F. Borondo, and S. Miret-Art\'es, {\it Quantum
 trajectories in atom-surface scattering with single adsorbates:
 The role of quantum vortices}, J. Chem. Phys. {\bf 120}, 8794 (2004).

\bibitem{as:sanz15}
 A. S. Sanz, F. Borondo, and S. Miret-Art\'es, {\it Role of quantum
 vortices in atomic scattering from single adsorbates},
 Phys. Rev. B {\bf 69}, 115413 (2004).

\bibitem{as:toennies-1}
 W. Sch\"ollkopf and J. P. Toennies, {\it Nondestructive mass selection
 of small van der Waals clusters}, Science {\bf 266}, 1345 (1994).

\bibitem{as:toennies-2}
 R. E. Grisenti, W. Sch\"ollkopf, J. P. Toennies, G. C. Hegerfeldt,
 and T. K\"ohler, {\it Determination of atom-surface van der Waals
 potentials from transmission-grating diffraction intensities},
 Phys. Rev. Lett. {\bf 83}, 1755 (1999).

\bibitem{as:toennies-3}
 R. E. Grisenti, W. Sch\"ollkopf, J. P. Toennies, G. C. Hegerfeldt,
 T. K\"ohler, and M. Stoll, {\it Determination of the bond length and
 binding energy of the helium dimer by diffraction from a transmission
 grating}, Phys. Rev. Lett. {\bf 85}, 2284 (2000).

\bibitem{as:toennies-4}
 R. E. Grisenti, W. Sch\"ollkopf, J. P. Toennies, J. R. Manson,
 T. A. Savas, and H. I. Smith, {\it He-atom diffraction from
 nanostructure transmission gratings: The role of imperfections},
 Phys. Rev. A {\bf 61}, 033608 (2000).

\bibitem{as:sanz14}
 A.S. Sanz, F. Borondo, and S. Miret-Art\'es, {\it Causal trajectories
 description of atom diffraction by surfaces},
 Phys. Rev. B {\bf 61}, 7743 (2000).

\bibitem{decoh}
 A. S. Sanz, F. Borondo, and M. Bastiaans, {\it Loss of coherence in
 double-slit diffraction experiments},
 Phys. Rev. A. {\bf 71}, 42103 (2005).

\bibitem{as:sanz:AOP:2015-1}
 A. S. Sanz, M. Davidovi\'c, and M. Bo\v zi\'c, {\it Full quantum mechanical analysis
 of atomic three-grating Mach-Zehnder interferometry},
 Ann. Phys. {\bf 353}, 205 (2015).

\bibitem{as:sanz:AOP:2015-2}
 A. Luis and A. S. Sanz, {\it What dynamics can be expected for mixed states
 in two-slit experiments?}, Ann. Phys. {\bf 357}, 95 (2015).

\bibitem{as:prezhdo}
 O. V. Prezhdo and C. Brooksby, {\it Relationship between quantum
 decoherence times and solvation dynamics in condensed phase chemical
 systems}, Phys. Rev. Lett. {\bf 86}, 3215 (2001).

\bibitem{as:beswick-1}
 E. Gindensperger, C. Meier, and J. A. Beswick, {\it Mixing quantum and
 classical dynamics using Bohmian trajectories},
 J. Chem. Phys. {\bf 113}, 9369 (2000).

\bibitem{as:beswick-2}
 E. Gindensperger, C. Meier, and J. A. Beswick, {\it Quantum-classical
 dynamics including continuum states using quantum trajectories},
 J. Chem. Phys. {\bf 116}, 8 (2002).

\bibitem{as:beswick-3}
 E. Gindensperger, C. Meier, J. A. Beswick, and M.-C. Heitz,
 {\it Quantum-classical description of rotational diffractive
 scattering using Bohmian trajectories: Comparison with full quantum
 wave packet results}, J. Chem. Phys. {\bf 116}, 10051 (2002).

\bibitem{as:beswick-4}
 C. Meier and J. A. Beswick, {\it Femtosecond pump-probe spectroscopy
 of I$_2$ in a dense rare gas environment: A mixed quantum/classical
 study of vibrational decoherence},
 J. Chem. Phys. {\bf 121}, 4550 (2004).

\bibitem{as:garashchuk:CPL:2002}
 S. Garashchuk and V. Rassolov, {\it Semiclassical dynamics based on quantum trajectories},
 Chem. Phys. Lett. {\bf 364}, 562 (2002).

\bibitem{as:garashchuk:JCP:2004}
 S. Garashchuk and V. Rassolov, {\it Modified quantum trajectory dynamics using a mixed wave function representation},
 J. Chem. Phys. {\bf 121}, 8711 (2004).

\bibitem{as:garashchuk:RCC:2011}
  S. Garashchuk, V. Rassolov, and O. Prezhdo, {\it Semiclassical Bohmian dynamics},
  Rev. Comput. Chem. {\bf 27}, 287 (2011).

\bibitem{as:bittner:JCP:2003}
 E. R. Bittner, {\it Quantum initial value representation using approximate Bohmian trajectories},
J. Chem. Phys. {\bf 119}, 1358 (2003).

\bibitem{as:makri:JCP:2003}
 Y. Zhao and N. Makri, {\it Bohmian versus semiclassical description of interference phenomena},
 J. Chem. Phys. {\bf 119}, 60 (2003).

\bibitem{as:makri:JPCA-1:2004}
 J. Liu and N. Makri, {\it Forward-backward quantum dynamics for time correlation functions},
 J. Phys. Chem. A {\bf 108}, 806 (2004).

\bibitem{as:makri:JPCA-2:2004}
 J. Liu and N. Makri, {\it Monte Carlo Bohmian dynamics from trajectory stability properties},
 J. Phys. Chem. A {\bf 108}, 5408 (2004).

\bibitem{as:franco:JCP:2017}
 B. Gu and I. Franco, {\it Partial hydrodynamic representation of quantum molecular dynamics},
 J. Chem. Phys. {\bf 146}, 194104 (2017).

\bibitem{as:goldberg:AJP:1968}
 A. Goldberg, H. M. Schey, and J. L. Schwartz, {\it Computer-generated
 motion pictures of one-dimensional quantum-mechanical transmission and
 reflection phenomena}, Am. J. Phys. {\bf 36}, 454 (1968).

\bibitem{as:kohn-nobel}
 \url{http://nobelprize.org/nobel_prizes/chemistry/laureates/1998/kohn-lecture.pdf}

\end{thebibliography}
\end{document}